\documentclass{ws-book9x6}
\usepackage{ws-book-thm}
\usepackage{ws-book-har}

\usepackage[pdfpagelabels=false,colorlinks=true,allcolors=black]{hyperref}

\usepackage{bm,mathrsfs,isotope,color,aas_macros}
\usepackage{lipsum}

\begin{document}

\chapter[Hot Neutron Star Matter and Proto Neutron Stars]{Hot Neutron Star
  Matter and Proto Neutron Stars}\label{ch.FWeber}

\index[aindx]{Farrell, D.}  \index[aindx]{Alp, A.}  \index[aindx]{
  Spinella, W.}  \index[aindx]{Weber, F.}  \index[aindx]{Malfatti, G.}
\index[aindx]{Orsaria, M.G.}  \index[aindx]{Ranea-Sandoval, I.F.}

\author[Delaney Farrell et al.]{Delaney Farrell$^{\dagger}$, Aksel Alp$^{\ddagger}$}
\address{Department of Physics, San Diego State University \\ 5500
  Campanile Drive, San Diego, CA 92182, USA\\ Email:
$^{\dagger}$dfarrell@sdsu.edu, $^{\ddagger}$aalp@sdsu.edu}

\centerline{Fridolin Weber} \address{Department of Physics, San Diego
  State University \\ 5500 Campanile Drive, San Diego, CA 92182, USA
  \\ Center for Astrophysics and Space Sciences \\ University of
  California at San Diego \\ La Jolla, CA 92093, USA \\ Email:
  fweber@sdsu.edu, fweber@ucsd.edu}

\centerline{William Spinella} \address{Department of Physical
  Sciences, Irvine Valley College \\ Irvine Center Drive, Irvine, CA
  92618, USA \\ Email: wspinella@ivc.edu}

\centerline{Germ{\'a}n Malfatti$^*$, Milva G. Orsaria$^{**}$}
\address{Grupo de Gravitaci\'on, Astrof\'isica y Cosmolog\'ia
  \\ Facultad de Ciencias Astron{\'o}micas y Geof{\'i}sicas
  \\ Universidad Nacional de La Plata \\ Paseo del Bosque S/N, La
  Plata (1900), Argentina\\ Email: $^*$gmalfatti@fcaglp.unlp.edu.ar,
  $^{**}$morsaria@fcaglp.unlp.edu.ar}

\centerline{Ignacio F. Ranea-Sandoval} \address{Grupo de
  Gravitaci\'on, Astrof\'isica y Cosmolog\'ia \\ Facultad de Ciencias
  Astron{\'o}micas y Geof{\'i}sicas \\ Universidad Nacional de La
  Plata \\ Paseo del Bosque S/N, La Plata (1900), Argentina \\ Consejo
  Nacional de Investigaciones Cient{\'{i}}ficas y T{\'{e}}cnicas
  (CONICET) \\ Godoy Cruz 2290, Buenos Aires (1425),
  Argentina\\ Email: iranea@fcaglp.unlp.edu.ar}

\begin{minipage}[!]{\textwidth}  
{\bf Abstract:} In this chapter, we investigate the structure and
composition of hot neutron star matter and proto-neutron stars.  Such
objects are made of baryonic matter that is several times denser than
atomic nuclei and tens of thousands of times hotter than the matter in
the core of our Sun.  The relativistic finite-temperature Green
function formalism is used to formulate the expressions that determine
the properties of such matter in the framework of the
density-dependent mean field approach. Three different sets of nuclear
parametrizations are used to solve the many-body equations and to
determine the models for the equation of state of ultra-hot and dense
stellar matter.  The meson-baryon coupling schemes and the role of the
$\Delta(1232)$ baryon in proto-neutron star matter are investigated in
great detail.  Using the non-local three-flavor Nambu–-Jona-Lasinio
model to describe quark matter, the hadron-quark composition of dense
baryonic matter at zero temperature is discussed.  General
relativistic models of non-rotating as well as rotating proto-neutron
stars are presented in part two of our study.  \\ \\ {\bf Keywords:}
proto-neutron stars; neutron stars; nuclear equation of state; nuclear
field theory; quark matter.
\end{minipage}

\tableofcontents


\renewcommand{\thefootnote}{\arabic{footnote}}

\section{Introduction}\label{sec:Intro}

Within a few million years after a massive star ($\gtrsim 8\,
M_\odot$) is born, its core undergoes nuclear fusion reactions that
will result in a dense, heavy iron center. Up until the formation of
an iron core, the massive star has been supported from collapsing by
the energy released from fusing lighter elements into iron and
electron degeneracy pressure
\citep{Mezzacappa05:a,Janka2012:a,Foglizzo:2016,Burrows21:a}. When an
iron core is formed, the fusion processes and subsequent energy cease;
at this point, the star can no longer support its mass against the
force of gravity and will begin to rapidly collapse in the span of
just a few milliseconds. At this moment, the core's temperature
skyrockets and the density surpasses the point of electron degeneracy,
sparking the formation of neutrons through electron capture,
\begin{eqnarray}
  p + e^- \rightarrow n^0 + \nu_e \, , 
  \end{eqnarray}
where $p$, a proton, and $e^-$, an electron, combine to form a
neutron, $n^0$, and an electron neutrino, $\nu_e$. These neutrinos are
released carrying large quantities of energy, contracting the core
further. The density of the core increases until it reaches nuclear
density (baryon number density of around $0.16~ {\rm fm}^{-3}$, mass
density of or $2.65 \times 10^{14}~{\rm g/cm}^3$), where nucleon
degeneracy pressure halts the collapse. Parts of the core surpassing
nuclear density, like the inner most part of the core, will rebound to
create a shock wave as the exterior core layers are expelled. Over the
next tens of seconds, the shock wave reverses the inward trajectory of
the collapsing stellar material as it moves through the stellar
envelope, partially cooling the extreme temperature and contracting
the material it passes through. The shock wave alone does not possess
enough energy to pass through the entire stellar envelope and complete
the supernova explosion; the shock wave is revived by the massive
quantities of neutrinos created alongside neutrons. While most
neutrinos are expelled, some remain trapped behind the shock wave,
increasing the pressure and pushing the wave outward
\citep{CamelioEvol:2017}. This portion of the star's collapse is
referred to as the Kelvin-Helmholtz phase, and the contracting core
during this phase is called a proto-neutron star (PNS)
\index{Proto-neutron star}
\citep{Prakash:1997,Pons:1999ApJ,CamelioEvol:2017}. Depending on the
final mass of the core after the short-lived life of a PNS, a black
hole or neutron star (NS) is left behind. This chapter will focus on
the structure and evolution of the compact stellar
objects (proto-neutron stars) produced by the collapse of massive
($8\, M_\odot$ to around $20\, M_\odot$) stars.

The macroscopic evolution of the PNS during the Kelvin-Helmholtz
phase, where a hot, lepton-rich PNS turns into a cold, deleptonized
neutron star \index{Neutron star}
\citep{Weber:1999book,Weber:2005PPNP,Sedrakian:2007PPNP,Becker:2009neutron,
  Glendenning:book2012, Rezzolla:2019physics,Orsaria.2019}, is
dependent on the microphysical ingredients of the star, the equation
of state (EOS) of the dense matter comprising the core, and neutrino
opacity \citep{Pons:1999ApJ}. Immediately (in a matter of 0.1 to 0.5
seconds) following the core bounce during a massive star's collapse
and just prior to the Kelvin-Helmholtz phase, the PNS radius rapidly
decreases from over 150~km to less than 20~km as pressure decreases as
a result of neutrinos being released from the outer envelope of the
star \citep{Pons:1999ApJ}. While the star's original matter rapidly
compresses, the supernova's shock causes accretion which results in a
substantial increase in mass and total neutrino emission.  These
conditions make it so the copious amount of neutrinos cannot escape
freely, and instead diffuse over the course of about a minute (the
deleptonization stage) while a large fraction of the gravitational
binding energy is released during the contraction of the stellar
envelope \citep{Foglizzo:2016}. After this minute-long period,
neutrinos can escape freely, and the PNS enters a cooling stage where
the entropy steadily decreases \citep{Pons:1999ApJ}. The completion of
the deleptonization and cooling stages signifying the end of the
Kelvin-Helmholtz phase and the beginning of the life of a neutron
star.

\begin{figure}[htb] 
\centering
\includegraphics[width=11.0cm]{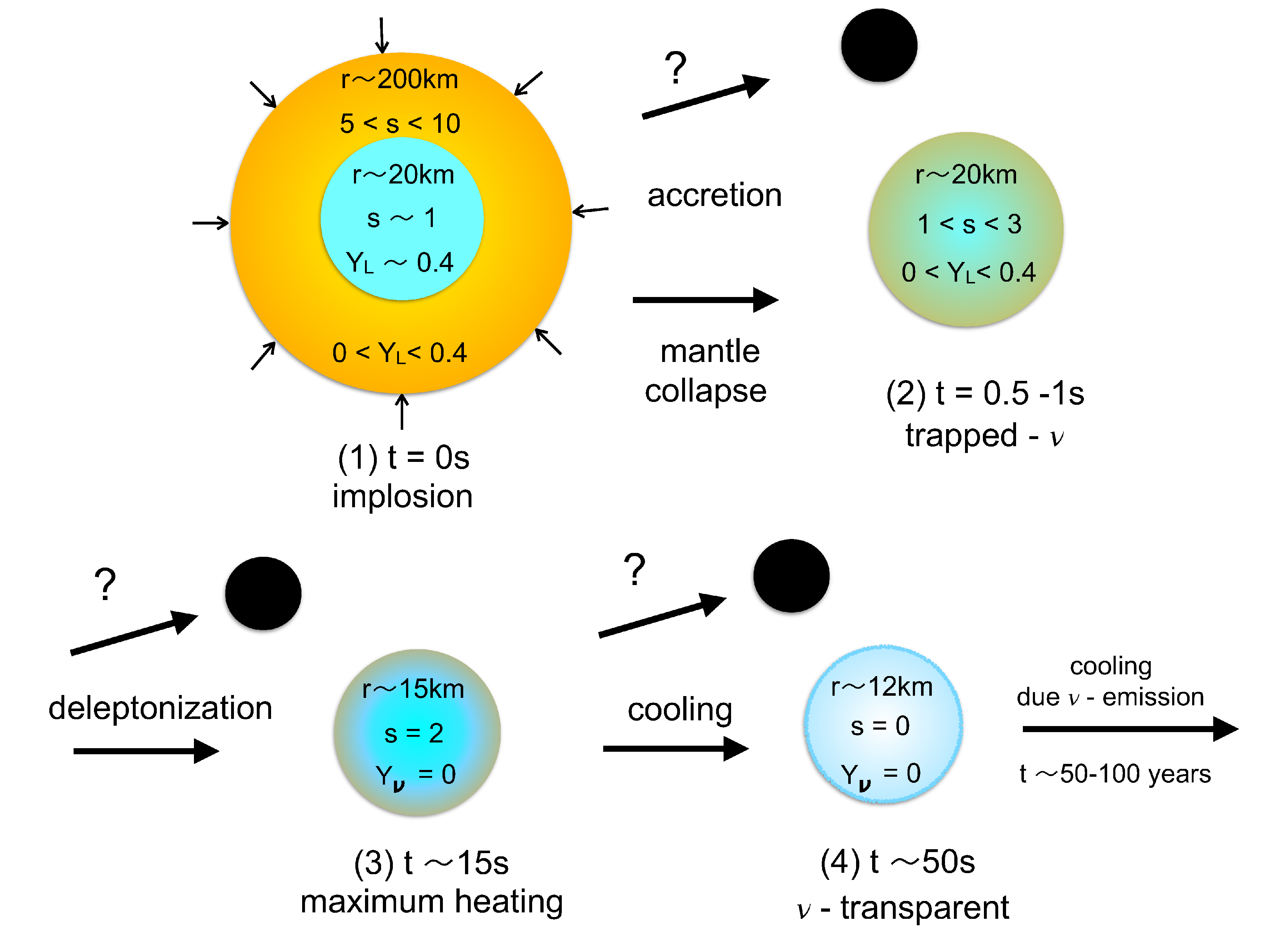}
\caption{Schematic illustration of different temporal stages in
  the evolution of proto-neutron stars to neutron stars
  \cite{Prakash:1997}. They are characterized by different values of
  entropy ($s$) and lepton number ($Y_L$). The formation of black
  holes (solid black spheres) is possible during different
  evolutionary stages, depending on the interplay between gravity and
  pressure. The transition of a hot PNS to a cold NS takes less than a
  minute. During the first few hundred years NSs cool quickly via
  neutrino emission from the core. Photon emission becomes the
  dominant cooling mechanism thereafter  \cite{PAGE2006497}.}
    \label{fig:stellar_evolution}
\end{figure}

The different stages in the evolution of hot proto-neutron stars to
cold neutron stars, as described above, are schematically illustrated
in Fig.~\ref{fig:stellar_evolution}. Proto-neutron stars are the
compact remnants produced at the end of the evolution of
intermediate-mass stars with masses of $M\gtrsim 10 M_\odot$ (see
\citet{mariani2020estrellas}, and references therein).  Their
structure and composition passes through different physical stages
within just a few seconds (see, for example,
\citet{Prakash:1997,Pons:1999ApJ}).  Stars with $M\gtrsim 10 M_\odot$
are known to evolve in a complex fashion via nuclear burning.  At the
end of their lives, when most of the nuclear fuel has been consumed
and massive cores of Fe (or O-Ne/Mg) have been built up, gravitational
collapse occurs. During this phase (stage ``1'' in
\fref{fig:stellar_evolution}) a rebound of the outer mantle of the
star occurs.  The core is surrounded by a mantle characterized by low
density but high entropy of the matter.  The mantle extends for around
200~km and is stable until it explodes due to the aforementioned
rebound. At this point, two different evolutionary tracks of the star
are possible, which essentially depend on how powerful the explosion
was.  If the explosion was not strong enough to deleptonize the outer
mantle, continued accretion of matter onto the star would give way to
the formation of a black hole. The other alternative is that the star
explodes successfully as a supernova (i.e., the mantle collapses and
accretion of matter becomes less important), giving birth to a hot PNS
where neutrinos are trapped in the stellar core (stage ``2'' in
\fref{fig:stellar_evolution}).  During the next stage of evolution,
the star begins to rapidly lose neutrinos. This leads to a reduction
of the pressure due to deleptonization, which would be followed by the
formation of a black hole if the gravitational pull on the matter
overcomes the pressure provided by the matter. If this does not
happen, the star will continue to deleptonize itself as it is being
heated-up by the Joule effect of the escaping neutrinos (stage ``3''
in \fref{fig:stellar_evolution}).  It is assumed that the maximum
heating of the star occurs immediately after the neutrinos have left
the star. Continued cooling via neutrino emission from the stellar
core (stage ``4'' in \fref{fig:stellar_evolution})
\cite{Malfatti:2020eyh} quickly reduces the star's temperature to just
a few MeV or less \cite{PAGE2006497}. At such temperatures the matter
in the core can be described by a cold nuclear EOS and the
corresponding star is referred to as a NS.

Understanding the physics behind a core collapse supernova and the
subsequent formation of a PNS has been of interest in the particle and
astrophysics communities for decades (see
\citet{Mezzacappa05:a,Janka2012:a,Burrows21:a}, and references
therein). The well documented explosion of a type II supernova in the
Large Magellanic Cloud in 1987 (SN1987a) was the first supernova event
of this kind that could be studied in detail.  Nineteen neutrinos have
been detected from this event, which may be too few to provide a
significant constraint on our understanding of the particle
composition and physics of the supernova, but do provide an important
milestone for these types of events. Since then, physicists have made
great strides using numerical models to simulate supernova explosions
\citep{CamelioEvol:2017}. More difficult to describe through numerical
codes is the lifespan of a PNS, but recent efforts as in
\citep{NeutrinoSignal:2010PRL,Fischer:2010AA,CamelioEvol:2017} have
been able to more accurately describe the quasi-stationary evolution
of a PNS.

In this book chapter we investigate the structure and composition of
(hot) proto-neutron stars. In part one of the paper, we introduce the
field-theoretic lagrangian that is used to compute models for the EOS
of the matter in the cores of such stars.  The relativistic mean field
approach is used to describe the interactions among nucleons mediated
by scalar, vector and iso-vector mesons. In our calculations, we focus
on the density-dependent SWL, DD2, and GM1L nuclear models
\citep{Typel:2009sy,Spinella2017:thesis, Malfatti:2019PRC100,
  Spinella:2020WSBook}. All three models account for the presence of
hyperons as well as of $\Delta$ baryons in hot and dense matter.  The
possible existence of deconfined quarks in such matter will be briefly
discussed in this paper too.  Investigations of this topic have also
been carried out by \citep{steiner2000quark, Shao_2011,
  Mariani:2016pcx, Malfatti:2019PRC100}.

The relativistic finite-temperature Green function formalism is used
to derive the equations that characterize ultra-hot and dense stellar
matter \citep{Weber:1999book}.  In part two of our study, the
properties of non-rotating as well as rotating proto-neutron stars are
studied by solving Einstein's field equation using the models for the
EOS derived in part one of the paper. The rotating stellar models are
computed fully self-consistently, as required by the general
relativistic expression for the Kepler (mass shedding) frequency.

\goodbreak
\section{Modeling Hot and Dense Neutron Star Matter}\label{sec:eos}

\subsection{The Non-Linear Nuclear Lagrangian}\label{ssec:lagrangian}

While a NS does get its namesake from the large quantities of neutrons
created in the core during its birth, a more accurate depiction of
interior composition a mixture of neutrons and protons whose electric
charge is balanced by leptons ($L = {e^-, ~\mu^-}$). Other particles
may also exist in the core like hyperons ($B = [n,
  ~p,~\Lambda,~\Sigma^\pm,~\Sigma^0,~\Xi^0,~ \Xi^-]$
\citep{Glendenning:1985a} and the electrically charged states of the
$\Delta$ isobar
\citep{Pandharipande:1971NPA,Sawyer:1972ApJ176,Boguta1982}. The
existence of these particles is made possible only if their Fermi
energies become large enough that existing baryon populations need to
be rearranged so that a lower energy state can be reached
\citep{Glendenning:1985a}. To understand how the baryons within the
core interact, we shall make use of the non-linear density-dependent
relativistic mean-field \index{DDRMF theory} (DDRMF) theory. This
theory describes the interactions between baryons in terms of meson
exchange. These mesons include a scalar meson ($\sigma$) which
describes attraction between baryons, a vector meson ($\omega$) which
describes repulsion, and an isovector meson ($\rho$) which is
important to describe the baryon-baryon interactions in isospin
asymmetric matter such as NS matter
\citep{Glendenning:1985a,Spinella2017:thesis}. Due to the pion's odd
parity, this particle does not contribute at the mean-field
description of dense matter.  The nuclear lagrangian \index{Nuclear
  lagrangian} of the theory is therefore given by (see also
\citet{Weber:1999book, Glendenning:book2012, Spinella:2020WSBook,
  Sedrakian:2022astrophysics})
\begin{eqnarray}
  \mathcal{L} =&& \sum_{B}\bar{\psi}_B \bigl[\gamma_\mu (
    i\partial^\mu - g_{\omega B} \omega^\mu - g_{\rho B}
    {\boldsymbol{\tau}} \cdot {\boldsymbol{\rho}}^\mu ) - (m_B -
    g_{\sigma B}\sigma) \bigr] \psi_B \nonumber \\
  &&+ \frac{1}{2}
  (\partial_\mu \sigma\partial^\mu \sigma - m_\sigma^2 \sigma^2) -
  \frac{1}{3} \tilde{b}_\sigma m_N (g_{\sigma N} \sigma)^3 -
  \frac{1}{4} \tilde{c}_\sigma (g_{\sigma N} \sigma)^4    \label{eq:Blag} \\
  &&- \frac{1}{4}\omega_{\mu\nu} \omega^{\mu\nu} +
  \frac{1}{2}m_\omega^2\omega_\mu \omega^\mu + \frac{1}{2}m_\rho^2
       {\boldsymbol{\rho\,}}_\mu \cdot {\boldsymbol{\rho\,}}^\mu -
       \frac{1}{4} {\boldsymbol{\rho\,}}_{\mu\nu} \cdot
            {\boldsymbol{\rho\,}}^{\mu\nu} \, ,\nonumber
\end{eqnarray}
where $\psi_B$ stands for the various baryon fields, $g_{\sigma B}$,
$g_{\omega B}$ and $g_{\rho B}$ are (density dependent) meson-baryon
coupling constants, and $\tilde b_\sigma$ and $\tilde c_\sigma$ denote
two additional coupling parameters associated with non-linear (cubic
and quartic) self-interactions introduced by \citet{Boguta1977}.  The
density dependent coupling constants are given by
\citet{Typel:2018cap}
\begin{eqnarray}
    g_{i B}(n) = g_{i B}(n_0) a_{i} \left[ 1+b_{i}({n}/{n_0} +
      d_{i})^{2} \right] \left[ 1+c_{i}({n}/{n_0} + d_{i})^{2}
      \right]^{-1} ,
    \label{eq:CC1}
\end{eqnarray}
for $\sigma$ and $\omega$ mesons ($i=\sigma,\, \omega$),  and by
\begin{eqnarray}
    g_{\rho B}(n) = g_{\rho B}(n_0)\,\mathrm{exp} \left[\,-a_{\rho}
      \left( {n}/{n_0} - 1\right)\,\right] \, , 
    \label{eq:CC2}
\end{eqnarray}
for $\rho$ mesons.  Here the choice of parameters $a_i$, $b_i$, $c_i$, and
$d_i$ account for nuclear medium effects, and are fixed by the binding
energies, charge, and diffraction radii, spin-orbit splittings, and
the neutron skin thickness of finite nuclei.

The quantities $m_B$, $m_\sigma$, $m_\omega$, $m_\rho$ in
\eref{eq:Blag} denote the masses of baryons and mesons and $m_N$ is
the nucleon mass.  The quantity ${\boldsymbol{\tau}} = (\tau_1, \tau_2,
\tau_3)$ are the Pauli isospin matrices.  The quantities
$\omega^{\mu\nu}$ and ${\boldsymbol{\rho\,}}^{\mu\nu}$ denote meson field
tensors, where $\omega^{\mu\nu} = \partial^\mu \omega^\nu -
\partial^\nu \omega^\mu$ and ${\boldsymbol{\rho\,}}^{\mu\nu} = \partial^\mu
        {\boldsymbol{\rho\,}}^{\nu} - \partial^\nu {\boldsymbol{\rho\,}}^{\mu}$.  The
        field equations of the baryon and meson fields are obtained by
        evaluating the Euler-Lagrange equations for the fields in
        \eref{eq:Blag}. This leads for the baryon fields to \index{Field equations}
\begin{equation} \label{eq:baryon-field-eqn}
  \left(i\gamma^{\mu}\partial_{\mu}-m_B\right) \psi_B = \bigl( g_{\omega
      B} \gamma^{\mu}\omega_{\mu} +\tfrac{1}{2}g_{\rho
      B}\gamma^{\mu}\boldsymbol{\tau}\cdot\boldsymbol{\rho_{\mu}}
    -g_{\sigma B}\sigma \bigr) \psi_B \, .
\end{equation}
The field equation of the scalar $\sigma$-meson is given by
\begin{equation} \label{eq:sigma-field-eqn}
  \begin{aligned}
    \left(\partial^{\mu}\partial_{\mu}+m^2_{\sigma}\right)\sigma =&
    \sum_B g_{\sigma B} \bar\psi_B \psi_B -b_{\sigma}m_n g_{\sigma N}
    \left( g_{\sigma N} \sigma \right)^2 -c_{\sigma}g_{\sigma N}
    \left(g_{\sigma N} \sigma\right)^3 \, ,
  \end{aligned}
\end{equation}
and the field equations of the vector mesons have the form
\begin{eqnarray} 
    \partial^{\mu}\omega_{\mu\nu} + m_{\omega}^2 \omega_{\nu} &=&
    \sum_B g_{\omega B} \bar{\psi}_B \gamma_{\nu}\psi_B \, ,
    \label{eq:omega-field-eqn} \\
  \partial^{\mu}\boldsymbol{\rho}_{\mu\nu}+m_{\rho}^2
  \boldsymbol\rho_{\nu} &=& \sum_B g_{\rho B} \bar\psi_B
  \boldsymbol\tau \gamma_{\nu}\psi_B \, .
   \label{eq:rho-field-eqn}
\end{eqnarray}
In relativistic mean-field approximation, the field equations
(\ref{eq:baryon-field-eqn}) through (\ref{eq:rho-field-eqn}) become
\begin{eqnarray}
m_{\sigma}^2 \bar{\sigma} &=& \sum_{B} g_{\sigma B} n_B^s -
\tilde{b}_{\sigma} \, m_N\,g_{\sigma N} (g_{\sigma N}\bar{\sigma})^2
 - \tilde{c}_{\sigma} \, g_{\sigma N} \, (g_{\sigma N}
\bar{\sigma})^3 \, , \label{eq:sigmafeq} \\ m_{\omega}^2 \bar{\omega} &=& \sum_{B}
g_{\omega B} n_{B}\, , \label{eq:omegafeq} \\ m_{\rho}^2\bar{\rho} &=& \sum_{B}g_{\rho
  B} I_{3B} n_{B} \, , \label{eq:rhofeq}
\end{eqnarray}
where $I_{3B}$ is the 3-component of isospin and $n_{B}^s$ and $n_{B}$
are the scalar and particle number densities for each baryon $B$. The
latter are given by
\begin{eqnarray}
  n_B &=&  \langle \psi_B^\dagger(x) \, \psi_B(x) \rangle \, , 
  \label{eq:nB1} \\
  n^s_B &=& \langle \bar\psi_B(x) \, \psi_B(x)\rangle \, ,
  \label{eq:nsB1}
\end{eqnarray}
respectively, where $\psi_B^\dagger$ denotes the conjugate Dirac
spinor and $\bar\psi \equiv \psi_B^\dagger \gamma^0$ stands for the
adjoint Dirac spinor. To keep the notation to a minimum, we use the
definition $x\equiv (x^0, \boldsymbol{x})$.

\subsection{Baryonic Field Theory at Finite Density and Temperature}

To calculate the densities (\ref{eq:nB1}) and (\ref{eq:nsB1}) for NS
matter at finite temperature, we use the finite-temperature Green
function formalism.  The starting point is the spectral function
representation of the two-point Green function \index{Two-point Green
  function} given by \citep{Dolan1974,Weber:1999book}
\begin{eqnarray}
g^B(p^0, \boldsymbol{p})=&&\int\! d\omega \, \frac{a^B(\omega , \boldsymbol{p})}
{\omega -(p^0 - \mu_B) (1+i\eta)} \label{eq:gBp} \\ &&- 2 i \pi \,
{\rm sign}(p^0-\mu_B) \; \frac{1}{\exp(|p^0 - \mu_B|/T) + 1} \;
a^B(p^0-\mu_B, \boldsymbol{p}) \, ,\nonumber
\end{eqnarray}
where $\mu_B$ denotes the chemical potential \index{Chemical
  potential} of a baryon of type $B$ and $a^B$ stands for the spectral
function of that baryon ($\eta >0$ and infinitesimally small). The
spin and isospin dependences of $g^B$ and $a^B$ are not shown
explicitly.   The spectral function is obtained by
  evaluating
\begin{eqnarray}
  a^B(\omega,\boldsymbol{p}) = \frac{1}{2 i \pi} \left( \tilde{
    g}^B(\omega+ i\eta,\boldsymbol{p}) - \tilde{g}^B(\omega-
  i\eta,\boldsymbol{p}) \right) \, .
\label{eq:aB.tildeg}
\end{eqnarray}
Here $\tilde g^B$ denotes the analytically continued two-point Green
function, which obeys the analytically continued \index{Dyson
  equation} Dyson equation,
  \begin{eqnarray}
    \left( \gamma^0 (z+\mu_B) - \boldsymbol{\gamma} \cdot
    \boldsymbol{p} - m_B - \tilde \Sigma^B(z,\boldsymbol{p}) \right)
    \tilde{g}^B(z,\boldsymbol{p}) = - 1 \, .
\label{eq:Dyson}
\end{eqnarray}

The spectral function \index{Spectral function} has a
scalar, vector and a time-like contribution generally written as
\citep{Weber:1999book}
\begin{equation}
  a^B (\boldsymbol{p}) = a_S^B(\boldsymbol{p}) + \boldsymbol{\gamma}
  \cdot \boldsymbol{\hat{p}} \, a_V^B(\boldsymbol{p}) +\gamma^0
  a_0^B(\boldsymbol{p}) \, ,
  \label{eq:a.B}
\end{equation}
where $a_S^B(\boldsymbol{p}) = m^*_B/(2 E^*_B(\boldsymbol{p}))$,
$a^B_V(\boldsymbol{p}) = - |\boldsymbol{p}|/(2
E^*_B(\boldsymbol{p}))$, and $a_0^B(\boldsymbol{p}) =1/2$.  For
thermally excited anti-baryon states one has
\begin{equation}
\bar a^B (\boldsymbol{p}) = \bar a_S^B(\boldsymbol{p}) +
\boldsymbol{\gamma} \cdot \boldsymbol{\hat{p}} \, \bar
a^B_V(\boldsymbol{p}) +\gamma^0 \bar a_0^B(\boldsymbol{p}) \, ,
\label{eq:a.barB}
\end{equation}
where $\bar a_S^B(\boldsymbol{p}) = - m^*_B/(2
E^*_B(\boldsymbol{p}))$, $\bar a^B_V(\boldsymbol{p}) =
|\boldsymbol{p}|/(2 E^*_B(\boldsymbol{p}))$, and
$a_0^B(\boldsymbol{p}) =1/2$.  The effective single-baryon energy,
$E^*_B$ and effective baryon mass, $m^*_B$, are given by
\begin{eqnarray}
E^*_B(\boldsymbol{p}) = \sqrt{\boldsymbol{p}^2 + {m^*_B}^2}  \label{eq:Estar}
\end{eqnarray}
and 
\begin{eqnarray}
m^*_B = m_B - g_{\sigma B} \bar\sigma\, , \label{eq:mstar}
\end{eqnarray}
respectively.

In terms of the two-point Green function, the expression for the
baryon number density (\ref{eq:nB1}) becomes
\begin{equation}
  n_B = i \; {\rm Tr} \, \gamma^0 \int\! d^3 \boldsymbol{x} \left(
  g^B(x,x^+) + g^B(x,x^-) \right) \, ,
  \label{eq:nBx}
\end{equation}
where the trace is to be taken over the spin and isospin matrix
indices. Transformation of \eref{eq:nBx} to momentum space leads to
\begin{equation}
  n_B = i \; {\rm Tr} \, \gamma^0 \int\! \frac{d^4 p}{(2 \pi)^4}
  \left( {\rm e}^{i \eta p^0} + {\rm e}^{-i \eta p^0} \right) g^B(p) \, .
  \label{eq:nBp4}
\end{equation}
Next we note that
\begin{eqnarray}
\int\! \frac{d^4 p}{(2\pi)^4} \, {\rm e}^{i \eta p^0} p^0 g^B(p^0,
\boldsymbol{p}) = -i \int\! \frac{d^3 \boldsymbol{p}}{(2\pi)^3} \,
a^B(\boldsymbol{p}) \omega_B(\boldsymbol{p}) f_{B^-}(\boldsymbol{p})
\, ,
       \label{eq:cint1}
\end{eqnarray}
and
\begin{eqnarray}
\int\! \frac{d^4 p}{(2\pi)^4} \, {\rm e}^{-i \eta p^0} p^0 g^B(p^0,
\boldsymbol{p}) = i \int\! \frac{d^3 \boldsymbol{p}}{(2\pi)^3} \, \bar{a}^B(\boldsymbol{p})
\bar{\omega}_B(\boldsymbol{p}) f_{B^+}(\boldsymbol{p}) \, ,
       \label{eq:cint2} 
\end{eqnarray}
which leads for \eref{eq:nBp4} to 
\begin{equation}
  n_B = \gamma_B \int\! \frac{d^3 \boldsymbol{p}}{(2\pi)^3} \left(
  f_{B^-}(\boldsymbol{p}) - f_{B^+}(\boldsymbol{p}) \right) \, ,
  \label{eq:nBfp}
\end{equation}
where $\gamma_B \equiv (2 J_B+1)$ accounts for the spin-degeneracy.
The quantities $f_{B^\pm}$ in \eref{eq:nBfp} denote Fermi-Dirac
\index{Fermi-Dirac distribution} distribution functions given by
\begin{equation}
  f_{B^-}(\boldsymbol{p}) = \frac{1}{ {\rm e}^{(E^*_B(\boldsymbol{p}) - \mu^*_B)/T} +1  } \, ,
  \label{eq:FD1}
\end{equation}
and
\begin{equation}
  f_{B^+}(\boldsymbol{p}) = \frac{1}{ {\rm e}^{(- \bar{E}^*_B(\boldsymbol{p}) + \mu^*_B)/T} +1  } \, .
    \label{eq:FD2}
\end{equation}
The quantity $\mu^*_B$ in Eqs.~(\ref{eq:FD1}) and (\ref{eq:FD2}),
given by
\begin{equation}
  \mu_B^* = \mu_B - g_{\omega B} \bar{\omega} - g_{\rho B} \bar{\rho}
  I_{3B} - \tilde{R} \, ,
  \label{eq:mustar.B}
\end{equation}
defines the effective baryon chemical potential in terms of the
standard chemical potential and the mean-fields of $\sigma$ and $\rho$
mesons. The quantity $\tilde{R}$ is the rearrangement term given by
\citep{Fuchs:1995as,Spinella:2020WSBook}
\begin{eqnarray}
\tilde{R} =\sum_B&&\left( \frac{\partial g_{\omega B}(n)}{\partial
  n} n_B \bar{\omega} + \frac{\partial g_{\rho B}(n)}{\partial n}
I_{3B} n_B \bar{\rho} - \frac{\partial g_{\sigma B}(n)}{\partial n}
n_B^s \bar{\sigma}\right) \, .
\label{eq:rear}
\end{eqnarray}
This term is mandatory for thermodynamic consistency, proven with
the Hugenholtz-van Hove theorem that relates the total baryonic
pressure (which contains the rearrangement term) of a particle to its
chemical potential \citep{Hofmann:2001}. The expression of the total baryonic
pressure of the standard non-linear relativistic mean-field theory
therefore contains the additional term $n \tilde{R}$ (see
\eref{eq:eos.pD}).

The single-baryon energies, $\omega_B(\boldsymbol{p})$, are given in terms of
these meson fields plus the effective single-baryon energies, $E^*_B$,
according to the relation
\begin{eqnarray}
  \omega_B(\boldsymbol{p}) = g_{\omega B} \bar\omega + g_{\rho B} I_{3B}
  \bar\rho + E^*_B(\boldsymbol{p})
  \label{eq:omega.B} \, .
\end{eqnarray}
Similarly, the single-particle energies of thermally excited
anti-baryon states, $\bar\omega_B(\boldsymbol{p})$, are given by
\begin{eqnarray}
  \bar\omega_B(\boldsymbol{p}) = g_{\omega B} \bar\omega + g_{\rho B} I_{3B}
  \bar\rho - E^*_B(\boldsymbol{p}) \, .
  \label{eq:omega.barB}
\end{eqnarray}
From the above relations, one sees that for baryons
\begin{equation}
  \omega_B(\boldsymbol{p}) - \mu_B = E^*_B(\boldsymbol{p}) - \mu^*_B \, ,
  \label{eq:omega1}
\end{equation}
and for states outside the Fermi sea of anti-particles
\begin{equation}
  - \bar\omega_B(\boldsymbol{p}) + \mu_B = E^*_B(\boldsymbol{p}) + \mu^*_B \, .
  \label{eq:omega2}
\end{equation}
With these definitions, the traces in \eref{eq:nBp4} and in the
  expressions for the energy density and pressure to be discussed
  below can be calculated. In particular, one obtains
  \begin{eqnarray}
{\rm Tr}\, a^B = \gamma_B m^*_B/E^*_B\, , \quad  
{\rm Tr} \, \bar{a}^B = - \gamma_B m^*_B/E^*_B\, ,
\label{eq:traces.1}
\end{eqnarray}
\begin{eqnarray}
{\rm Tr} \, \gamma^0 a^B = \gamma_B \, , \quad  
{\rm Tr} \, \gamma^0 \bar{a}^B = -\gamma_B \, .
\label{eq:traces.2}
\end{eqnarray}

Next we turn to the scalar density, $n^s_B$, defined in
\eref{eq:nsB1}.  Expressed in terms of the two-point Green function,
\eref{eq:nsB1} reads
\begin{equation}
  n^s_B = i \;  {\rm Tr} \int\! d^3 \boldsymbol{x} \left( g^B(x,x^+) +
  g^B(x,x^-) \right) \, .
\label{eq:nsB2x}
\end{equation}
Transforming this expression to momentum space gives
\begin{equation}
  n^s_B = i \; {\rm Tr} \int\! \frac{d^4 p}{(2 \pi)^4} \left( {\rm
    e}^{i \eta p^0} + {\rm e}^{-i \eta p^0} \right) g^B(p) \, .
  \label{eq:nsBp4}
\end{equation}
By making use of \eref{eq:cint2}, the integration of $p^0$ can be
carried out analytically. The Green functions then get replaced by the
baryon spectral functions and the Fermi-Dirac distribution functions,
leading to the final result for the scalar density given by
\begin{equation}
  n^s_B = \gamma_B \int\! \frac{d^3 \boldsymbol{p}}{(2\pi)^3}
  \frac{m^*_B}{E^*_B(\boldsymbol{p})} \left( f_{B^-}(\boldsymbol{p}) - f_{B^+}(\boldsymbol{p})
  \right) \, .
  \label{eq:nsBfp}
\end{equation}

\section{Composition and EOS of Hot and Dense (Proto-) Neutron Star Matter}
\label{sec:comp.EOS}

The total energy density and pressure of the stellar matter are
calculated from the energy-momentum tensor \index{Energy-Momentum
  tensor}
\begin{eqnarray}
  T_{\mu\nu}(x) = g_{\mu\nu}\, \mathcal{L}(x) + \sum_B \,
  \frac{\partial\mathcal{L}(x)}{\partial\, \partial^\mu\psi_B(x)} \;
  \partial_\nu\psi_B(x) \, ,
\label{eq:Tmunu}
\end{eqnarray}
with the lagrangian $\mathcal{L}$ given by \eref{eq:Blag}. The energy
density and pressure are given by $\epsilon =\, \langle
  T^{00}\rangle$ and $P = \frac{1}{3} \sum_k \langle
    T^{kk}\rangle$, respectively. Using the Green function
formalism, the expression for the energy density is given by
\citep{Weber:1999book}
\begin{eqnarray} 
    \epsilon = &&\; i\, \sum_B\, {\rm Tr} \int\! \frac{d^4 p}{(2\pi)^4}
    \, \Bigl( {\rm e}^{i \eta p^0} +{\rm e}^{-i \eta p^0} \Bigr) \nonumber \\
    && ~~~~~~~~\times
    \Bigl( p^0 \gamma^0 - \frac{1}{2} \left( g_{\sigma B} \bar\sigma +
    \gamma^0 \left(g_{\omega B} \bar\omega + g_{\rho B} I_{3B}
    \bar\rho \right) \right)\Bigr) g^B(p) \nonumber \\ && - \frac{1}{6} \tilde
    b_{\sigma} m_N \bigl(g_{\sigma N} \sigma\bigr)^3 - \frac{1}{4}
    \tilde c_{\sigma} \bigl(g_{\sigma N}\sigma\bigr)^4 \, .
  \label{eq:eos.epsilonA}
\end{eqnarray}
The integration over $p^0$ in \eref{eq:eos.epsilonA} can be carried
out analytically via contour integration, which leads to
\begin{eqnarray}
\int\! \frac{d^4 p}{(2\pi)^4} \, {\rm e}^{i \eta p^0} g^B(p^0, \boldsymbol{p})
= -i \int\! \frac{d^3 \boldsymbol{p}}{(2\pi)^3} \, a^B(\boldsymbol{p}) f_{B^-}(\boldsymbol{p})
       \label{eq:cint3}
\end{eqnarray}
and 
\begin{eqnarray}
\int\! \frac{d^4 p}{(2\pi)^4} \, {\rm e}^{-i \eta p^0} g^B(p^0,
\boldsymbol{p}) = i \int\! \frac{d^3 \boldsymbol{p}}{(2\pi)^3} \, \bar{a}^B(\boldsymbol{p})
f_{B^+}(\boldsymbol{p}) \, .
       \label{eq:cint4} 
\end{eqnarray}
The energy density is then given as a momentum integral over
single-baryon energies, baryon spectral functions, and Fermi-Dirac
distribution functions, as shown below:
\begin{eqnarray} 
    \epsilon =&&\sum_B \, {\rm Tr} \int\! \frac{d^3 \boldsymbol{p}}{(2\pi)^3}
    \, \Bigl( \omega^B(\boldsymbol{p}) \gamma^0 a^B(\boldsymbol{p}) f_{B^-}(\boldsymbol{p}) -
    \bar\omega^B(\boldsymbol{p}) \gamma^0 \bar{a}^B(\boldsymbol{p})
     f_{B^+}(\boldsymbol{p}) \Bigr) \nonumber  \\ &&- \frac{1}{2}
    \sum_B \, {\rm Tr} \int\! \frac{d^3 \boldsymbol{p}}{(2\pi)^3} \, \Bigl(
    \left( - g_{\sigma B} \bar\sigma + \gamma^0 \left( g_{\omega B}
      \bar\omega + g_{\rho B} I_{3B} \bar\rho \right) \right) a^B(\boldsymbol{p})
    f_{B^-}(\boldsymbol{p}) \nonumber \\ && ~~~~~~~~~~~~~~~~~~~~~~~ - \bigl( - g_{\sigma
        B} \bar\sigma + \gamma^0 \left( g_{\omega B} \bar\omega +
      g_{\rho B} I_{3B} \bar\rho \right) \bigr) \bar{a}^B(\boldsymbol{p})
    f_{B^+}(\boldsymbol{p}) \Bigr) \nonumber \\  &&- \frac{1}{6} \tilde b_{\sigma} m_N
    \bigl(g_{\sigma N}\sigma\bigr)^3 - \frac{1}{4} \tilde
    c_{\sigma} \bigl(g_{\sigma N}\sigma\bigr)^4 \, .
  \label{eq:eos.epsilonB}
\end{eqnarray}
By making use of Eqs.~(\ref{eq:omega.barB}), (\ref{eq:traces.1}) and
(\ref{eq:traces.2}), this expression can be written as
\begin{equation} 
  \begin{aligned} 
    \epsilon = &\sum_B \gamma_B \int\! \frac{d^3 \boldsymbol{p}}{(2\pi)^3} \,
    E^*_B(\boldsymbol{p}) \left( f_{B^-}(\boldsymbol{p}) + f_{B^+}(\boldsymbol{p}) \right)
    \\ &+ \sum_B \gamma_B \int\! \frac{d^3 \boldsymbol{p}}{(2\pi)^3} \, \left(
      g_{\omega B} \bar\omega + g_{\rho B} I_{3B} \bar\rho \right)
    \left( f_{B^-}(\boldsymbol{p}) - f_{B^+}(\boldsymbol{p}) \right) \\ &- \frac{1}{2}
    \sum_B \gamma_B \int\! \frac{d^3 \boldsymbol{p}}{(2\pi)^3} \, \Bigl( -
      \frac{m^*_B}{E^*_B(\bf{p})} g_{\sigma B} \bar\sigma + \gamma^0
      \left( g_{\omega B} \bar\omega + g_{\rho B} I_{3B} \bar\rho
      \right) \Bigr) \\ & \qquad\qquad\qquad \qquad ~~~ \times \left(
    f_{B^-}(\boldsymbol{p}) - f_{B^+}(\boldsymbol{p}) \right) \\ & - \frac{1}{6}
    \tilde b_{\sigma} m_N \bigl(g_{\sigma N}\sigma\bigr)^3 -
    \frac{1}{4} \tilde c_{\sigma} \bigl(g_{\sigma N}\sigma\bigr)^4 \, .
  \end{aligned}
  \label{eq:eos.epsilonC}
\end{equation}
It is customary to express \eref{eq:eos.epsilonC} in a more compact
way. This is accomplished by noticing that, according to
\eref{eq:nsBfp}, the integral over the first term in the third line
above can be written as
\begin{eqnarray}
  \sum_B \gamma_B \int\! \frac{d^3 \boldsymbol{p}}{(2\pi)^3} \,
  \frac{m^*_B}{E^*_B(\bf{p})} g_{\sigma B} \bar\sigma
  \left( f_{B^-}(\boldsymbol{p}) + f_{B^+}(\boldsymbol{p}) \right)
  =  \sum_B  g_{\sigma B} n^s_B \bar\sigma \, .
  \label{eq:aux1}
\end{eqnarray}
Making use of the $\sigma$-meson field equation (\ref{eq:sigmafeq}) to
replace $\sum_B g_{\sigma B} n^s_B$ in \eref{eq:aux1} leads after some
algebra to the final result for the energy density given by
\citep{Weber:1999book}
\begin{equation} 
  \begin{aligned} 
    \epsilon = &\sum_B \gamma_B \int\! \frac{d^3 \boldsymbol{p}}{(2\pi)^3} \,
    E^*_B(\boldsymbol{p}) \left( f_{B^-}(\boldsymbol{p}) + f_{B^+}(\boldsymbol{p}) \right) +
    \frac{1}{2} m^2_\sigma \bar\sigma^2 + \frac{1}{2} m^2_\omega
    \bar\omega^2 \\ & + \frac{1}{2} m^2_\rho \bar\rho^2 + \frac{1}{3}
    \tilde b_{\sigma} m_N \bigl(g_{\sigma N}\sigma\bigr)^3 +
    \frac{1}{4} \tilde c_{\sigma} \bigl(g_{\sigma N}\sigma\bigr)^4\, .
  \end{aligned}
  \label{eq:eos.epsilonD}
\end{equation}

The expression for the pressure of hot NS matter has the form
\citep{Weber:1999book}
\begin{eqnarray} 
    P =&& i\, \sum_B\, {\rm Tr} \int\! \frac{d^4 p}{(2\pi)^4} \,
    \left( {\rm e}^{i \eta p^0} +{\rm e}^{-i \eta p^0} \right)
    \nonumber \\ && ~~~~~~~~~~~\times \Bigl( \frac{1}{3} \boldsymbol{\gamma
      \cdot \hat p} + \frac{1}{2} \left(- g_{\sigma B} \bar\sigma +
    \gamma^0 \left(g_{\omega B} \bar\omega + g_{\rho B} I_{3B}
    \bar\rho \right) \right) \Bigr) g^B(p) \nonumber \\ && +
    \frac{1}{6} \tilde b_{\sigma} m_N \bigl(g_{\sigma N}\sigma\bigr)^3
    + \frac{1}{4} \tilde c_{\sigma} \bigl(g_{\sigma N}\sigma\bigr)^4
    \, .
  \label{eq:eos.pA}
\end{eqnarray}
As for the energy density, the integration of $p^0$ can be carried out
analytically using the mathematical relations shown in
Eqs.~(\ref{eq:cint3}) and (\ref{eq:cint4}). This leads to
\begin{eqnarray}
     P =&&\frac{1}{3} \sum_B \, {\rm Tr} \int\! \frac{d^3
       \boldsymbol{p}}{(2\pi)^3} \, \Bigl( \boldsymbol{ \gamma \cdot \hat p}\,
     a^B(\boldsymbol{p}) f_{B^-}(\boldsymbol{p}) - \boldsymbol{ \gamma \cdot \hat p}\, \bar
     a^B(\boldsymbol{p}) f_{B^+}(\boldsymbol{p}) \Bigr) \nonumber \\ &&+\frac{1}{2}
     \sum_B \, {\rm Tr} \int\! \frac{d^3 \boldsymbol{p}}{(2\pi)^3} \, \Bigl(
     \left( - g_{\sigma B} \bar\sigma + \gamma^0 \left( g_{\omega B}
     \bar\omega + g_{\rho B} I_{3B} \bar\rho \right) \right)
     a^B(\boldsymbol{p}) f_{B^-}(\boldsymbol{p}) \nonumber \\ &&
     \qquad\qquad\qquad\qquad ~ - \left( - g_{\sigma B} \bar\sigma +
     \gamma^0 \left( g_{\omega B} \bar\omega + g_{\rho B} I_{3B}
     \bar\rho \right) \right) \bar{a}^B(\boldsymbol{p}) f_{B^+}(\boldsymbol{p}) \Bigr)
     \nonumber \\ && + \frac{1}{6} \tilde b_{\sigma} m_N
     \bigl(g_{\sigma N}\sigma\bigr)^3 + \frac{1}{4} \tilde c_{\sigma}
     \bigl(g_{\sigma N}\sigma\bigr)^4 \, .
  \label{eq:eos.pB}
\end{eqnarray}
With the help of Eqs.~(\ref{eq:a.B}) and (\ref{eq:a.barB}) for the
spectral functions and Eqs.~({\ref{eq:traces.1}}) and
({\ref{eq:traces.2}}) for the traces, expression \eref{eq:eos.pB} can
be written as
\begin{eqnarray} 
     P =&&\frac{1}{3} \sum_B \gamma_B \int\! \frac{d^3
       \boldsymbol{p}}{(2\pi)^3} \, \frac{\boldsymbol{p}^2}{E^*_B(\boldsymbol{p})} \left(
     f_{B^-}(\boldsymbol{p}) + f_{B^+}(\boldsymbol{p}) \right) \nonumber
     \\ &&+\frac{1}{2} \sum_B \gamma_B \! \int\! \frac{d^3
       \boldsymbol{p}}{(2\pi)^3} \, \Bigl( - \frac{m^*_B}{E^*_B(\bf{p})}
     g_{\sigma B} \bar\sigma + g_{\omega B} \bar\omega + g_{\rho B}
     I_{3B} \bar\rho \Bigr) \nonumber \\ &&
     ~~~~~~~~~~~~~~~~~~~~~~~~\times \left( f_{B^-}(\boldsymbol{p}) -
     f_{B^+}(\boldsymbol{p}) \right) \nonumber \\ &&+ \frac{1}{6} \tilde
     b_{\sigma} m_N \bigl(g_{\sigma N}\sigma\bigr)^3 + \frac{1}{4}
     \tilde c_{\sigma} \bigl(g_{\sigma N}\sigma\bigr)^4 \, .
  \label{eq:eos.pC}
\end{eqnarray}
The second line in this equation can be written in terms of the scalar
and baryon number densities. To see this we begin with
\eref{eq:nsBfp}, from which it follows that
\begin{eqnarray}
 \sum_B \gamma_B \! \int\! \frac{d^3 \boldsymbol{p}}{(2\pi)^3} \,
 \frac{m^*_B}{E^*_B(\bf{p})} g_{\sigma B} \bar\sigma \left(
 f_{B^-}(\boldsymbol{p}) - f_{B^+}(\boldsymbol{p}) \right) = \sum_B g_{\sigma B} n^s_B
 \bar\sigma \, .
 \label{eq:aux.p1a}    
\end{eqnarray}
On the other hand, it is known from the $\sigma$-meson field equation
(\ref{eq:sigmafeq}) that
\begin{eqnarray}
 \sum_B g_{\sigma B} n^s_B \bar\sigma = m_\sigma^2 {\bar\sigma}^2
 +\tilde b_{\sigma} m_N \bigl(g_{\sigma N}\sigma\bigr)^3 + \tilde
 c_{\sigma} \bigl(g_{\sigma N}\sigma\bigr)^4 \, .
 \label{eq:aux.p1b}    
\end{eqnarray}
Similarly, for the $\omega$-meson dependent term in \eref{eq:eos.pC} we have
\begin{eqnarray}
 \sum_B \gamma_B \! \int\! \frac{d^3 \boldsymbol{p}}{(2\pi)^3} \, g_{\omega B}
 \bar\omega \left( f_{B^-}(\boldsymbol{p}) - f_{B^+}(\boldsymbol{p}) \right)
 &=& \sum_B g_{\omega B} n_B  \bar\omega \, ,\nonumber \\
 &=& m_\omega^2 {\bar\omega}^2 \, ,
 \label{eq:aux.p2}
\end{eqnarray}
and for the $\rho$-meson dependent term 
\begin{eqnarray}
 \sum_B \gamma_B \! \int\! \frac{d^3 \boldsymbol{p}}{(2\pi)^3} \, g_{\rho B}
 I_{3B} \bar\rho \left( f_{B^-}(\boldsymbol{p}) - f_{B^+}(\boldsymbol{p}) \right) &=&
 \sum_B g_{\rho B} I_{3B} n^B \bar\rho \, , \nonumber \\ &=& m_\rho^2
     {\bar\rho}^2
 \label{eq:aux.p3}    
\end{eqnarray}
Substituting Eqs.~(\ref{eq:aux.p1a}) through (\ref{eq:aux.p3}) into
\eref{eq:eos.pC} leads for the pressure of NS matter to
\citep{Weber:1999book}
\begin{eqnarray} 
     P = &&\frac{1}{3} \sum_B \gamma_B \int\! \frac{d^3
       \boldsymbol{p}}{(2\pi)^3} \, \frac{\boldsymbol{p}^2}{E^*_B(\boldsymbol{p})} \left(
     f_{B^-}(\boldsymbol{p}) + f_{B^+}(\boldsymbol{p}) \right) - \frac{1}{2}
     m^2_\sigma \bar\sigma^2 + \frac{1}{2} m^2_\omega \bar\omega^2
     \nonumber \\ && + \frac{1}{2} m^2_\rho \bar\rho^2 - \frac{1}{3}
     \tilde b_{\sigma} m_N \bigl(g_{\sigma N}\sigma\bigr)^3 -
     \frac{1}{4} \tilde c_{\sigma} \bigl(g_{\sigma N}\sigma\bigr)^4 +
     n \tilde{R} \, .
  \label{eq:eos.pD}
\end{eqnarray}

\subsection{Leptons and Neutrinos}\label{ssec:leptons}

Leptons are treated as free Fermi gases with the grand canonical
potential \index{Grand canonical potential} given by
\citep{Weber:1999book,Malfatti:2019PRC100}
\begin{equation}
\Omega_L = - \sum_L \frac{\gamma_L}{3} \int \frac{d^3\boldsymbol{p}}{(2
  \pi)^3} \frac{ \boldsymbol{p}^2}{E_L(\boldsymbol{p})} \left(f_{L^-}(p) + f_{L^+}(p)
\right) \, ,
\label{eq:leptons}
\end{equation}
where $\gamma_L = (2J_L+1)$ is the lepton degeneracy factor.  The sum
over $L$ in Eq.~(\ref{eq:leptons}) runs over $e^-$ and $\mu^-$, with
masses $m_L$, and massless neutrinos, $\nu_{e}$, in the case they are
trapped in a PNS (see \ref{ssec:chemicaleq},
\ref{ssec:hot.dense}). The lepton distribution function is given by
\begin{eqnarray}
  f_{L^\mp}(p)= \frac{1} {{{\rm e}^{(E_L(p) \mp \mu_L)/T}} + 1}\, ,
\label{eq:fL} 
\end{eqnarray}
where $E_L(p)=\sqrt{p^2 + m_L^{2}}$ denotes the energy-momentum
relation of free leptons.

\subsection{Chemical Equilibrium and Electric Charge Neutrality}\label{ssec:chemicaleq}

Three important constraints must be taken into account when
determining the EOS of PNS matter: electric charge neutrality,
\index{Electric charge neutrality} baryon number conservation, and
chemical equilibrium\index{Chemical equilibrium}. Neutron star matter
must be charge neutral, satisfying
\citep{Glendenning:1985a,Weber:1999book,Malfatti:2019PRC100}
\begin{equation}
\sum_B q_B\,n_B + \sum_L q_L\,n_L = 0 \, ,
\end{equation}
where $q_B$ and $q_L$ are baryon and lepton electric charge,
respectively. Baryon number must also be conserved, which leads to
\begin{equation}
\sum_B \,n_B - n = 0 \, ,
\end{equation}
Finally, the constraint of chemical equilibrium for hadronic matter
can be defined as \citep{Prakash:1997}
\begin{equation}
\mu_B = \mu_n + q_B (\mu_e - \mu_{\nu_e}) \, ,
\label{chem}
\end{equation}
where $\mu_n$, $\mu_e$ and $\mu_{\nu_e}$ are the neutron, electron and
\begin{figure}[htb]
\centering
{
    \includegraphics[width=7.5cm]{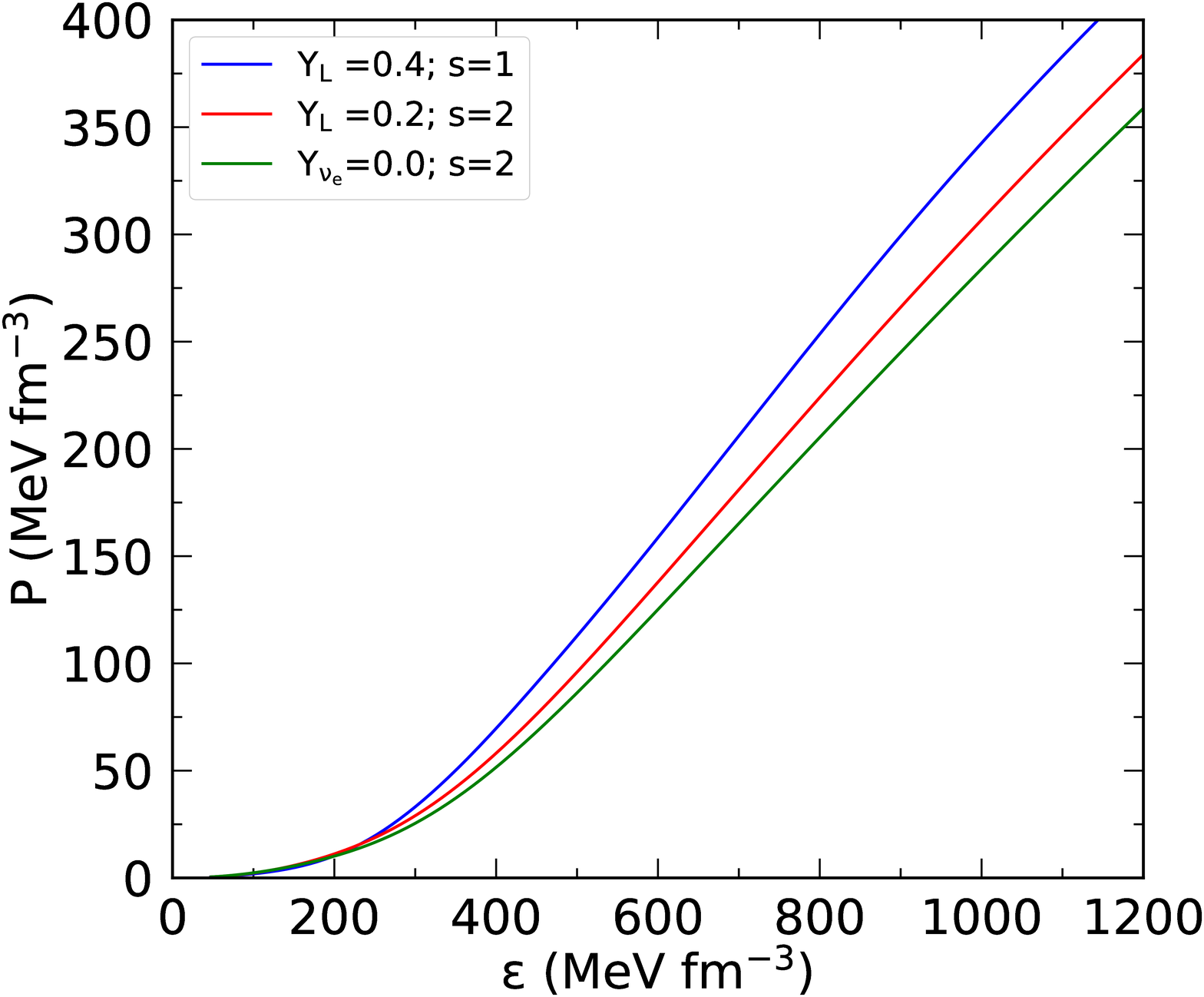}
    \caption{Pressure as a function of energy density for the DD2
      parameter set.}
    \label{fig:EoS_DD2_combined}
}{
    \includegraphics[width=7.5cm]{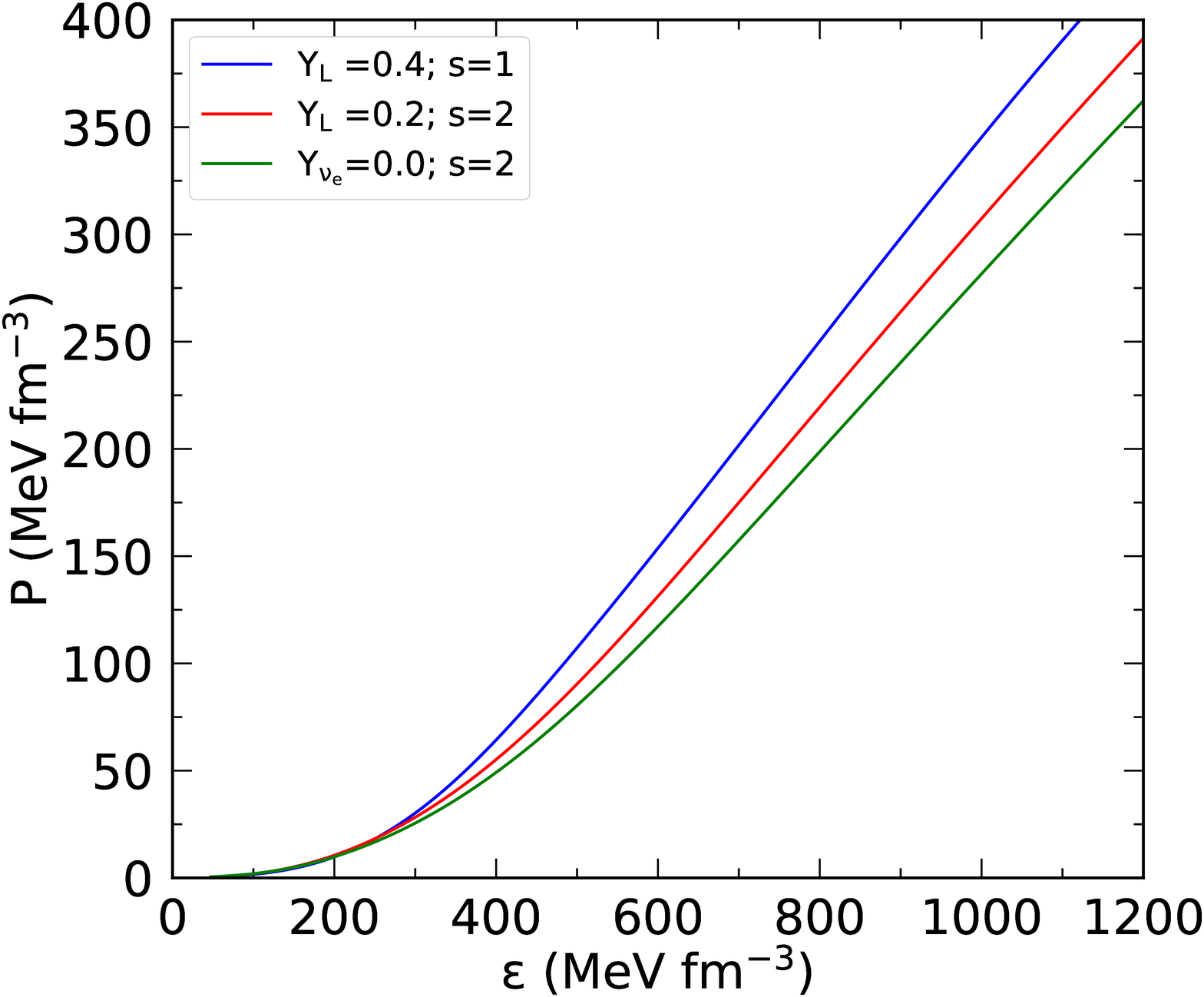}
    \caption{Pressure as a function of energy density for the GM1L
      parameter set.}
    \label{fig:EoS_GM1L_combined}
}
\end{figure}
\begin{figure}[htb]
\centering
{
    \includegraphics[width=7.5cm]{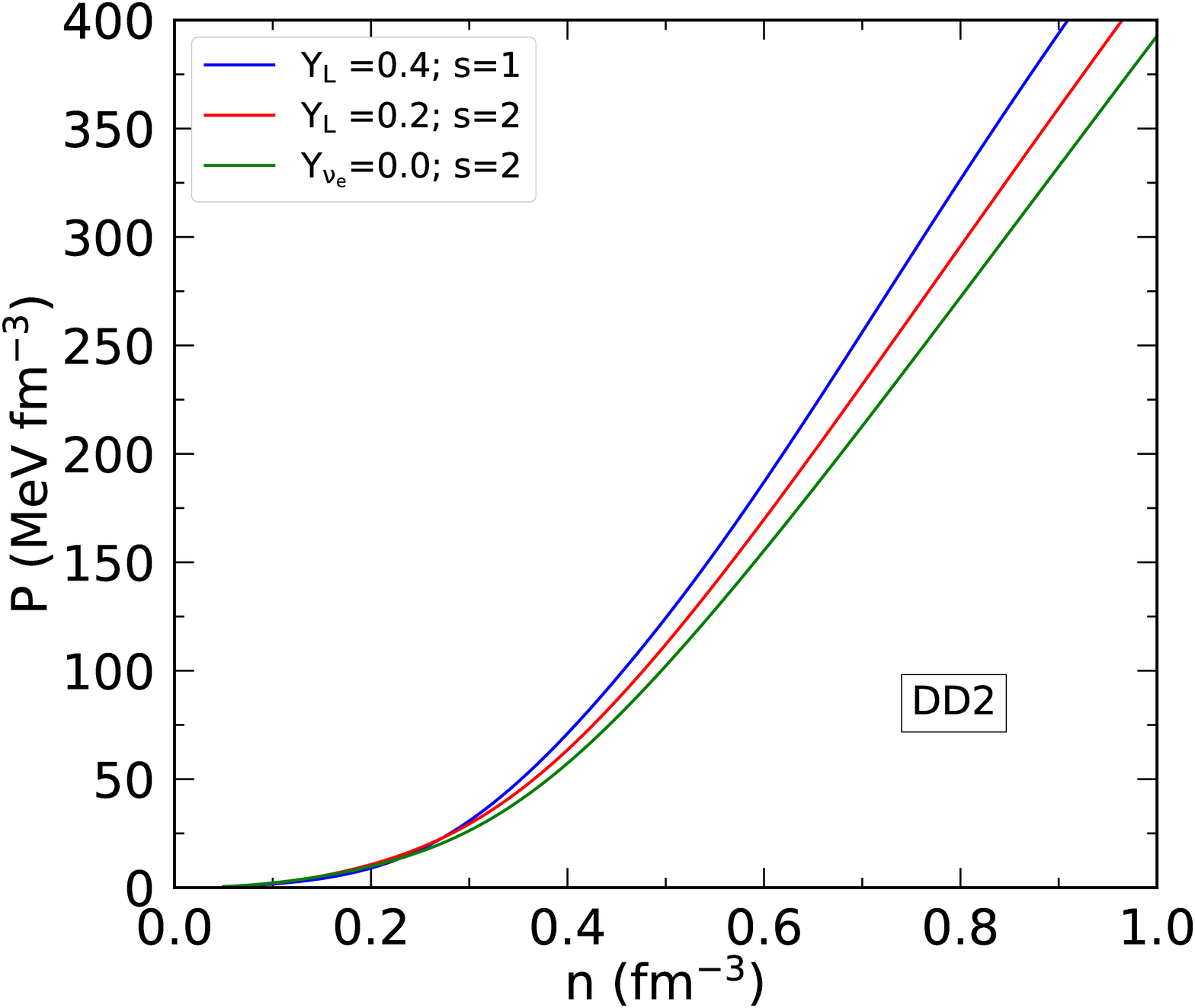}
    \caption{Pressure as a function of baryon number density for the DD2
      parameter set.}
    \label{fig:Pn.DD2}
}{
    \includegraphics[width=7.5cm]{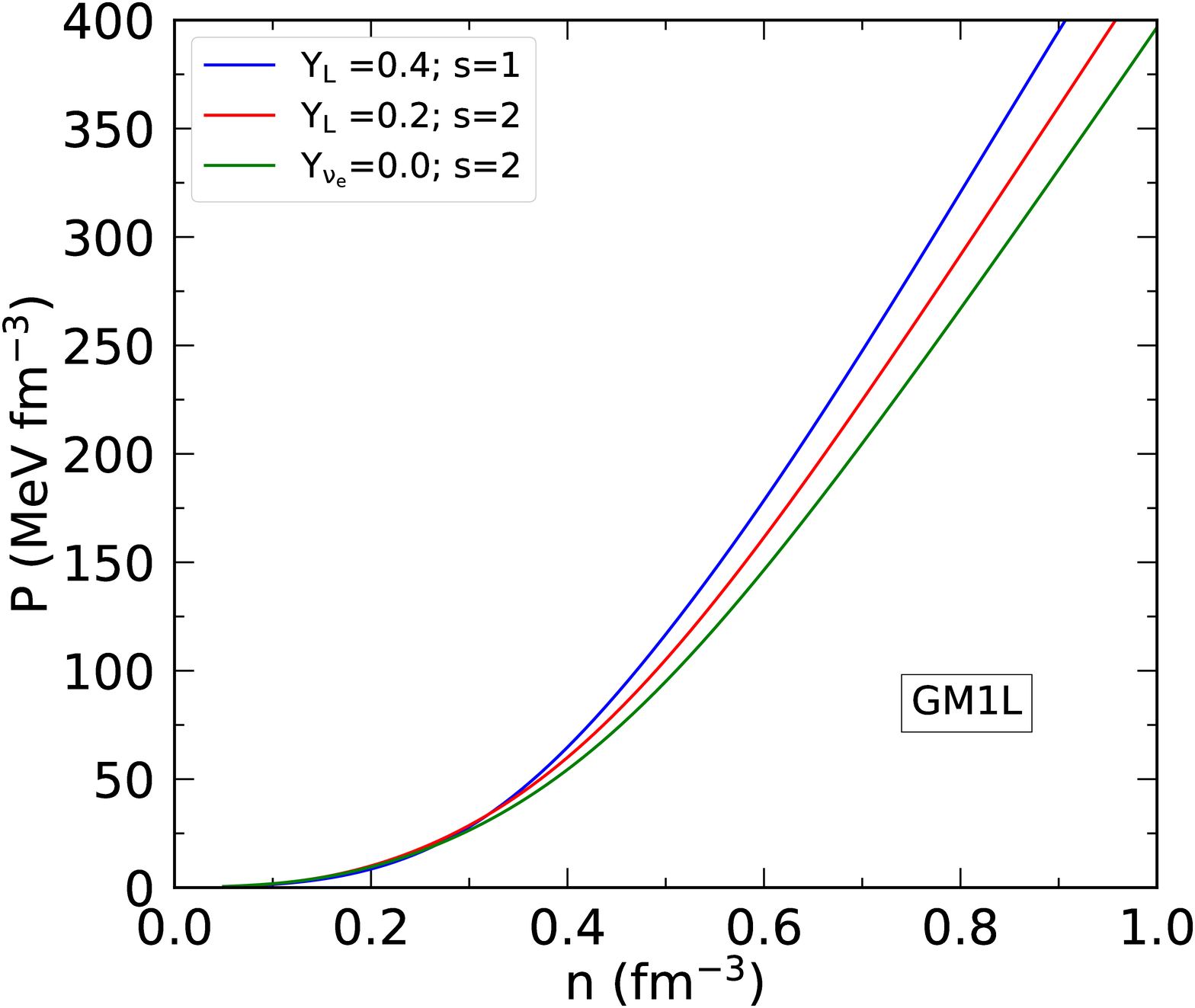}
    \caption{Same as Fig.~\ref{fig:Pn.DD2}, but for the GM1L parameter
      set.}
    \label{fig:Pn.GM1L}
}
\end{figure}
neutrino chemical potentials, respectively.  The chemical potential of
the latter follows from the equilibrium reaction
\begin{equation}
e^- \leftrightarrow \mu^- + \nu_e + \bar{\nu}_{\mu} \, ,
\end{equation}
which leads for the corresponding chemical potentials to the condition
\begin{equation}
\mu_e = \mu_{\mu} + \mu_{\nu_e} + \mu_{\bar{\nu}_{\mu}} \, . \label{eq:mue}
\end{equation}
Neutrinos are trapped inside of a proto-neutron star immediately after
its formation. Mathematically this is expressed as \citep{Prakash:1997,Malfatti:2019PRC100}
\begin{eqnarray}
  Y_{e} &=& \frac{n_e + n_{\nu_e}}{ n } \, ,\nonumber \\
  Y_{\mu} &=&
 \frac{n_{\mu} + n_{\nu_{\mu}}}{ n } = 0 \, ,
\end{eqnarray}
where $n_e$, $n_{\mu}$, $n_{\nu_e}$, and $n_{\nu_\mu}$ denote the
number densities of electrons, muons, electron neutrinos, and muon
neutrinos, respectively.  During this very early stellar phase, the
matter is opaque to neutrinos and the composition of the matter is
characterized by three independent chemical potentials, namely
$\mu_n$, $\mu_e$, and $\mu_{\nu_e}$. The condition $Y_{\mu}=0$
indicates that no muons are present in the matter.  The actual value
of $Y_{\mu}$ ($\lesssim 0.4$) depends on the efficiency of electron
capture reactions during the initial state of the formation of
proto-neutron stars \citep{Prakash:1997}. The quantity $Y_L$, which
shows the fraction of electrons and neutrinos at a given density, is
defined as
\begin{eqnarray}
  Y_L = Y_e + Y_{\nu_e} \, .
  \label{eq:YL}
\end{eqnarray}

In Figs.~\ref{fig:EoS_DD2_combined} and \ref{fig:EoS_GM1L_combined} we
show how pressure varies as a function of energy density for the DD2
and GM1L parameter sets. Figures~\ref{fig:Pn.DD2} and
\ref{fig:Pn.GM1L} show the pressure as a function baryon number
density for DD2 and GM1L.  Details of these parameter sets including
the coupling values used to compute the DD2 and GM1L EOSs will be
discussed in \sref{sec:parameters}. The EOSs are shown for different
\begin{figure}[htb]
\centering
{
    \includegraphics[width=7.5cm]{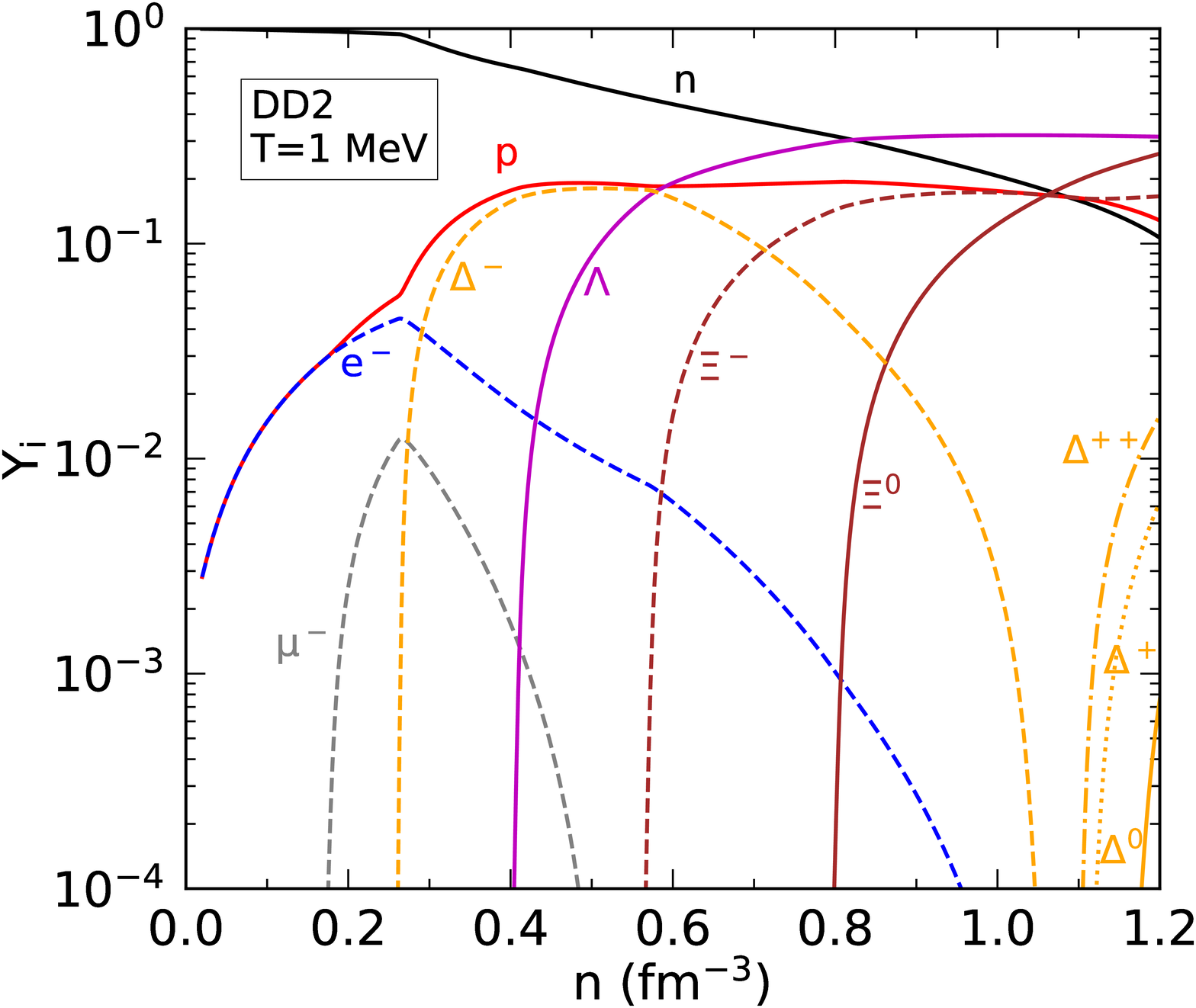}
    \caption{Composition of dense stellar matter computed for DD2
      parametrization and a temperature of $T=1$~MeV.}
    \label{fig:hadronpop_DD2_T1}
}{
    \includegraphics[width=7.5cm]{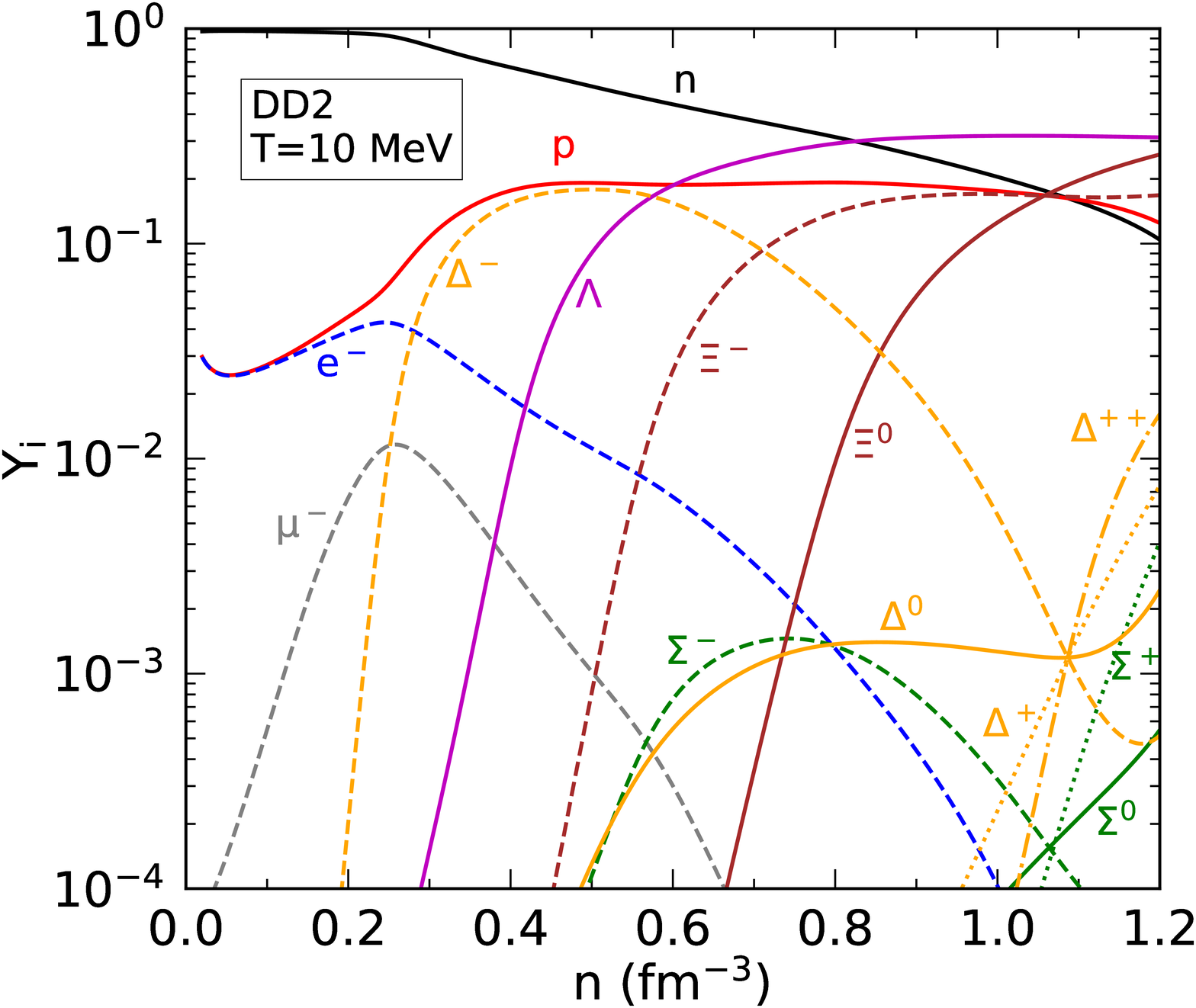}
    \caption{Same as Fig.\ \ref{fig:hadronpop_DD2_T1}, but for a stellar
      temperature of $T=10$~MeV.}
    \label{fig:hadronpop_DD2_T10}
}
\end{figure}
lepton fractions, $Y_L$, and entropies (per baryon), $s$ which
\begin{figure}[tb]
\centering
{
    \includegraphics[width=7.5cm]{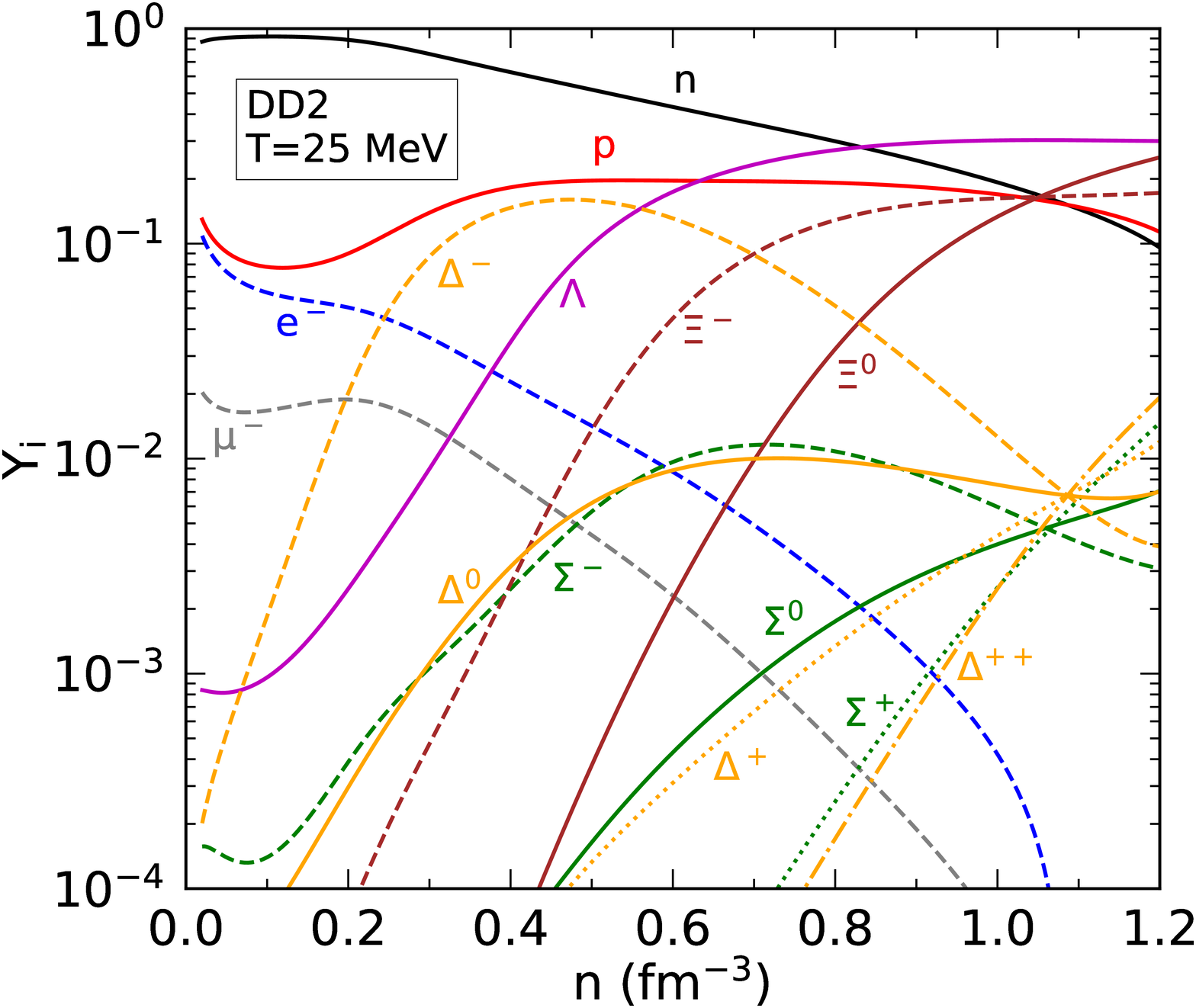}
    \caption{Same as Fig.\ \ref{fig:hadronpop_DD2_T1}, but for a stellar 
      temperature of $T=25$~MeV.} \label{fig:hadronpop_DD2_T25}
}{
    \includegraphics[width=7.5cm]{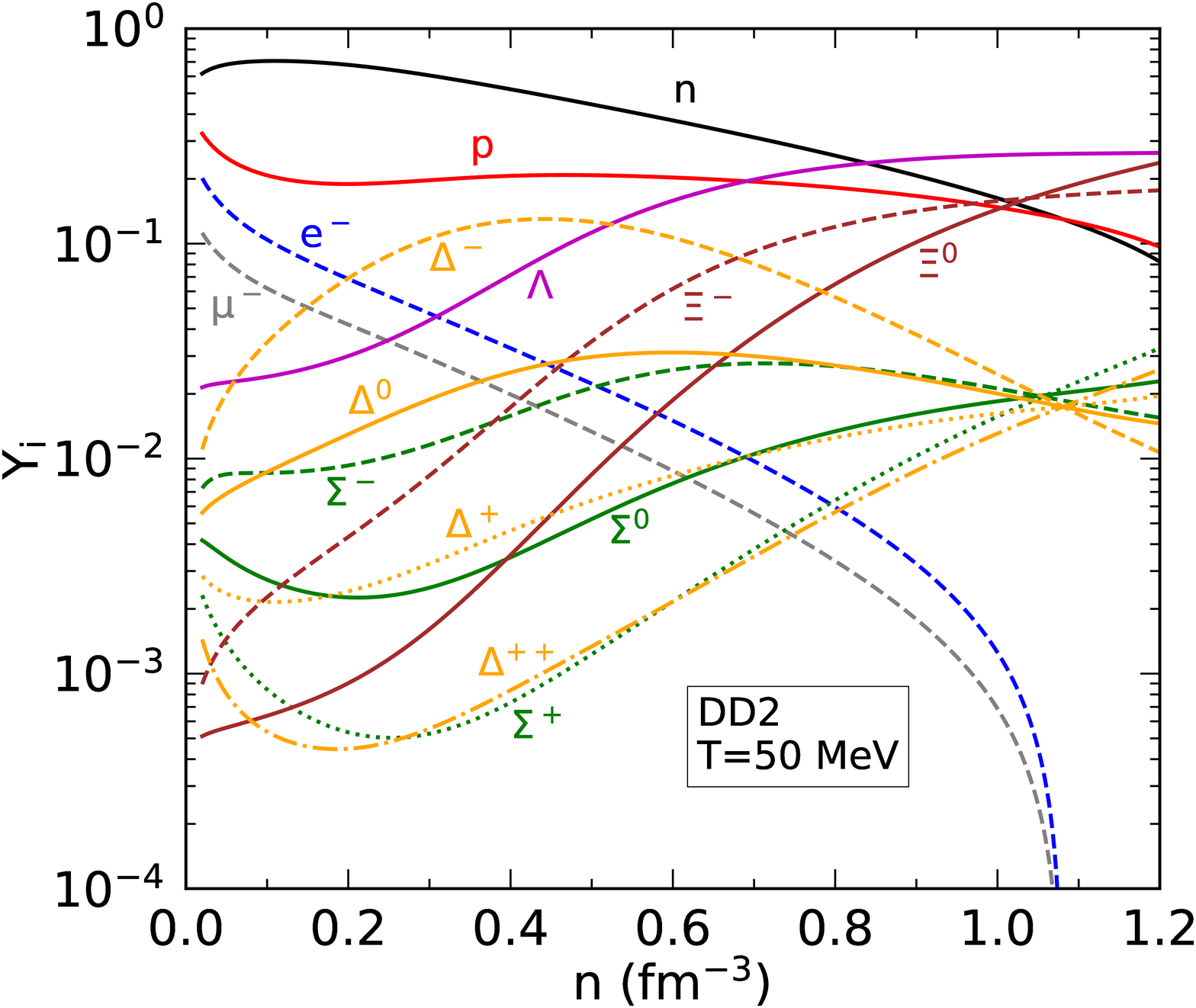}
    \caption{Same as Fig.\ \ref{fig:hadronpop_DD2_T1}, but for a stellar
      temperature of $T=50$~MeV.}\label{fig:hadronpop_DD2_T50}
}
\end{figure}
correspond to the main characteristic stages in the life of a
proto-neutron star (briefly summarized at the beginning of
\sref{ssec:hot.dense}). The EOSs (as well as all other dense-matter
properties presented in this chapter) shown in
Figs.~\ref{fig:EoS_DD2_combined} and \ref{fig:EoS_GM1L_combined} have
been computed for $B=n,p , \Lambda, \Sigma^\pm, \Sigma^0, \Xi^0,
\Xi^-$, all electrically charged states of the $\Delta(1232)$ baryon,
and $L=e, \mu, \nu_e$.
\begin{figure}[htb]
\centering
{
    \includegraphics[width=7.5cm]{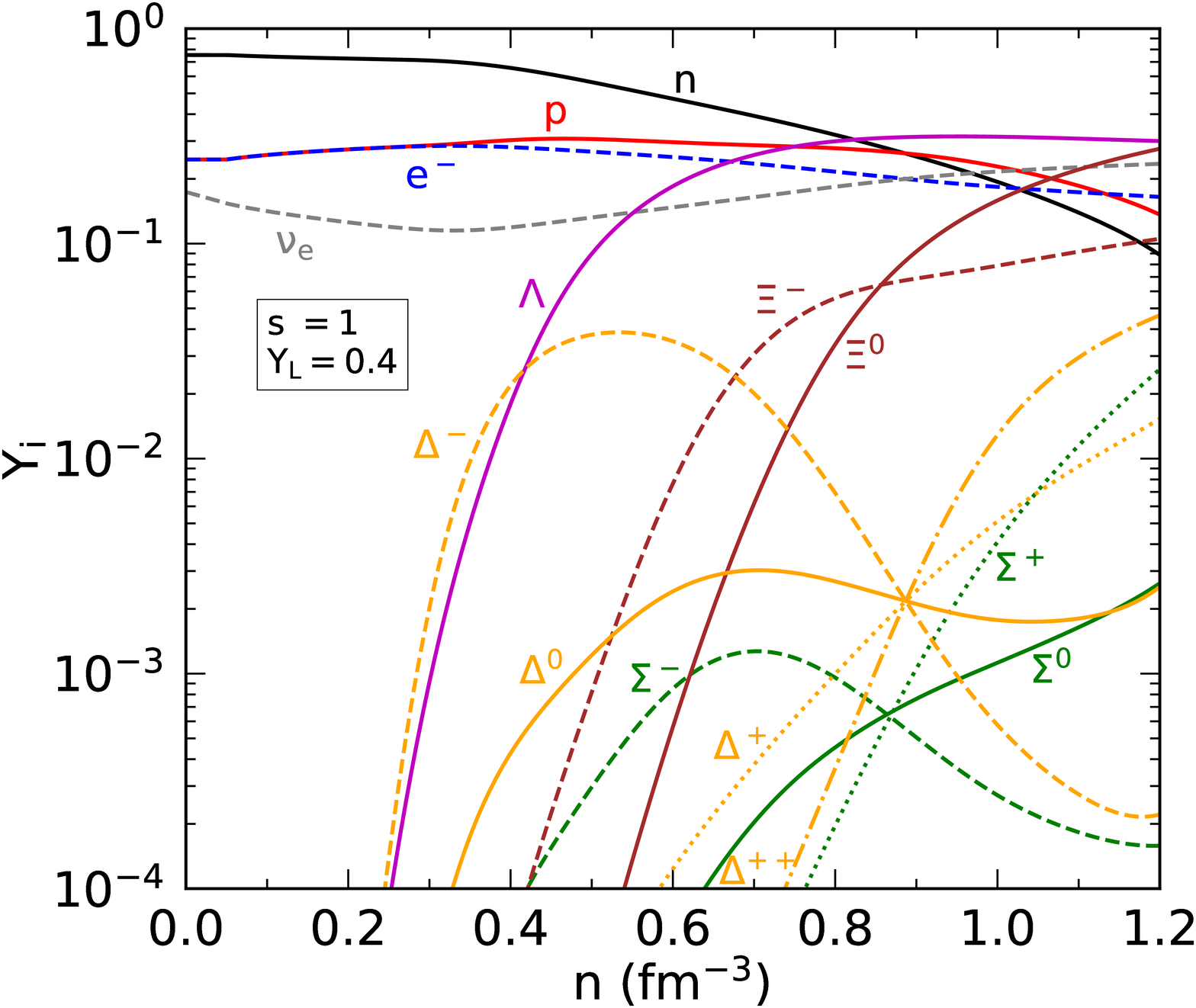}
    \caption{Baryon-lepton composition of PNS matter obtained for DD2 model with
       $s=1$ and $Y_L=0.4$ \citep{Malfatti:2019PRC100}.}
    \label{fig:Hadronpop_DD2_YL04s1}
}{
    \includegraphics[width=7.5cm]{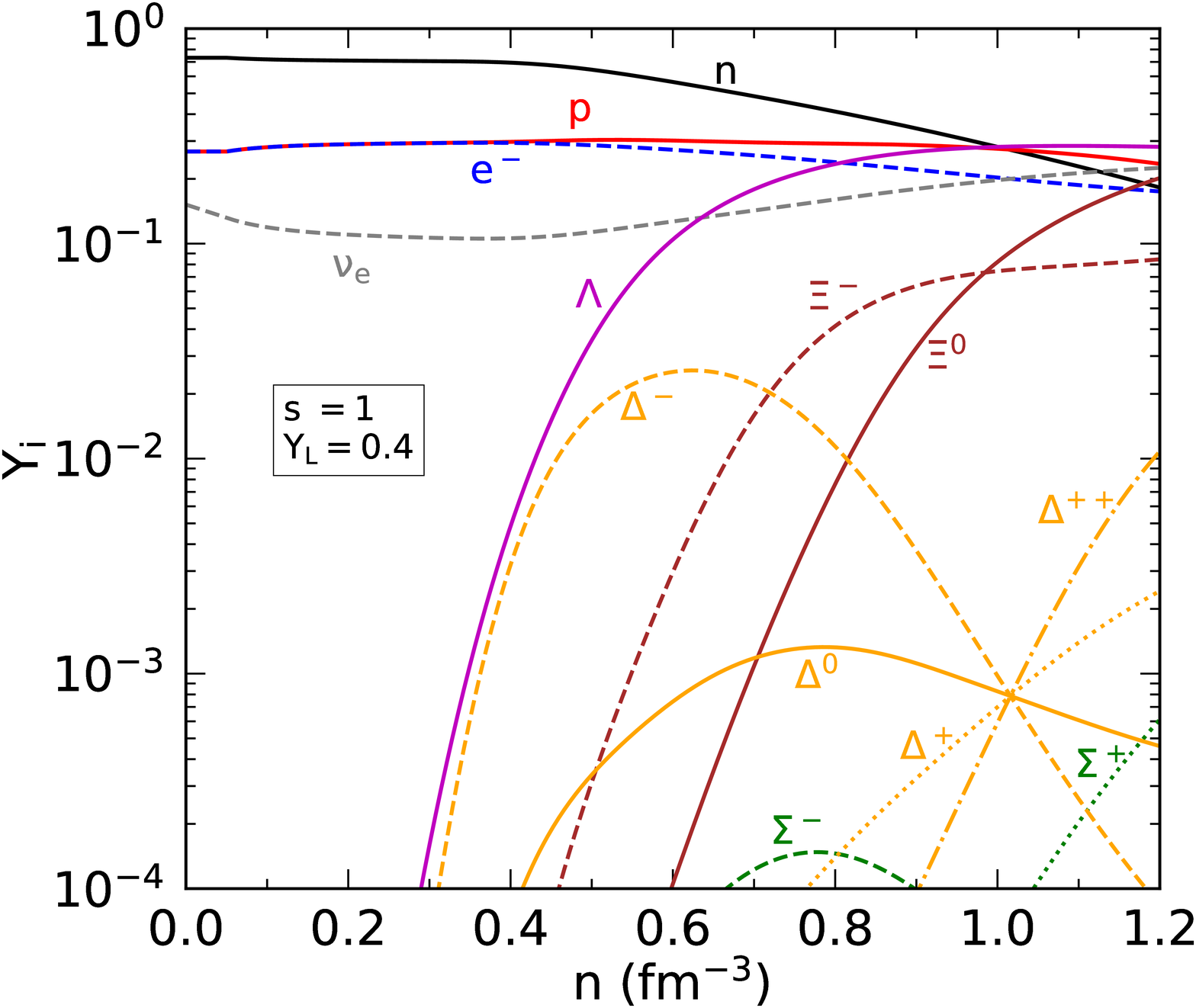}
    \caption{Baryon-lepton composition of PNS matter obtained for GM1L model with
       $s=1$ and  $Y_L=0.4$ \citep{Malfatti:2019PRC100}.}
    \label{fig:Hadronpop_GM1L_YL04s1}
}
\end{figure}

\begin{figure}[tb]
\centering
{
    \includegraphics[width=7.5cm]{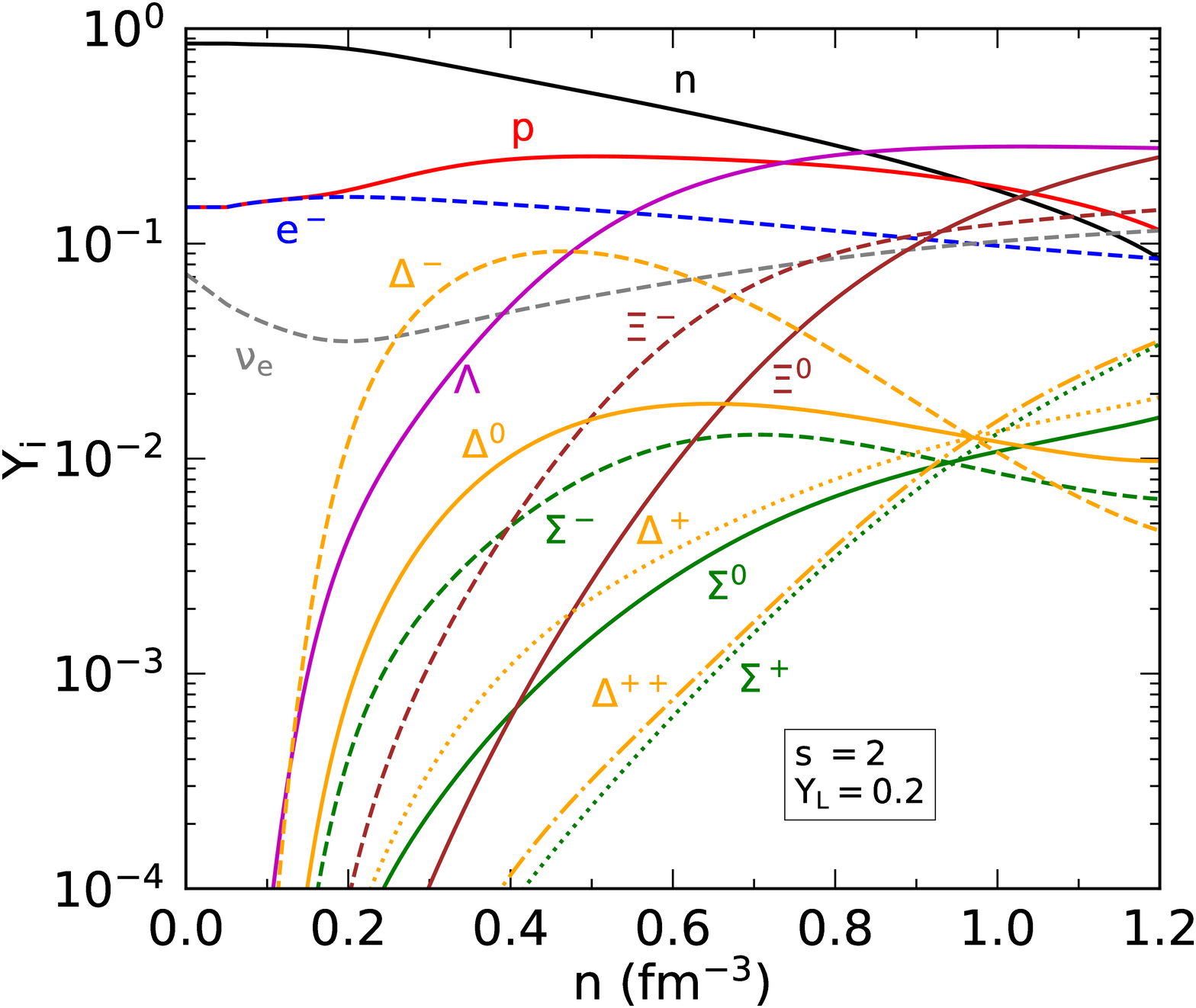}
    \caption{Baryon-lepton composition of PNS matter obtained for DD2 model with
       $s=2$ and $Y_L=0.2$ \citep{Malfatti:2019PRC100}.}
    \label{fig:Hadronpop_DD2_YL02s2}
}{
    \includegraphics[width=7.5cm]{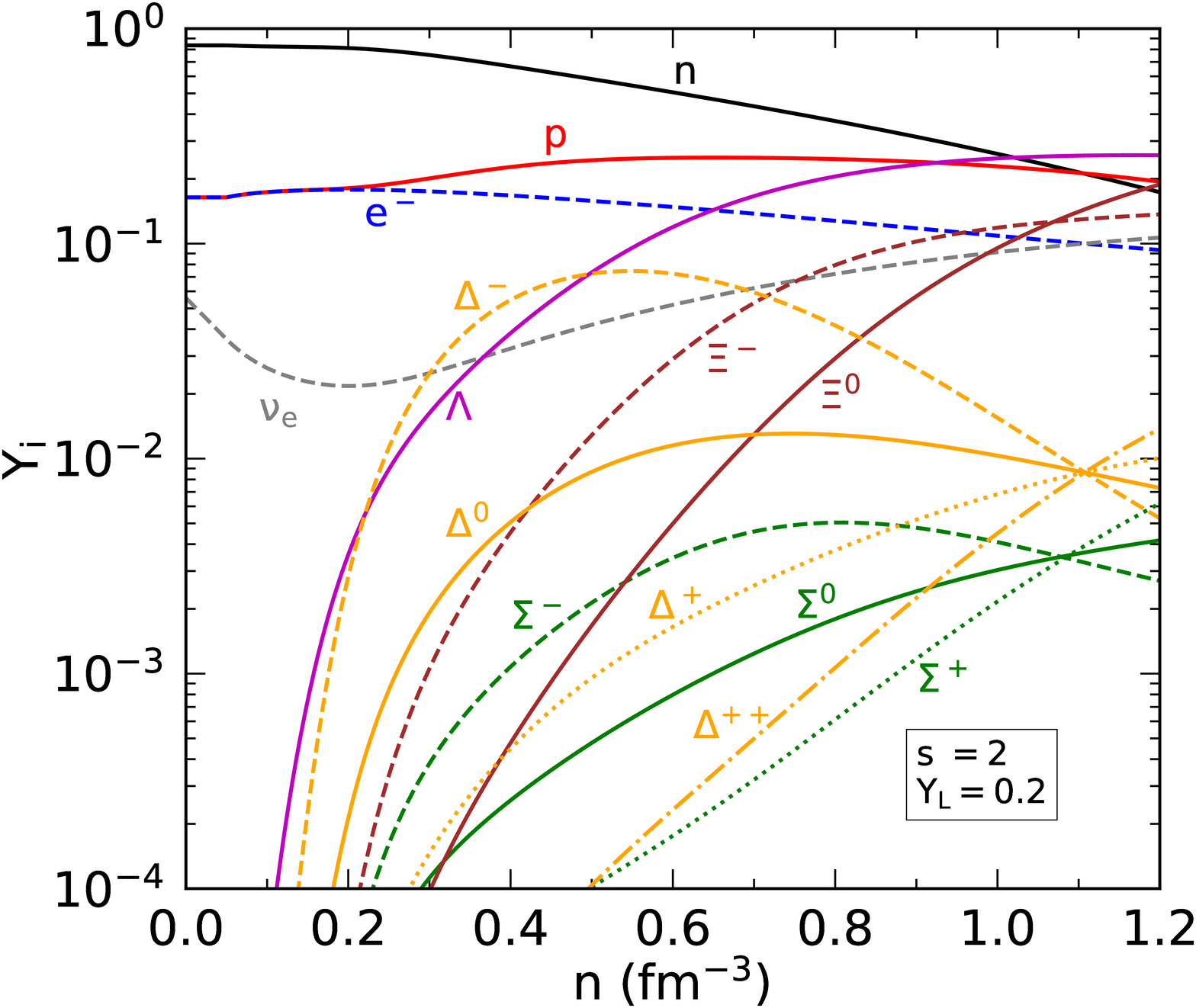}
    \caption{Baryon-lepton composition of PNS matter obtained for GM1L model with
       $s=2$ and $Y_L=0.2$ \citep{Malfatti:2019PRC100}.}
    \label{fig:Hadronpop_GM1L_YL02s2}
}
\end{figure}
The values of the baryon--hyperon coupling constants will be discussed
in detail in \sref{ssec:meson.hyperon}. The values chosen for the
$\Delta$--hyperon couplings are $x_{\sigma\Delta} = x_{\omega\Delta}
=1.1$ and $x_{\rho\Delta} = 1.0$ as described in
\sref{ssec:delta.isobars}. A general investigation of the
$\Delta(1232)$ coupling spaces is provided in \sref{ssec:meson.delta}.
\begin{figure}[htb]
\centering
    \includegraphics[width=7.5cm]{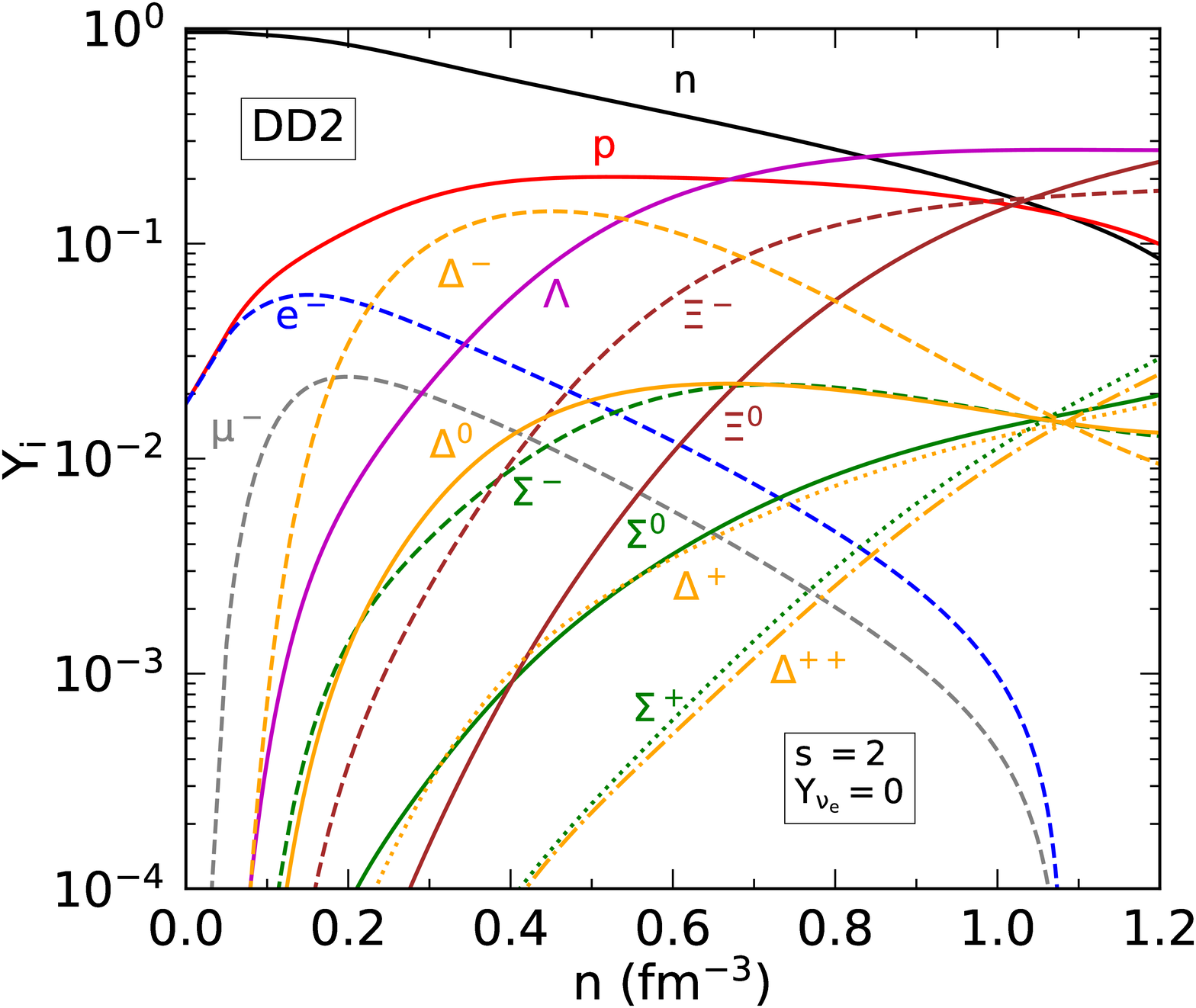}
    \caption{Baryon-lepton composition of PNS matter obtained for DD2 model with
       $s=2$ and $Y_{\nu_e}=0$ \citep{Malfatti:2019PRC100}.}
    \label{fig:Hadronpop_DD2_YL0s2}
\end{figure}

\subsection{Composition of Hot and Dense  Matter}\label{ssec:hot.dense}

In this section we show the composition of hot and dense matter as
it exists in the cores of proto-neutron stars.  Following the core
bounce post supernova explosion, PNSs experience a deleptonization
stage where hot, lepton-rich matter becomes lepton-poor over the
course of about a minute. During this time, the entropy per baryon and
lepton fraction of the dense matter within the core change
quickly. These values start at around $s = 1$ and $Y_L = 0.4$, then
change to $s = 2$ and $Y_L = 0.2$, and become $s = 2$ and $Y_{\nu_e}
= 0$ \citep{Prakash:1997,Malfatti:2019PRC100}. As neutrinos and
photons continue to diffuse over the next several minutes and
\begin{figure}[tb]
\centering
  \includegraphics[width=7.5cm]{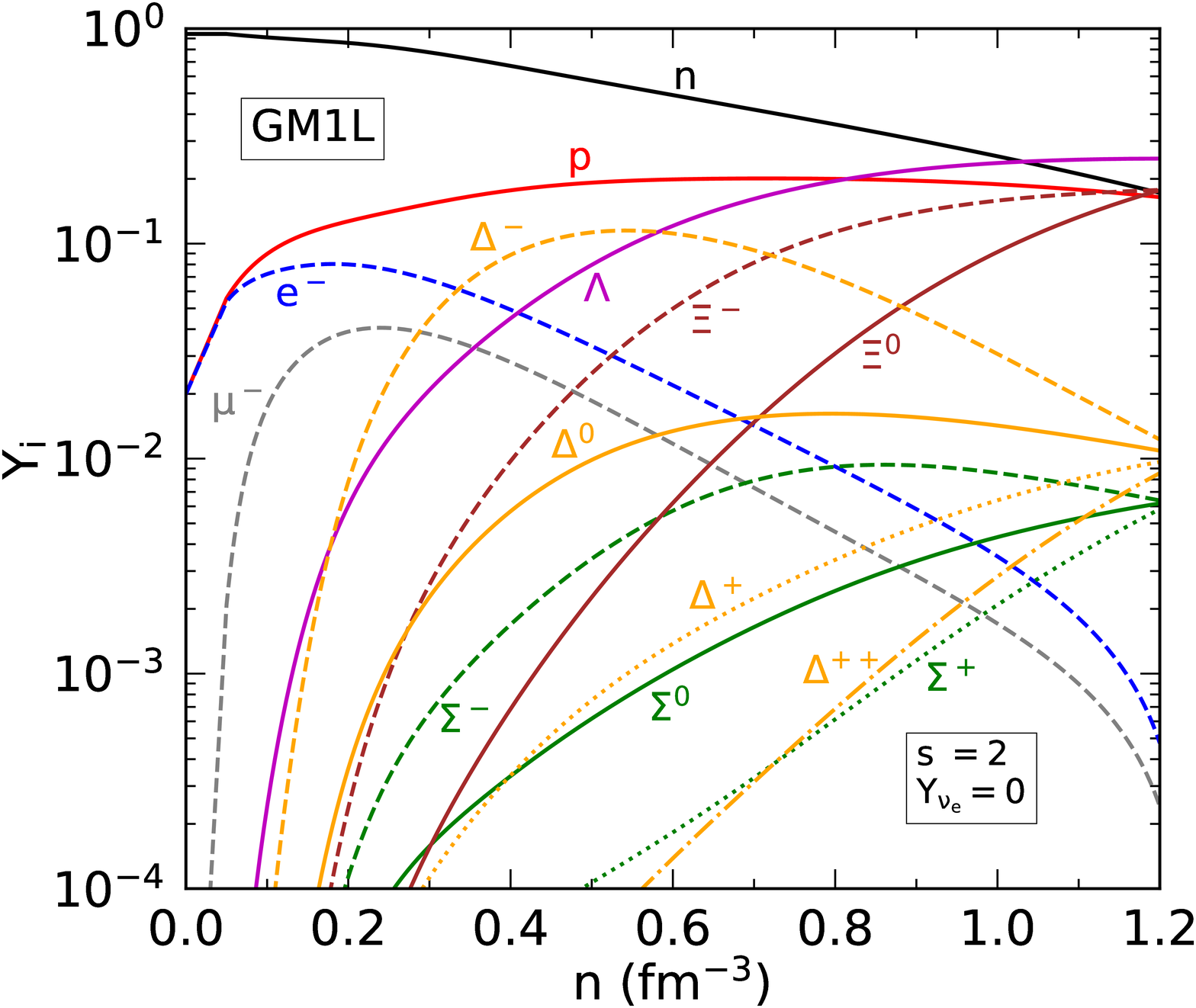}
    \caption{Baryon-lepton composition of PNS matter obtained for GM1L
      model with $s=2$ and $Y_{\nu_e}=0$ \citep{Malfatti:2019PRC100}.}
    \label{fig:Hadronpop_GM1L_YL0s2}
\end{figure}
the temperature drops to less than 1~MeV, the hot PNS becomes a cold NS.

Figures~\ref{fig:hadronpop_DD2_T1} through \ref{fig:hadronpop_DD2_T50}
illustrate how drastically the particle composition in the core of a
neutron star changes with temperature.  In fact, as can be seen by
comparing the compositions shown in Figs.~\ref{fig:hadronpop_DD2_T1}
and \ref{fig:hadronpop_DD2_T25} with each other, the particle
composition at a temperature of 25~MeV no longer resembles the
\begin{figure}[tb]
\centering
  \includegraphics[width=7.5cm]{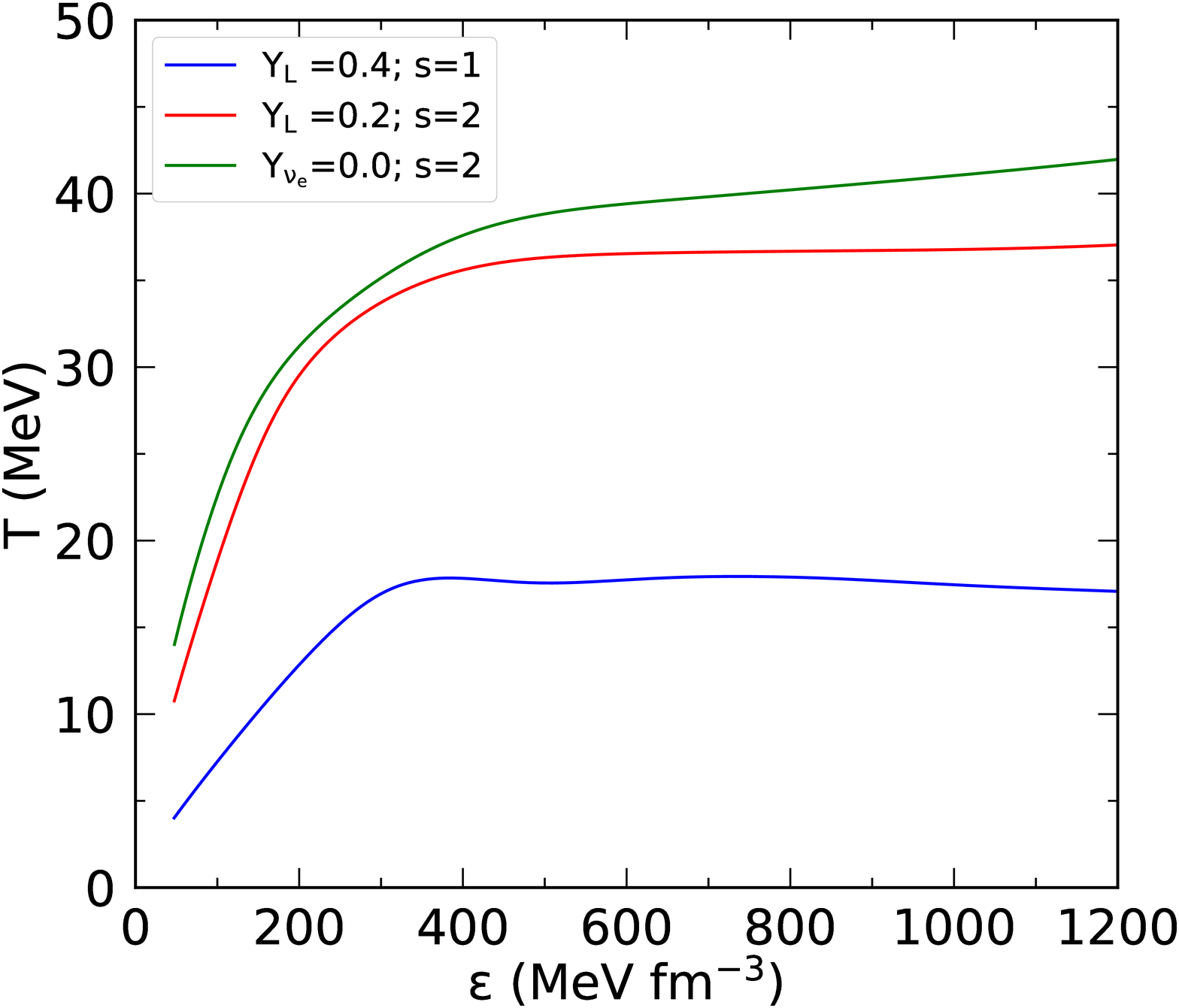}
    \caption{Temperature as a function of energy density of PNS matter
      obtained for the DD2 model.}
    \label{fig:temp_vs_density_DD2_combined}
\end{figure}
zero-temperature (i.e., 1~MeV) composition at all.  Moreover, in
matter at even higher temperatures the threshold densities of all the
baryons have changed so much that all baryonic particle states taken
into account in our calculations are present at all densities, as
shown in \fref{fig:hadronpop_DD2_T50}.

The next set of figures show the composition of proto-neutron star
matter for different combinations of entropy and lepton number, which
characterize several different stages in the evolution of a hot, newly
formed PNS to a cold NS.  Proto-neutron stars in their earliest phases
of evolution have $s = 1$ and $Y_L = 0.4$ followed by $s = 2$ and $Y_L
= 0.2$. The particle compositions
of such matter are shown in Figs.~\ref{fig:Hadronpop_DD2_YL04s1}
through \ref{fig:Hadronpop_GM1L_YL02s2} for the DD2 and GM1L
parametrizations. The matter in proto-neutron stars with $s = 2$ and
$Y_L = 0.2$ undergoes deleptonization and becomes lepton poor. Such
\begin{figure}[tb]
\centering  
    \includegraphics[width=7.5cm]{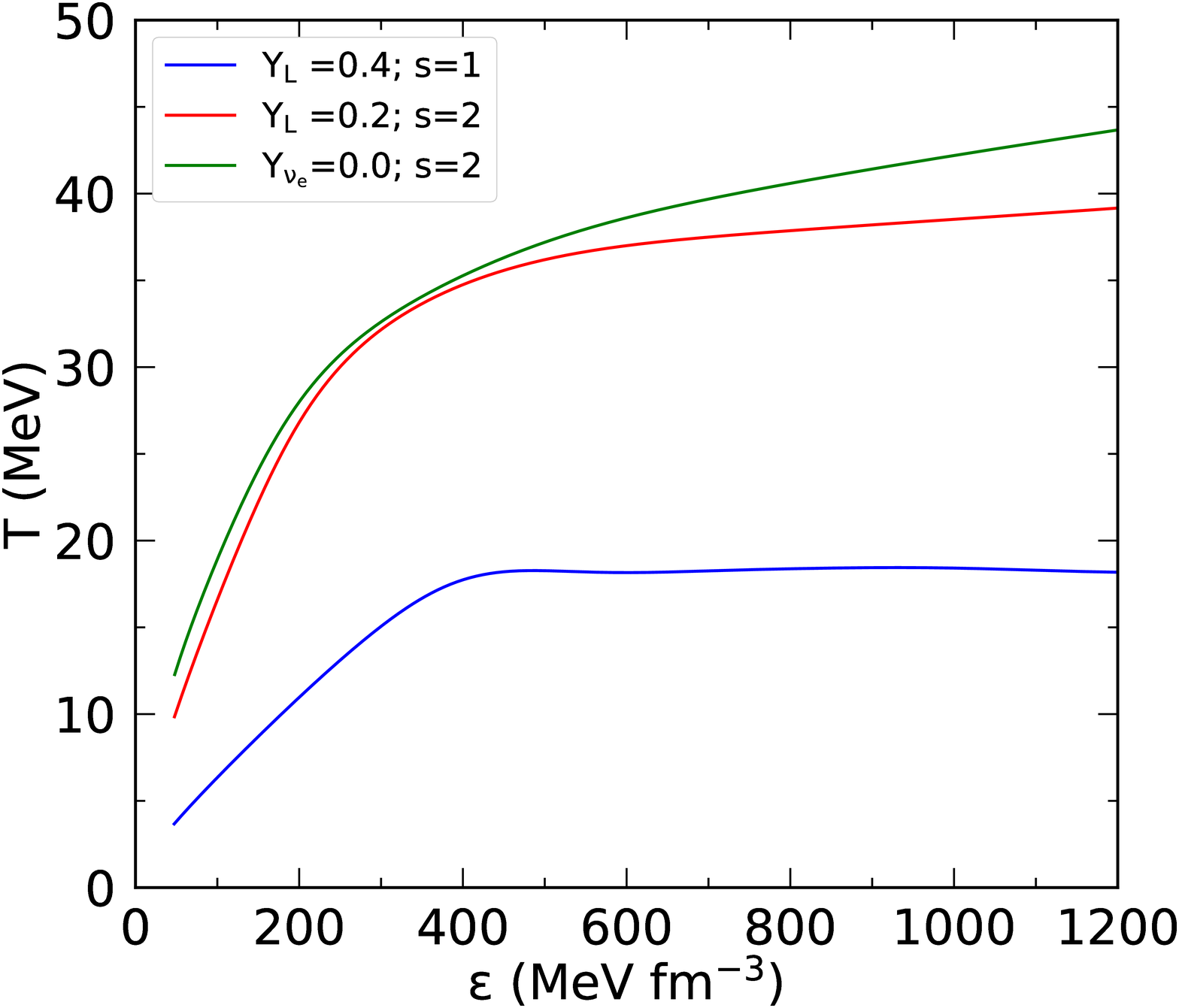}
    \caption{Same as \fref{fig:temp_vs_density_DD2_combined}, but for
      the GM1L model.}
    \label{fig:temp_vs_density_GM1L_combined}
\end{figure}
matter is characterized by $s = 2$ and $Y_{\nu_e} = 0$ (neutrinos are
no longer present) and its compositions are shown in
Figs.~\ref{fig:Hadronpop_DD2_YL0s2} and \ref{fig:Hadronpop_GM1L_YL0s2}
for DD2 and GM1L, respectively. Finally, after several minutes stars
with $s = 2$ and $Y_{\nu_e} = 0$ have cooled down to just $\sim
1$~MeV, containing core compositions similar to the one shown in
\fref{fig:hadronpop_DD2_T1}.

One sees from Figs.~\ref{fig:hadronpop_DD2_T1} through
\ref{fig:Hadronpop_GM1L_YL0s2} that both nuclear models, GM1L and DD2,
predict the same overall particle compositions in hot and dense
(proto-) neutron stars, despite the fact that the coupling constants
of the models are treated quite differently.  We recall that for the
DD2 model the coupling constants of all ($\sigma, \omega, \rho$)
mesons are density dependent, while for the GM1L model this is only
the case for the coupling constant of the $\rho$ meson. A noteworthy
difference, however, concerns the particle threshold densities which
tend to be somewhat lower for the DD2 model.

The temperature of PNS matter is shown in
Figs.~\ref{fig:temp_vs_density_DD2_combined} and
\ref{fig:temp_vs_density_GM1L_combined} for different combinations of
entropy and lepton number. For a constant entropy density of $s = 1$
the temperature varies between around 15 and 18~MeV over most of the
density ranges. The temperatures is more than twice as high for PNS
matter with $s = 2$.  Temperatures significantly higher than 40~MeV
are not reached in PNS matter. The increase in temperature shown in
these figures explains the ever more complex particle compositions
shown in Figs.~\ref{fig:Hadronpop_DD2_YL04s1} through
\ref{fig:Hadronpop_GM1L_YL0s2}.

\section{The Hadron-Quark Phase Transition}

In this chapter, we briefly turn to the study of quark matter in
compact stars.  The possible existence of such matter in compact stars
was already discussed in the 1960s by \citet{Ivanenko:1965} and in the
1970s by \citet{Itoh:1970uw,Fritzsch:1973, Baym:1976a,Keister:1976,
  Chapline:1977, Fechner:1978}.  Since then, a large number of
scientific papers have been published describing the possible
existence of quark matter in neutron stars with increasingly improved
theoretical models (see, for instance, \citet{Page:ARNPS2006,
  Alford:RMP2008, Burgio:PRD2008, Bonnano:2012AA, Orsaria:2013,
  Orsaria:2014, Baym:2017whm, Blaschke:Springer2018, Tolos:2020PPNP},
and references therein).  In the following we concentrate on the
hadron-quark phase transition as described by the Nambu--Jona-Lasinio
(NJL) model\index{Nambu--Jona-Lasinio model} \index{NJL model}
\cite{Klevanski:RMP1992, Hatsuda:1994PRept, Buballa:2005PhysRept,
  Fukushima:2010bq,Fukushima:2013ThePD}.  We shall use a non-local
variant of the NJL model, denoted 3nPNJL, which includes vector
interactions as well as the Polyakov loop. The lagrangian of this
model is given by \cite{Malfatti:2019PRC100}
\begin{eqnarray}
\mathcal{L} &=& \bar{\psi} (-i \gamma^\nu D_\nu + \hat{m} )\psi +
\frac{G_V}{2}j_{a}^{\mu} \, j_{a}^{\mu} - \frac{G_S}{2}\bigl[\,
  j_{a}^s \, j_{a}^s + j_{a}^p \, j_{a}^p \, \bigr] \nonumber \\ &-&
\frac{H}{4} A_{abc} \bigl[ \, j_{a}^s \, j_{b}^s \, j_{c}^s - 3\,
  j_{a}^s \, j_{b}^p \, j_{c}^p \, \bigr] + {\cal U}\,[\, {\cal A}\, ]
\, , \label{qm:lagrangian}
\end{eqnarray}
where ${\cal U}[{\cal A}]$ accounts for the Polyakov loop
  \index{Polyakov loop} dynamics and the $H$-dependent term is the
  't~Hooft term \index{'t~Hooft term} responsible for quark flavor
  mixing. The quark fields are described by $\psi \equiv (u,d,s)^T$
  and $\hat{m} = {\rm diag}(m_u, m_d, m_s)$ is the current quark mass
  matrix. The quantities $j_a^\mu$, $j_a^s$, and $j_a^p$ denote scalar
  ($s$), pseudo-scalar ($p$), and vector ($\mu$) interaction currents,
  respectively, and $G_S$ and $G_V$ are the scalar and vector coupling
  constants. It is customary to express $G_V$ in multiples of $G_S$
  and to write their ratio as $\mathrm{\zeta_v} \equiv G_V/G_S$.  The
  covariant derivative is given by $D_\nu \equiv \partial_\nu - i g
  A^a_\nu t^a$, where $A^a_\nu$ are the gluon fields and $t^a =
  \lambda^a/2$ the generators of SU(3) (for more details, see
  \citet{Malfatti:2019PRC100}).

To model the phase equilibrium between hadronic matter and quark
matter in a neutron star, we assume here that this equilibrium is of
first order and Maxwell-like, that is, the pressure in the mixed
\begin{figure}[tb]
\centering
{\includegraphics[width=9.5cm]{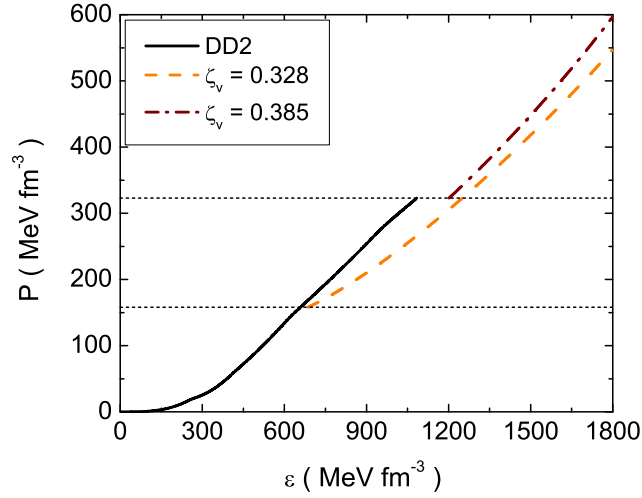}
  \caption{Pressure as a function of the energy density for different
    values of the vector coupling constant $\mathrm{\zeta_v}$ (see
    text) \citep{Malfatti:2019PRC100}. The black line represents the
    hadronic DD2 EOS and the dash-dotted and dashed lines are the EOSs
    of the quark (3nPNJL) phase for different $\mathrm{\zeta_v}$
    values. The horizontal lines mark the hadron-quark phase
    transitions.}  \label{eosh_dd2}}
\end{figure}
hadron-quark phase is constant. Theoretically the transitions could be
Gibbs-like as well, depending on the surface tension \index{Surface
  tension} at the hadron-quark interface. The value of the surface
tension is however only poorly known. Lattice gauge calculations, for
instance, predict surface tension values in the range of $0-100$
MeV~fm$^{-2}$ \cite{Kajantie:1991}. Using different theoretical models
for quark matter, a range of values for the hadron-quark surface
tension have been obtained in the literature (see, for example,
Refs.~\cite{2001PhRvD..64g4017A,2013PhRvC..88d5803L,2014PhRvD..89g4041K},
and references therein). According to theoretical studies, surface
tensions above around $70$~MeV~fm$^{-2}$ favor the occurrence of a
sharp (Maxwell-like) hadron-quark phase transition rather than a
softer Gibbs-like transition \cite{sotani2011, Yasutake2014}.

The EOS of both the hadronic phase and the quark phase is obtained
from the Gibbs relation
\begin{equation}
 \epsilon = - P + T S + \sum_i \mu_i \, n_i \, ,
 \label{eq:EoS}
\end{equation}
where pressure, entropy, and the particle number densities are given
by $P=-\Omega$, $S = \frac{\partial P}{\partial T}$, and $n_i =
\frac{\partial P}{\partial \mu_i}$, respectively.
\begin{figure}[tb]
\centering
  {\includegraphics[width=9.5cm]{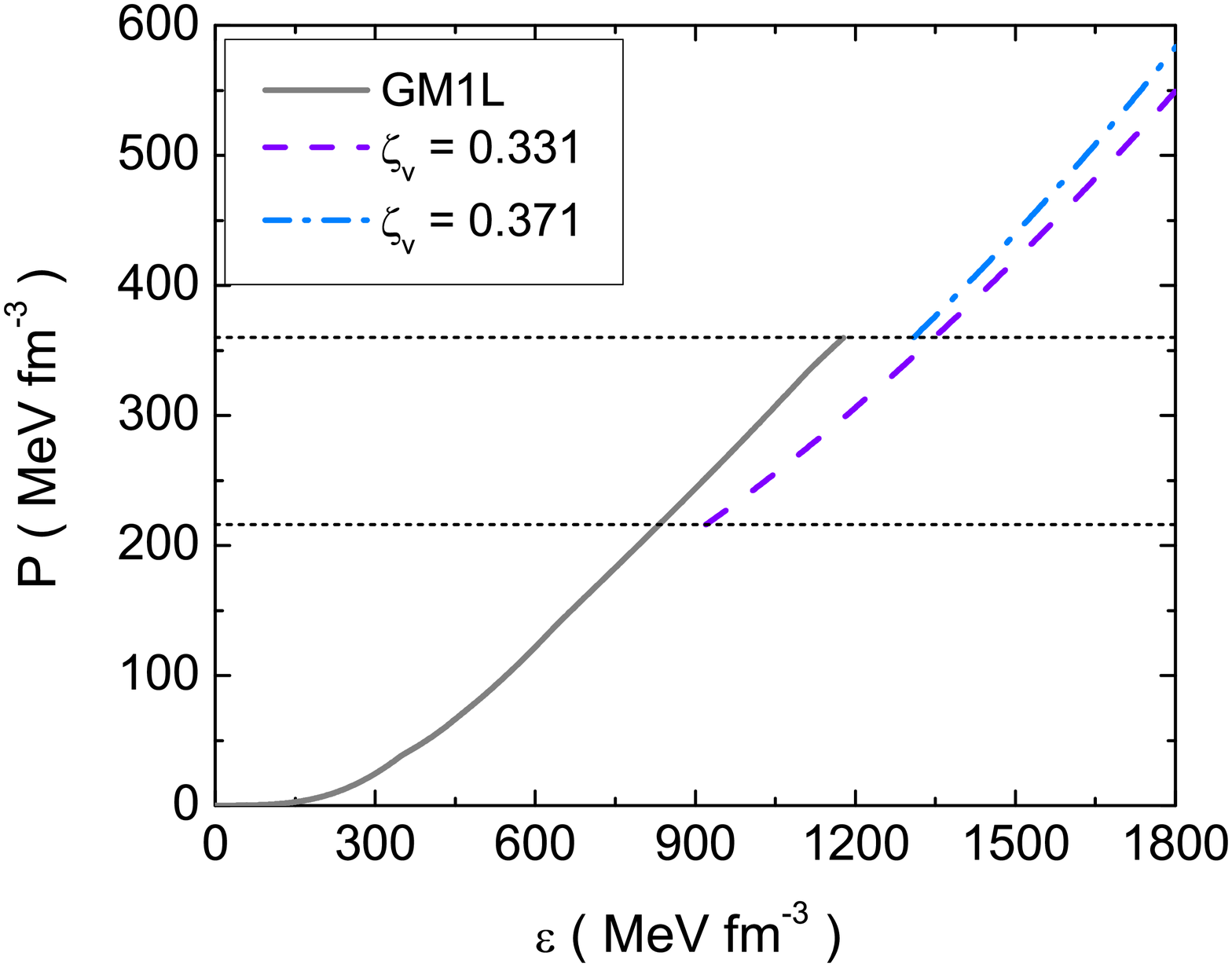}
    \caption{Same as \fref{eosh_dd2}, but for the hadronic GM1L
      EOS \citep{Malfatti:2019PRC100}.}  \label{eosh_gm1l}}
\end{figure}
To construct the hadron-quark phase transition we adopt the Gibbs
condition \index{Gibbs condition} for equilibrium between both phases,
expressed as
\begin{eqnarray}
  G_H(P, T) = G_Q (P, T) \, ,
  \label{eq:GHGQ}
\end{eqnarray}
where $G_{H}$ and $G_{Q}$ are the Gibbs free energies per baryon of
the hadronic ($H$) and the quark ($Q$) phase, respectively, to be
determined at a given pressure and transition temperature. The
crossing of $G_H$ and $G_Q$ in the $G-P$ plane then determines the
pressure and density at which the phase transition occurs for a given
transition temperature.  The expressions of $G_H$ and $G_Q$ are given
by
\begin{eqnarray}
G_i (P,T) = \sum_j \frac{n_j}{n} \mu_j \, ,
\end{eqnarray}
where $i=H$ or $Q$ and the sum over $j$ is over all the particles
present in each phase. For the hadron-quark phase transition, the
particle chemical potentials in each phase are different, so that is
becomes necessary to calculate the Gibbs free energy as a function of
pressure to construct the phase transition. Results for the
hadron-quark phase transitions are shown in \fref{eosh_dd2} for the
DD2 nuclear model and in \fref{eosh_gm1l} for the GM1L nuclear
model. Two phase transitions are visible in each figure, depending on
the value of the vector coupling constant, $\mathrm{\zeta_v} ~(=
G_V/G_S)$.  The solid black and gray lines in these figures represent
the hadronic DD2 and GM1L EOSs, respectively, and the dash-dotted and
dashed lines are the EOSs of the quark phase computed for the 3nPNJL
model. The horizontal lines indicate the locations of the hadron-quark
phase transitions where $G_H(P, T=0) = G_Q (P, T=0)$ according to
\eref{eq:GHGQ}.
\begin{figure}[tb]
\centering
{\includegraphics[width=10.5cm]{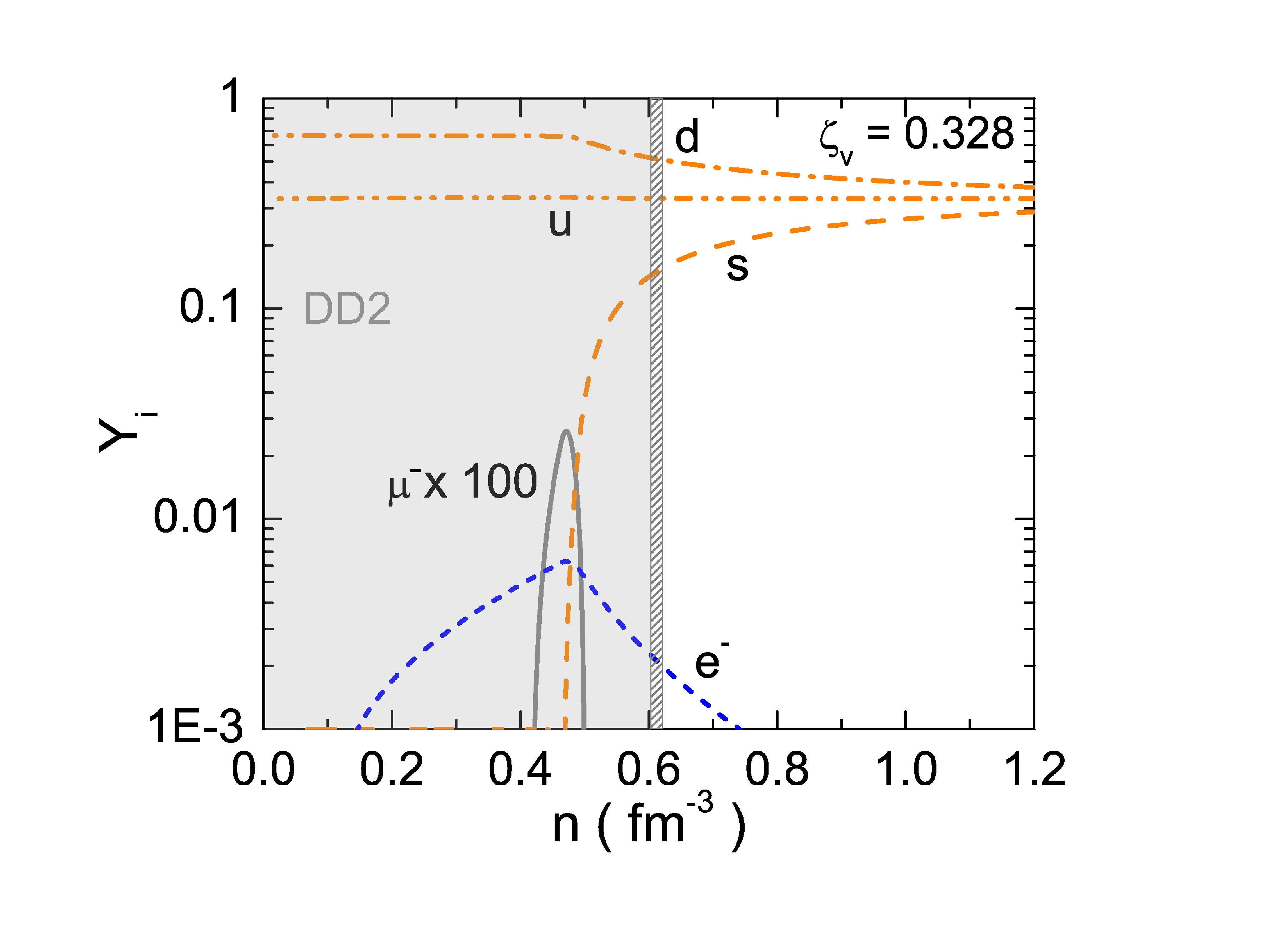}
\caption{Particle population of stellar quark matter at zero
  temperature as a function of baryon number density
  \citep{Malfatti:2019PRC100}. The gray area indicates the density
  regime where matter described by the hadronic DD2 model exists. The
  hadron phase ends abruptly at the vertical line slightly above
  $0.6~{\rm fm}^{-3}$.  The population of muons is increased by a
  factor of 100 to make it visible. The strength of the vector
  repulsion among quarks is
  $\mathrm{\zeta_v}=0.328$.}  \label{quark_pop_dd2}}
\end{figure}
The hadronic and the quark matter EOS are very similar at pressures
where $G_H(P, T=0) \approx G_Q (P, T=0)$ \cite{Malfatti:2019PRC100}).
\begin{figure}[tb]
\centering
{\includegraphics[width=10.5cm]{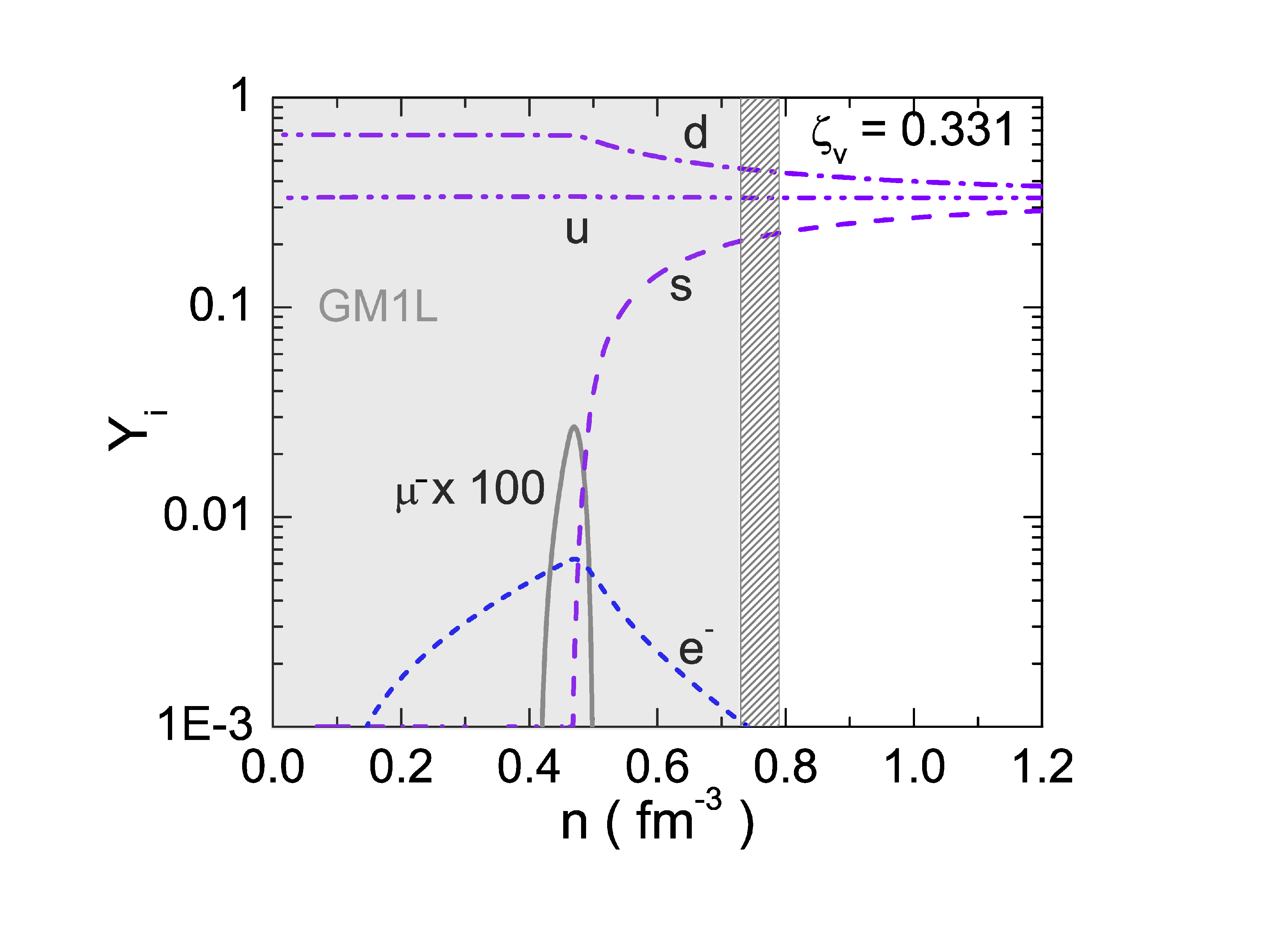}
\caption{Same as \fref{quark_pop_dd2}, but for a vector repulsion
  among quarks of $\mathrm{\zeta_v}=0.331$
  \citep{Malfatti:2019PRC100}.}  \label{quark_pop_gm1l}}
\end{figure}
This makes it difficult to distinguish between the two phases in the
relevant pressure regions, $P \sim 100 - 400$~MeV/fm$^3$. This can be
interpreted as a masquerading behavior of dense matter, different from
pure deconfined quark matter (see \citet{Malfatti:2019PRC100}, and
references therein for details).

The $2 \, M_\odot$ constraint of PSR J1614-2230 and PSR J0348+0432
\cite{Demorest:2010bx,Lynch:2012vv,Antoniadis:2013pzd,Arzoumanian_2018}
and the assumption that quark matter exists in the cores of neutron
stars have been used to determine the range of the vector coupling
constant $\mathrm{\zeta_v}$ in the quark matter phase. This leads to
$0.331 < \mathrm{\zeta_v} < 0.371$ for GM1L, and $0.328 <
\mathrm{\zeta_v} < 0.385$ for DD2, where the lower bounds are
determined by the $2\, M_\odot$ mass constraint and the upper bounds
by the requirement that quark matter exists in the cores of neutron
stars. In Figs.~\ref{quark_pop_dd2} and \ref{quark_pop_gm1l} we show
the quark compositions of cold neutron stars computed for GM1L in
combination with 3nPNJL and DD2 in combination with 3nPNJL,
respectively. 

\section{The Parameters of the Hadronic Theory}\label{sec:parameters}

For this study, we will consider three popular nuclear parametrization
sets which are denoted SWL\index{SWL parameter set}, GM1L\index{GM1L
  parameter set} and DD2\index{DD2 parameter set}
\citep{Spinella2017:thesis,Spinella:2018bdq, Typel:2009sy}. The
parameter values of these sets are shown in
\tref{table:parametrizations} and the corresponding saturation
properties of symmetric nuclear matter are shown in
\tref{table:properties} \citep{Malfatti:2019PRC100}. These are the
nuclear saturation density $n_0$, energy per nucleon $E_0$, nuclear
compressibility $K_0$, effective nucleon mass $m^*_N/m_N$, asymmetry
energy $J$, asymmetry energy slope $L_0$, and the value of the nucleon
potential $U_N$. The values of $L_0$ listed in \tref{table:properties}
are in agreement with the value of the slope of the symmetry energy
deduced from nuclear experiments and astrophysical observations
\cite{RevModPhys.89.015007}.
\begin{table}[htb]
  \tbl{Parameters of the SWL and GM1L
    \citep{Spinella2017:thesis,Spinella:2018bdq} and DD2
    \citep{Typel:2009sy} parametrizations used in this work.}
    {\begin{tabular}{ccccc} \toprule $~~$Parameters$~~$ &$~~$Units$~~$ & $~~$SWL$~~$ &
        $~~$GM1L$~~$ &DD2$~~$\\ \colrule
        $m_{\sigma}$ &GeV &0.550  & 0.550 & 0.5462 \\
        $m_{\omega}$ &GeV &0.783  &0.783 & 0.783 \\
        $m_{\rho}$  &GeV  &0.763  & 0.7700 & 0.7630 \\
        $g_{\sigma N}$ &$-$    &9.7744 & 9.5722 & 10.6870 \\
        $g_{\omega N}$ &$-$     &10.746 & 10.6180 & 13.3420 \\
        $g_{\rho N}$  &$-$     &7.8764 & 8.1983  & 3.6269 \\
        $\tilde{b}_{\sigma}$ &$-$  &0.003798 &0.0029 & 0  \\
        $\tilde{c}_{\sigma}$ &$-$  &$-0.003197$  &$-0.001068$ & 0 \\
        $a_{\sigma}$  &$-$     &0 & 0 &1.3576 \\
        $b_{\sigma}$  &$-$     &0 & 0 & 0.6344 \\
        $c_{\sigma}$  &$-$     &0 & 0 & 1.0054 \\
        $d_{\sigma}$  &$-$     &0 & 0 & 0.5758 \\
        $a_{\omega}$  &$-$     &0 &0 &1.3697 \\
        $b_{\omega}$  &$-$     &0 & 0 &0.4965 \\
        $c_{\omega}$  &$-$     &0 & 0 & 0.8177 \\
        $d_{\omega}$ &$-$      &0 &0 & 0.6384 \\
        $a_{\rho}$  &$-$      &0.3796  &0.3898 &0.5189 \\ \botrule
\end{tabular}}
\label{table:parametrizations}
\end{table}
The DD2 parametrization is designed such that it eliminates the need
for the nonlinear self-interactions of the $\sigma$ meson shown in
Eqs.~(\ref{eq:Blag}) and (\ref{eq:sigma-field-eqn}) \citep{Malfatti:2019PRC100}. The
nonlinear terms are therefore only considered for the GM1L model.
As already mentioned in \sref{ssec:lagrangian}, the baryons considered
in this study to populate NS matter include all states of
the spin-$\frac{1}{2}$ baryon octet comprised of the nucleons $(n,p)$
and hyperons $(\Lambda,\Sigma^+,\Sigma^0,\Sigma^-,\Xi^0,\Xi^-)$. In
addition, all states of the spin-$\frac{3}{2}$ delta isobar
$\Delta(1232)$ ($\Delta^{++},\Delta^+,\Delta^0,\Delta^-$) are taken
into account as well.

\subsection{The Meson-Hyperon Coupling Space}\label{ssec:meson.hyperon}

A detailed discussion of the meson-hyperon coupling constants $g_{iH}$
(where $i=\sigma, \omega, \rho$) can be found in
\citep{Spinella2017:thesis,Malfatti:2019PRC100,Malfatti:2020PRC102,
  Spinella:2020WSBook}. As usual we express the values of the
meson-hyperon coupling constant, $g_{iH}$, in terms of the
meson-nucleon coupling strength, $g_{i N}$, that is, $x_{i H} = g_{i
  H}/g_{iN}$.  The meson-hyperon couplings are not well constrained
experimentally compared to those of the nucleons. However, the scalar
meson-hyperon couplings ($x_{\sigma H}$) can be constrained by the
available experimental data on hypernuclei, but their calculation
first requires the determination of the vector meson-hyperon couplings
($x_{\omega H}$). The coupling scheme used in our study is based on
the Nijmegen extended-soft-core (ESC08) model
\citep{Rijken:2010PTPS.185}. The scalar meson-hyperon coupling
constants ($x_{\sigma H}$) can be fit to the hyperon potential depths,
$U_H$, at nuclear saturation density, $n_0$. Our parameters sets are
fitted to potential depths of $U_\Lambda = -28$~MeV, $U_\Xi =
-18$~MeV, and $U_\Sigma = +30$~MeV (see
\citet{Schaffner:2000PRC62,Spinella:2020WSBook,Tolos:2020PPNP,Friedman:2021.PLB},
and references cited therein). The values of the isovector
meson-hyperon coupling constants are chosen as $x_{\rho H} = 2
|I_{3H}|$ \citep{Schaffner:2012PRC85,Miyatsu:2013PRC88,Maslov:2016NPA}.

\subsection{$\Delta(1232)$ Isobars}\label{ssec:delta.isobars}

The potential presence of the delta isobar \index{$\Delta$ isobar}
$\Delta$(1232) in neutron star matter
\citep{Pandharipande:1971NPA,Sawyer:1972ApJ176,Boguta1982,Huber:1998a}
has been relatively ignored, especially when compared to the attention
\begin{table}[tb]
\tbl{Properties of symmetric nuclear matter at saturation density for
  the SWL and GM1L \citep{Spinella2017:thesis,Spinella:2018bdq} and DD2
  \citep{Typel:2010a} parametrizations.}
{\begin{tabular}{ccccc}\toprule $~~$Saturation property$~~$ &$~~$Units$~~$
    &$~~$SWL$~~$ &$~~$GM1L$~~$ &$~~$DD2$~~$ \\ \colrule
    $n_0$ &fm$^{-3}$ &0.150  & 0.153 & 0.149\\
    $E_0$ &MeV      &$-16.0$ & $-16.3$ & $-16.02$ \\
    $K_0$ &MeV      &260.0   & 300.0 & 242.7 \\
    ${m^*_N}/{m_N}$  &$-$    &0.70    & 0.70 &0.56 \\
    $J$   &MeV       &31.0    & 32.5  &31.67 \\
    $L_0$ &MeV     &55.0     & 55.0   & 55.04\\
    $U_N$ &MeV     &$-64.6$  &$-65.5$ &$-75.2$ \\ \botrule
    \end{tabular}}
    \label{table:properties}
\end{table}
that hyperons have received in the literature. It is reasonable to
assume $\Delta$s would not be favored in NS matter for a number of
reasons.  First, their rest mass is greater than both the $\Lambda$
and $\Sigma$ hyperons.  Second, negatively charged baryons are
generally favored as their presence reduces the high Fermi momenta of
the leptons, but the $\Delta^-$ has triple the negative isospin of the
neutron ($I_{3\Delta^-}=-3/2$), and thus its presence should be
accompanied by a substantial increase in the isospin asymmetry of the
system.  However, these arguments now appear to be largely invalid
since recent many-body calculations paint a different picture
\citep{Dexheimer:ApJ2008,Chen:2009PC79, Schuerhoff:2010ApJ,
  Lavagno:2010PRC81, Drago:2014PRC90,
  Cai:2015PRC92,Zhu:2016PRC,Spinella2017:thesis,Li:2018PLB783,Malfatti:2020PRC102}.

Recent theoretical works have suggested conflicting constraints on the saturation
potential of the $\Delta$s in symmetric nuclear matter given by
\begin{equation} \label{eq:delta-potential}
  U_{\Delta}(n_0) = x_{\omega\Delta}g_{\omega N} \bar\omega
  - x_{\sigma\Delta}g_{\sigma N} \bar\sigma + \tilde R\, ,
\end{equation}
where $\tilde R$ denotes the rearrangement term of \eref{eq:rear}.
\citet{Drago:2014PRC90} incorporated a number of experimental and
\begin{table}[b]
\tbl{Saturation potentials of nucleons and $\Delta$s in symmetric
  nuclear matter with $x_{\sigma\Delta}=x_{\omega\Delta}=1.1$ and
  $x_{\rho\Delta} = 1.0$ \citep{Spinella2017:thesis}.}
{\begin{tabular}{cccc}\toprule
      Potential & SWL & GM1L & DD2\\ 
      $U_{N}(n_0)$ (MeV) & $-64.6$ & $-65.5$ & $-75.2$  \\
      $U_{\Delta}(n_0)$ (MeV) & $-71.1$ & $-72.1$ & $-86.0$  \\
\botrule
\end{tabular}}
\label{table:delta-potentials}
\end{table}
theoretical results to deduce the following range for
$U_{\Delta}$ at $n_0$,
\begin{equation} \label{eq:drago-constraint}
  -30\,\mathrm{MeV}+U_N(n_0) \lesssim U_{\Delta}(n_0) \lesssim U_N(n_0)\,,
\end{equation}
indicating a slightly more attractive potential than that of the
nucleons.  Included in these analyses was an analysis of the
photo-excitation of nucleons to $\Delta$s that suggested the following
relation between the scalar and vector couplings,
\begin{equation} \label{eq:wehrberger-constraint}
  0 < x_{\sigma\Delta}-x_{\omega\Delta} < 0.2\,.
\end{equation}
\citet{Kolomeitsev:2017NPA} cited numerous studies of $\Delta$
production in heavy ion collisions to suggest a less attractive
potential in the range $U_N(n_0) \lesssim U_{\Delta}(n_0) \lesssim
\frac{2}{3}U_N(n_0)$, finally settling on $U_{\Delta}(n_0) \approx
-50$~MeV as a best estimate \citep{Riek:2009PRC80}.  However,
it is worth noting that constraining the potential does not directly
constrain $x_{\sigma\Delta}$ or $x_{\omega\Delta}$, rather the
relationship between the two.

The meson-$\Delta$ coupling space will be systematically investigated
in Section \ref{ssec:meson.delta}, but first the particle number
densities in the presence of both hyperons and $\Delta$s will be
examined with the following set of couplings,
\begin{equation}
  x_{\sigma\Delta}=x_{\omega\Delta}=1.1,\;\;x_{\rho\Delta} = 1.0 \, .
  \label{eq:meson-delta-couplings}
\end{equation}
These lead to saturation potentials more attractive than that of the
nucleons as shown in Table \ref{table:delta-potentials}.
The scalar and isovector meson-hyperon coupling constants will
continue to be determined as described in \sref{ssec:meson.hyperon}
and the vector meson-hyperon couplings will be given by the SU(3)
ESC08 model.

The properties of maximum mass NSs made of hyperonic matter with and
without the $\Delta$ states are shown in
\tref{table:deltas-max-mass}. The properties include the stellar mass
$M$, the radius $R$, and the baryon number density $n_{\mathrm{c}}$ at
the center of the stars. Also shown are the radii $R_{1.4}$ of neutron
stars with a canonical mass of $1.4\, M_\odot$.
\begin{table}[htb]
    \tbl{Properties of maximum mass NSs with $\Delta$s and
      hyperons with vector meson-hyperon coupling constants given in
      SU(3) symmetry with the ESC08 model \citep{Spinella2017:thesis}.}
        {\begin{tabular}{ccccccccccc}\toprule &
            \multicolumn{3}{c}{$~~~~~~~~~$Hyperons} &$~$ &
            \multicolumn{4}{c}{$~~~~~~~~~~~~$Hyperons plus $\Delta$s}
            \\ \cline{2-5} \cline{7-10} \noalign{\smallskip} $~~$EOS
            $~~$ & $M$ & $R$ & $n_{\mathrm{c}}$ &$R_{1.4}$ & & $M$ &
            $R$ & $n_{\mathrm{c}}$ & $R_{1.4}$\\ &$(M_{\odot})$ &(km)
            &(1/fm$^3$) &(km) & &$(M_{\odot})$ & (km) &(1/fm$^3$)
            &(km)\\ \colrule
            SWL  & 2.01 & 11.5 & 0.98 &12.80 & & 2.02  & 11.4 & 1.00 & 12.85 \\            
            GM1L & 2.04 & 11.6 & 0.95 &12.82 & & 2.04  & 11.5 & 0.97 & 12.90 \\
            DD2  & 2.09 & 12.1 & 0.89 &13.45 & & 2.11  & 11.9 & 0.92 & 13.28 \\ \botrule
    \end{tabular}}
    \label{table:deltas-max-mass}
\end{table}
As can be seen, including the $\Delta$ baryon actually leads to equal
or marginally greater maximum masses in both cases.  While the maximum
masses are very similar, the mass-radius curves differ slightly due to
the low density appearance of the $\Delta^-$ that causes a bend toward
lower radii reducing the radius of the canonical $1.4\,M_{\odot}$
NS. We also note that specifying the vector meson-hyperon
couplings with the SU(3) ESC08 model rather than SU(6) is necessary in
order to satisfy the $\sim 2 \,M_\odot$ mass constraint with the GM1L
and DD2 parametrizations \citep{Spinella2017:thesis}.

\begin{figure}[htb]
\centering
{\includegraphics[width=7.5cm]{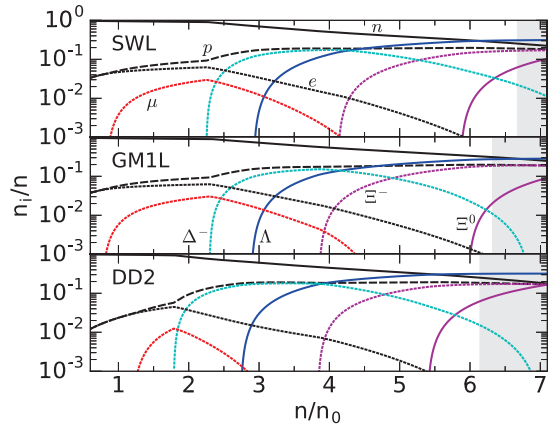}}
\caption{The relative number density of particles in cold NS
  matter as a function of baryon number density (in units of the
  saturation density) \citep{Spinella2017:thesis}. The meson-$\Delta$
  coupling constants are $x_{\sigma\Delta}=x_{\omega\Delta}=1.1$, and
  $x_{\rho\Delta} = 1.0$, and the vector meson-hyperon coupling
  constants are given by the SU(3) ESC08 model.  The gray shading
  indicates baryon number densities beyond the maximum for the given
  parametrization.}
    \label{fig:number-density-deltas}
\end{figure}

The relative particle number densities for the SU(3) coupling scheme
are presented in \fref{fig:number-density-deltas} for the GM1L and DD2
parametrizations. For GM1L the $\Delta^-$ is the first additional
baryon to be populated at $\sim 2.3\,n_0$ and reaches nearly the same
number density as the proton before it starts being replaced by the
$\Xi^-$ at around $4\,n_0$. In DD2 the $\Delta^-$ again precedes the
onset of hyperonization but appears at an extremely low baryon number
density of around $1.8\,n_0$ and again reaches densities comparable to
that of the proton before beginning to decline due to the population
of the $\Xi^-$ at around $4\,n_0$. At low densities the $\Sigma^-$ and
$\Xi^-$ may be disfavored in comparison to the $\Delta^-$ due to a
repulsive potential and significantly higher rest mass respectively.
However, the low $\Delta^-$ critical density in DD2 is primarily due
to the density dependence of the isovector meson-baryon coupling that
greatly reduces the isovector contribution to the $\Delta$ chemical
potential compared to standard relativistic mean-field calculations.
The early appearance of the $\Delta^-$ is directly related to the
slope of the asymmetry energy, $L_0$, as discussed by
\citet{Drago:2014PRC90}.  The extreme low density appearance of the
$\Delta^-$ has important consequences for the mass-radius curve of a
NS, since it bends toward smaller radii much more substantially
compared to EOSs where $\Delta$s are absent. The effect is the most
drastic for the DD2 parametrization, where the presence of $\Delta$s
reduces the canonical NS radius by about a kilometer compared to the
nucleonic and hyperonic EOSs.

\Fref{fig:number-density-deltas} shows that the isospin neutral
$\Lambda$ is the first hyperon to appear at around $3\,n_0$, reducing
the high Fermi momentum of the neutron. The isospin negative $\Xi^-$
($I_{3\Xi^-}\!=\!-1/2$) follows at around $4\,n_0$ as stated,
replacing the more isospin negative $\Delta^-$
($I_{3\Delta^-}\!=\!-3/2$), reducing isospin asymmetry.  The isospin
positive $\Xi^0$ ($I_{3\Xi^0}=+1/2$) is populated at around $6\,n_0$
in GM1L and $5.5\,n_0$ in DD2, contributing to the replacement of the
isospin negative neutron further reducing isospin asymmetry. Finally,
the presence of the $\Delta^-$ excludes the previous appearance of the
$\Sigma^-$ in DD2.

\subsection{The Meson-$\Delta(1232)$ Coupling Spaces}\label{ssec:meson.delta}

Little empirical data exists that can unambiguously constrain the
meson-$\Delta$ coupling\index{Meson-$\Delta$ coupling} constants.  As
a result, in order to study the presence of $\Delta$s in NS matter and
their consequent effect on NS properties most studies choose just a
few sets of coupling constants to analyze, typically in the vicinity
of universal coupling
($x_{\sigma\Delta}=x_{\omega\Delta}=x_{\sigma\Delta}=1$). In this
section we seek to explore a large portion of the meson-$\Delta$
coupling space to more thoroughly investigate $\Delta$s in cold as
well as hot (proto-) neutron star matter and their effect on stellar
properties \citep{Spinella2017:thesis,Malfatti:2020PRC102}.

\subsubsection{The $\sigma\omega\Delta$ Coupling Space}
\label{sec:sigma-omega-Delta-coupling-space}

\begin{figure}[htb]
\centering
{\includegraphics[width=7.5cm]{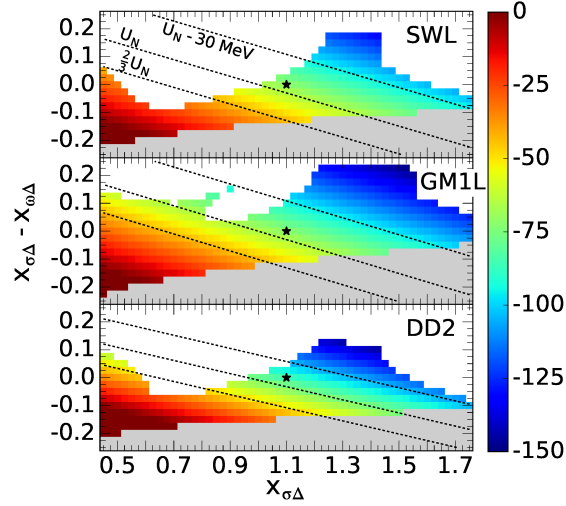}}
\caption{Nuclear saturation potential (in MeV) of $\Delta$s in
  symmetric nuclear matter in the $\sigma\omega\Delta$ coupling space
  \citep{Spinella2017:thesis}. Hyperons were included with the vector
  meson-hyperon given by the SU(3) ESC08 model.  The star marker
  indicates the location of  $x_{\sigma\Delta}=x_{\omega\Delta}=1.1$ and
  $x_{\rho\Delta} = 1.0$.  Dashed contours are lines of constant
  potential as labeled and represent possible constraints. Gray pixels
  indicate that no $\Delta$s were populated for the given set of
  couplings. White pixels indicate couplings for which the effective
  mass of at least one baryon became negative before the maximum
  baryon number density of the NS was reached.}
    \label{fig:deltas-potential-heatmap}
\end{figure}

The exploration of the $\sigma\omega\Delta$ coupling space begins with
a heatmap for the $\Delta$ saturation potential $U_{\Delta}(n_0)$ in
symmetric nuclear matter given in 
\fref{fig:deltas-potential-heatmap}. First, it is important to note that
including the $\Delta$ baryon can result
in a rapid increase of the scalar field ($\bar{\sigma}$) causing a
correspondingly rapid decrease in the effective baryon masses (see
\eref{eq:mstar}), some becoming negative before the maximum
central density is reached.  This invalidates the EOS and the
associated couplings; as a result, much of the $\sigma\omega\Delta$
coupling space is not accessible with a given EOS model and
parametrization.  These areas are identified in 
\fref{fig:deltas-potential-heatmap} and the heatmaps to follow as empty
(white) pixels. Further, $\Delta$s do not populate for a significant
region of the coupling space due to the presence of hyperons, and
these couplings are identified by the gray pixels.  In particular, we
find that for the chosen parametrizations $\Delta$s are largely
absent when $x_{\sigma\Delta}-x_{\omega\Delta} \lesssim -0.1$, and do
not populate at all when $x_{\sigma\Delta}-x_{\omega\Delta} < -0.2$.
A study by \citet{Zhu:2016PRC} investigated $\Delta$s in the
density-dependent relativistic Hartree-Fock (DDRHF) approach, and much
of the analysis therein was conducted with $x_{\sigma\Delta} = 0.8$
and $x_{\omega\Delta} = 1.0$.  However, they did not
account for hyperonization and our results suggest that $\Delta$s may
not even appear with the  given choice of couplings, illustrating
the importance of simultaneously considering hyperons. Finally, our
investigation of the coupling range spanning $-0.25 \lesssim
x_{\sigma\Delta}-x_{\omega\Delta} \lesssim 0.25$ appears sufficient,
as outside this range $\Delta$s either do not populate or their
presence results in an EOS that is erroneous due to either a negative
effective baryon mass or a pressure that is not monotonically
increasing.  Thus, if $\Delta$s are to appear in NS matter,
$x_{\sigma\Delta}$ and $x_{\omega\Delta}$ are likely relatively close
in value.

\Fref{fig:deltas-potential-heatmap} indicates that an increase in
either $x_{\sigma\Delta}$ or $x_{\sigma\Delta}-x_{\omega\Delta}$
results in a decrease in $U_{\Delta}$, the potential becoming more
attractive.  The region between the top two contours is
consistent with the potential constraint suggested by
\citet{Drago:2014PRC90} given in \eref{eq:drago-constraint}.
Satisfaction of this constraint requires that $1.0 \lesssim
x_{\sigma\Delta} \lesssim 1.7$ in SWL and DD2, and $0.9 \lesssim
x_{\sigma\Delta} \lesssim 1.6$ in GM1L.  Requiring that
\eref{eq:wehrberger-constraint} be simultaneously satisfied completely
excludes the bottom half of the coupling space and leaves only a
limited region in the top-middle that is consistent with the
constraints, this region including the previously employed couplings
indicated by the star marker and given in
\eref{eq:meson-delta-couplings}.  The region between the bottom two
contours is consistent with the potential constraint suggested in
\citet{Kolomeitsev:2017NPA}.  However, if we simultaneously require
the satisfaction of \eref{eq:wehrberger-constraint} here the SWL and
DD2 parametrizations are completely excluded, and the $\Delta$
couplings are limited to a very small range in the GM1L
parametrization.
\begin{figure}[htb]
\centering
{\includegraphics[width=7.5cm]{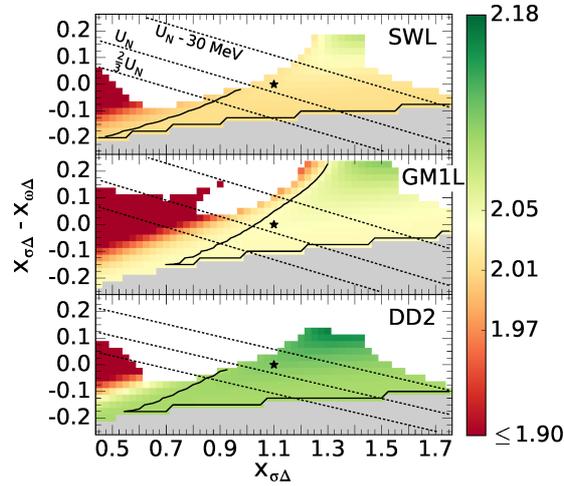}}
\caption{Maximum mass (in solar mass units $M_{\odot}$) in the
  $\sigma\omega\Delta$ coupling space
  \citep{Spinella2017:thesis}. Hyperons were included with the vector
  meson-hyperon coupling constants given by the SU(3) ESC08
  model. Solid lines are maximum mass contours for the associated
  hyperonic EOS (no $\Delta$s) in the ESC08 model.  Colorbar tick
  marks represent the maximum mass constraints set by PSR J0348+0432
  ($1.97-2.05\,M_{\odot}$ at $1\sigma$, and $1.90-2.18\,M_{\odot}$ at
  $3\sigma$).  Markers, contours, and pixels are as described for
  \fref{fig:deltas-potential-heatmap}.}
    \label{fig:deltas-mass-heatmap}
\end{figure}

The maximum mass of NSs in the $\sigma\omega\Delta$ coupling
space is shown in \fref{fig:deltas-mass-heatmap}.  The maximum mass
constraint is satisfied by the majority of the meson-$\Delta$ coupling
space in all parametrizations, with large regions producing a maximum
mass greater than that of the purely hyperonic EOS with ESC08 vector
couplings indicated by the solid contours.  Consequently the maximum
mass constraint alone does not serve to constrain $x_{\sigma\Delta}$
and $x_{\omega\Delta}$ significantly.  The highest maximum masses
appear where both the $\Delta$ saturation potential is the most
\begin{figure}[htb]
\centering
{\includegraphics[width=7.5cm]{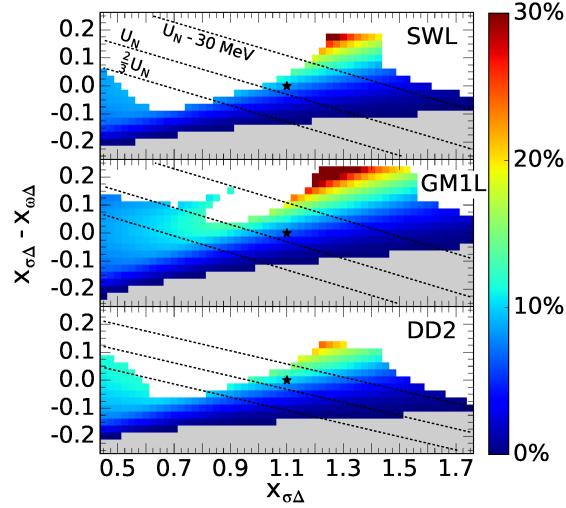}}
\caption{Delta isobar fraction (percentage) of the maximum mass NS in
  the $\sigma\omega\Delta$ coupling space \citep{Spinella2017:thesis}.
  Hyperons were included with the vector meson-hyperon coupling
  constants given by the SU(3) ESC08 model.  Markers, contours, and
  pixels are as described for \fref{fig:deltas-potential-heatmap}.
  The $\Delta$ fractions for $x_{\sigma\Delta}= x_{\omega\Delta} =
  1.1$ are as follows: $f_{\Delta}^{\mathrm{SWL}} = 8.41\%$,
  $f_{\Delta}^{\mathrm{GM1L}} = 6.31\%$, and
  $f_{\Delta}^{\mathrm{DD2}} = 10.2\%$.}
    \label{fig:deltas-fraction-heatmap}
\end{figure}
attractive and the difference between the scalar and vector
meson-$\Delta$ couplings is the greatest. Satisfaction of both
\eref{eq:wehrberger-constraint} and the mass constraint requires
$x_{\sigma\Delta} > 1.0$ for SWL, $x_{\sigma\Delta} > 0.9$ for GM1L,
and $x_{\sigma\Delta} > 0.975$ for DD2. \citet{Kolomeitsev:2017NPA}
concluded that the most likely value for $U_{\Delta}(n_0) \approx
-50\,\mathrm{MeV}$, and we find that the maximum mass constraint can
only be satisfied with this potential provided $x_{\omega\Delta} >
x_{\sigma\Delta}$, violating \eref{eq:wehrberger-constraint}
\citep{Riek:2009PRC80,Kolomeitsev:2017NPA}.

The total number $N_\Delta$ of delta isobars present in a given
NS model can be calculated from
\begin{equation} 
  \frac{dN_{\Delta}}{dr} = \frac{4\pi r^2}{\sqrt{1-2m(r)/r}}
  \sum_{\Delta} n_{\Delta}(r)\, ,
  \label{eq:delta-fraction}
\end{equation}
which is to be solved in combination with the TOV equation that will
be introduced in \sref{sec:GR}.  $N_\Delta$ is given in
\fref{fig:deltas-fraction-heatmap} as a fraction of the total baryon
number, $f_{\Delta} = N_{\Delta}/N_B$. The $\Delta$ fraction varies
considerably in the range $2\% \lesssim f_{\Delta} \lesssim 18\%$ when
the $\sigma\omega\Delta$ couplings are consistent with the constraints
given in \eref{eq:drago-constraint} and
\eref{eq:wehrberger-constraint}. However, a quick examination of the
same region in \fref{fig:deltas-mass-heatmap} reveals that this
variance in $f_{\Delta}$ has little to no effect on the maximum
stellar mass. It appears that there is a $f_{\Delta}$ hot-spot that is
centered in a region of the $\sigma\omega\Delta$ coupling space
\begin{figure}[htb]
\centering
{\includegraphics[width=7.5cm]{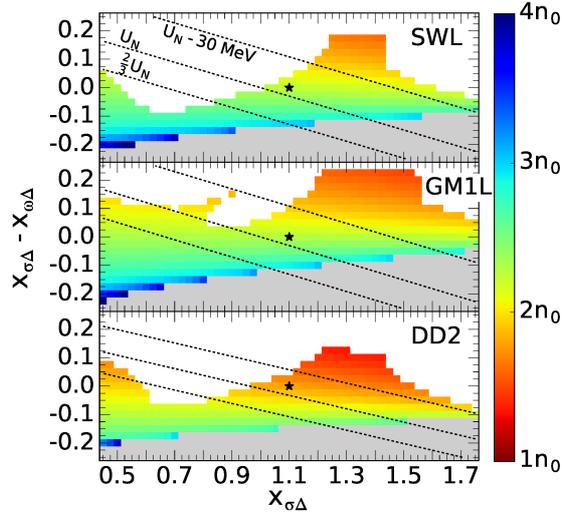}}
\caption{Critical baryon number density (in units of $n_0$) for the
  appearance of $\Delta$s in the $\sigma\omega\Delta$ coupling space
  \citep{Spinella2017:thesis}.  Hyperons were included with the vector
  meson-hyperon coupling constants given by the SU(3) ESC08 model.
  Markers, contours, and pixels are as described for Figure
  \ref{fig:deltas-potential-heatmap}.}
\label{fig:deltas-onset-heatmap}
\end{figure}
inaccessible to the GM1L and DD2 EOS models, but
the predictable result is that $f_{\Delta}$ increases with an increase
in $x_{\sigma\Delta}-x_{\omega\Delta}$ when $x_{\sigma\Delta}\gtrsim
0.8$.  The DD2 parametrization presents with the highest $f_{\Delta}$
for the smallest difference $x_{\sigma\Delta}-x_{\omega\Delta}$,
followed by SWL and then GM1L.

\begin{figure}[htb]
\centering
{\includegraphics[width=7.5cm]{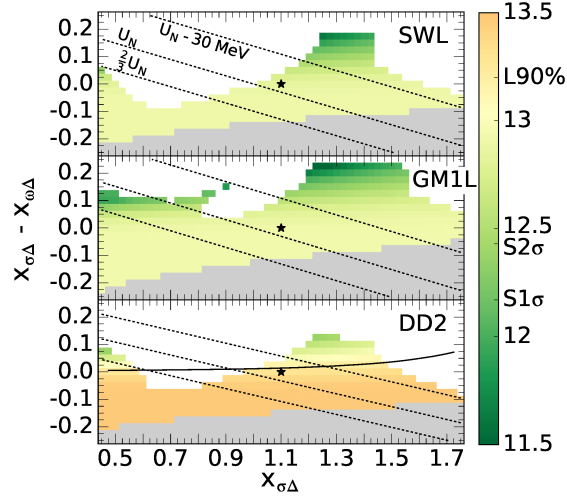}}
\caption{Radius (in km) of the canonical $1.4\,M_{\odot}$ NS in the
  $\sigma\omega\Delta$ coupling space \citep{Spinella2017:thesis}.
  Hyperons were included with the vector meson-hyperon coupling
  constants given by the SU(3) ESC08 model.  The solid contour in the
  bottom (DD2) panel represents the 13.2~km upper limit of the radial
  constraint from \citet{Lattimer:2014EPJA} represented as L90\% on
  the colorbar. The $1\sigma$ and $2\sigma$ upper limits from
  \citet{Steiner:2010ApJ} are represented on the colorbar as
  S$1\sigma$ and S$2\sigma$ respectively.  Markers, dashed contours,
  and pixels are as described for
  \fref{fig:deltas-potential-heatmap}.}
    \label{fig:deltas-radius-heatmap}
\end{figure}

The critical density $n_{\mathrm{cr}}$ for the appearance of $\Delta$s
is shown in \fref{fig:deltas-onset-heatmap} for the
$\sigma\omega\Delta$ coupling space.  As long as
\eref{eq:wehrberger-constraint} is satisfied, $\Delta$s appear prior
to the onset of hyperonization, and $n_{\mathrm{cr}}^{\mathrm{SWL}}
\lesssim 2.3\,n_0$, $n_{\mathrm{cr}}^{\mathrm{GM1L}} \lesssim
2.3\,n_0$, and $n_{\mathrm{cr}}^{\mathrm{DD2}} \lesssim 1.9\,n_0$.  If
we also enforce simultaneous satisfaction of
\eref{eq:drago-constraint} the critical densities could be as low as
$n_{\mathrm{cr}}^{\mathrm{SWL}} \approx 2\,n_0$,
$n_{\mathrm{cr}}^{\mathrm{GM1L}} \approx 1.9\,n_0$, and
$n_{\mathrm{cr}}^{\mathrm{DD2}} \approx 1.6\,n_0$.  Increasing
$x_{\sigma\Delta}$ leads to a gradual decrease in $n_{\mathrm{cr}}$
when $x_{\sigma\Delta}-x_{\omega\Delta} \gtrsim -0.1$, and a gradual
increase in $n_{\mathrm{cr}}$ when $x_{\sigma\Delta}-x_{\omega\Delta}
\lesssim -0.1$, the increase in the repulsive vector coupling
overcoming the increasingly attractive potential in the latter
case. However, increasing $x_{\sigma\Delta}-x_{\omega\Delta}$ leads to
an obvious and rapid decrease in $n_c$ for the entire
$\sigma\omega\Delta$ coupling space, and if
\eref{eq:wehrberger-constraint} is simultaneously satisfied an
increase in $x_{\sigma\Delta}-x_{\omega\Delta}$ also leads to a
significant reduction in the radius of the canonical $1.4 \,M_{\odot}$
NS as shown in \fref{fig:deltas-radius-heatmap}. For example,
the DD2 parametrization with nucleons (and hyperons) produces a
13.5~km radius for
the canonical $1.4 \,M_{\odot}$ NS and thus fails to satisfy the
13.2~km upper-radial constraint from \citet{Lattimer:2014EPJA}, but if
$\Delta$s are included and $x_{\sigma\Delta} > x_{\omega\Delta}$ the
radius is reduced sufficiently to satisfy the constraint. If we allow
\eref{eq:drago-constraint} to be violated resulting in a very
attractive $U_{\Delta}$, the inclusion of $\Delta$s makes it possible
for all three parametrizations to satisfy at least part of the
$2\sigma$ upper limit on the radial constraints from
\citet{Steiner:2010ApJ}.

\subsubsection{The $x_{\rho\Delta}$ Coupling}\label{sec:rho-Delta-coupling-space}

To examine the dependence of the NS mass, the $\Delta$ critical
density $n_{\mathrm{cr}}$, and the $\Delta$ fraction $f_{\Delta}$ on
the isovector meson-$\Delta$ coupling $x_{\rho\Delta}$ we set
$x_{\sigma\Delta} = x_{\omega\Delta} = 1.1$ and varied
$x_{\rho\Delta}$ in the range $0.5 < x_{\rho\Delta} < 2.5$
\citep{Spinella2017:thesis}. (Note that the saturation potential of the
$\Delta$ is determined in symmetric nuclear matter and is therefore
independent of $x_{\rho\Delta}$.)  The NS maximum mass was found to
not be terribly sensitive to $x_{\rho\Delta}$, decreasing over the
entire range by less than $1\%$ for the GM1L and DD2
parametrizations.  However, $n_{\mathrm{cr}}$ and $f_{\Delta}$ turned
out to be much more sensitive to changes in $x_{\rho\Delta}$ for the
GM1L parametrization due to the fact that the isovector contribution
to the chemical potential is much higher than for DD2.  The critical
density for GM1L increases from $\sim 2\,n_0$ to $\sim 3\,n_0$ across
the entire $x_{\rho\Delta}$ range, with a corresponding drop in
$f_{\Delta}$ of $\sim 8-9\%$ as this EOS reverts back to a nearly
purely hyperonic EOS. The $n_{\mathrm{cr}}$ of the DD2
parametrization increases very little from around $1.7\,n_0$ to
$1.9\,n_0$, but with an accompanying drop in $f_{\Delta}$ of almost
3\% down to about 8\%.  Overall, lower values of $x_{\rho\Delta}$ lead
to a lower critical density, resulting in higher fractions of
$\Delta$s to replace hyperons and lower the strangeness fraction,
increasing the NS maximum mass.

\section{General Relativistic Stellar Structure Equations}\label{sec:GR}

Neutron stars are objects of highly compressed matter so that the
geometry of surrounding space-time is changed considerably from flat
space.  Einstein's theory of general relativity is therefore to be
used when modeling the properties of NSs rather than Newtonian
mechanics. Einstein's field equation \index{Einstein's field equation}
is given by (we use units where the gravitational constant and the
speed of light are $G=c=1$)
\begin{eqnarray}
R^{\mu\nu} - \frac{1}{2} g^{\mu\nu} R = 8 \pi T^{\mu\nu} \, ,
\label{eq:einstein}
\end{eqnarray}
where $R^{\mu\nu}$ is the Ricci tensor, $g^{\mu\nu}$ the metric
tensor, $R$ the scalar curvature, and $T^{\mu\nu}$ the energy-momentum
tensor of matter.  The latter is given by
\begin{eqnarray}
T^{\mu\nu} = \left( \epsilon + P(\epsilon) \right) u^\mu u^\nu +
g^{\mu\nu} P(\epsilon) \, .
\label{eq:Tmunu.GR}
\end{eqnarray}
Models for the EOS, $P(\epsilon)$, which are input quantities in the
energy-momentum tensor equation, have been derived in
\sref{sec:eos}. These models will be used in this section to study the
properties of NSs.

\subsection{Non-rotating Proto-Neutron Stars}\label{ssec:non-rotPNS}
  
\begin{figure}[tb]
\centering
{
    \includegraphics[width=7.5cm]{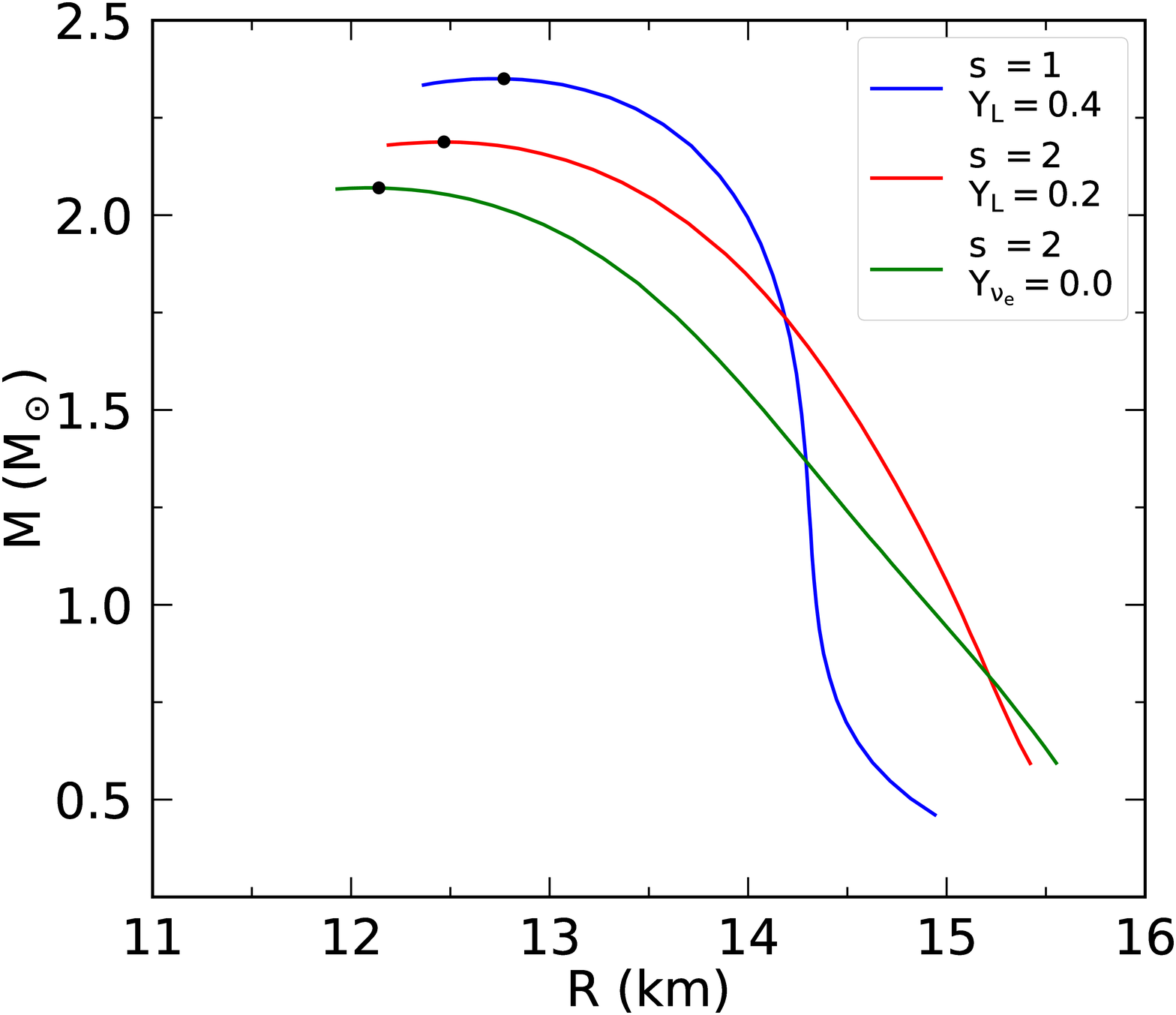}
    \caption{Mass-radius relationships of PNS computed for the DD2
      parametrization.}
    \label{fig:MR.PNS.DD2}
}{
    \includegraphics[width=7.5cm]{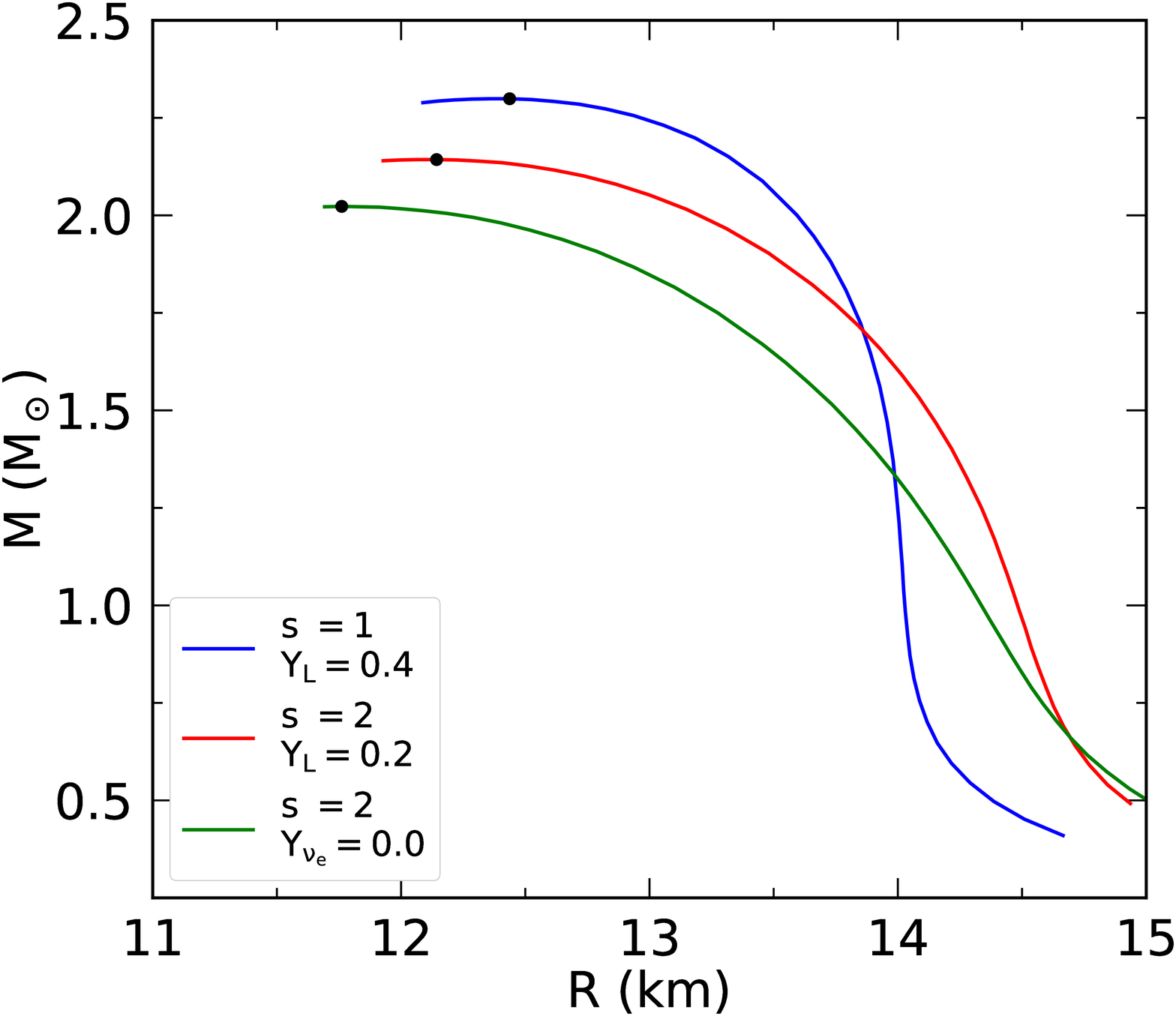}
    \caption{Same as \fref{fig:MR.PNS.DD2}, but computed for the GM1L
      nuclear parametrization.}
    \label{fig:MR.PNS.GM1L}
}
\end{figure}

We begin with non-rotating, spherically symmetric
NSs. They are relatively easy to study since the metric of such
objects depends only on the radial coordinate. The line element $ds^2$
in this case is given by the Schwarzschild metric \index{Schwarzschild
  metric} \citep{schwarzschild16:a,misner73:a,shapiro83:a}
\begin{eqnarray}
  ds^2 = -{\rm e}^{2\,\Phi(r)} dt^2 + {\rm e}^{2\,\Lambda(r)} dr^2 + r^2 \,
  (d\theta^2+{\rm sin}^2\theta \, d\phi^2) \, ,
\label{eq:schwarzschild}
\end{eqnarray}
where $\Phi(r)$ and $\Lambda(r)$ denote unknown metric functions
\index{Metric functions} whose mathematical form is determined by
Einstein's field equation \eref{eq:einstein} and the conservation of
energy-momentum, $\nabla_\mu T^{\mu\nu}=0$, and have the form
\begin{eqnarray}
{\rm e}^{2\,\Lambda(r)} &=&  \left( 1 - \frac{2 m(r)}{r} \right)^{-1} \quad
({\rm inside ~and~ outside~ of~star)}\, , 
\label{eq:f26}  \\
{\rm e}^{2\,\Phi(r)} &=& \left( 1 - \frac{2 m(r)}{r} \right) \qquad\; ({\rm
  only~ outside~of ~star})\, .
\label{eq:f27}
\end{eqnarray}
The solution of $\Phi(r)$ for the stellar interior
is given by
\begin{eqnarray}
  \frac{d\Phi(r)}{dr} = - \frac{1}{\epsilon + P(\epsilon)}
  \frac{dP(r)}{dr} \, ,
  \label{eq:dpdr}
\end{eqnarray}
where the pressure gradient is given by the Tolman-Oppenheimer-Volkoff
(TOV) \index{TOV equation} equation
\citep{oppenheimer39,tolman39,misner73:a},
\begin{eqnarray}
\frac{dP}{dr} = - \,  \frac{ \left( \epsilon(r) + P(r) \right) \, \left(
m(r) + 4 \pi r^3 P(r) \right)} { r^2 \left( 1 - 2 m(r)/r \right)} \, .
\label{eq:TOV}
\end{eqnarray}
The quantity $m(r)$ in \eref{eq:TOV} denotes the gravitational stellar
mass given by
\begin{eqnarray}
  m(r) = 4 \pi \int_0^r \! dr \; r^2 \, \epsilon(r) \, .
  \label{eq:mTOV}
\end{eqnarray}
The boundary condition associated with \eref{eq:TOV} specifies the
pressure at the stellar center, $P(r=0)$. \Eref{eq:TOV} is integrated,
for a given EOS, outward to a radial distance where the pressure
vanishes (turns negative). This defines the radius, $R$, of the
stellar model and the star's total gravitational mass is then given by
$M\equiv m(R)$. Figures~\ref{fig:MR.PNS.DD2} and \ref{fig:MR.PNS.GM1L}
show $M$ as a function of $R$ of proto-neutron stars computed for the
DD2 and GM1L models for the nuclear EOS. The most massive stars of our
sample are those with an entropy per baryon of $s = 1$ and a lepton
fraction of $Y_L = 0.4$, since the matter in their cores provides the
most pressure (see Figs.~\ref{fig:EoS_DD2_combined} and
\ref{fig:EoS_GM1L_combined}) of all the entropy--lepton number
combinations investigated in this work. The EOS computed for $s = 2$
and $Y_{\nu_e} = 0$ provides the least amount of pressure (i.e., is
the softest EOS of our collection) and therefore leads to the least
massive stars. We note that the mass-radius curves of all stars
lighter than $\sim 1\, M_\odot$ are not reliable since temperature
effects on the matter of the stellar crust have not been taken into
account in our calculations.

\subsection{Rotating Proto-Neutron Stars}\label{ssec:RPNS}

The properties of rotating compact objects are much more complicated to
study than those of non-rotating compact objects. This has its origin
in the fact that due to rotation the stars are deformed (so that the
metric functions depend on the polar angle $\theta$) as well as the
dragging of the local inertial frames\index{Frame dragging}, which is
caused by rotating systems in the general theory of relativity. The
line element accounting for these effects is given by
\citep{butterworth76:a,friedman86:a}
\begin{eqnarray}
 ds^2 = &-& {\rm e}^{2\,\nu(r,\theta,\Omega)}\, dt^2 +
 {\rm e}^{2\,\psi(r,\theta,\Omega)} \, (d\phi - \omega(r,\Omega) \,dt)^2 +
 {\rm e}^{2\,\mu(r,\theta,\Omega)} \, d\theta^2 \nonumber \\ &+&
 {\rm e}^{2\,\lambda(r,\theta,\Omega)}\, dr^2 \, .
\label{eq:metric.rot}
\end{eqnarray}
Here, $\omega$ is the angular frequency of the local inertial frames
which depends on the radial coordinate, $r$, and on the star's
rotational frequency $\Omega$. We take $\Omega$ to be constant
throughout the star's fluid (rigid-body rotation).  Of particular
interest in the discussion of rotating compact objects is the
effective angular frequency $\bar\omega \equiv \Omega -\omega$, which
is the angular frequency of the stellar fluid relative to the local
inertial frames.

\subsubsection{The General Relativistic Kepler Frequency}\label{sssec:Kepler}

No simple stability criteria are known for rotating star
configurations in general relativity. An absolute upper limit on rapid
rotation, is however set by the Kepler frequency \index{Kepler
  frequency} $\Omega_{\rm K}$, at which mass shedding from a star's
equator sets in. The expression of the general relativistic Kepler
frequency is derived from the line element shown in
\eref{eq:metric.rot}, evaluated at the equator of a compact stellar
object. Since $dr=0$ and $d\theta=0$ for a mass element rotating at
the equator, one obtains from \eref{eq:metric.rot} for the proper time
$d\tau^2$ $(= - ds^2)$ \index{Proper time} the relation
\begin{eqnarray}
  d \tau = \bigl( {\rm e}^{2\,\nu(r,\theta,\Omega)} -
  {\rm e}^{2\,\psi(r,\theta,\Omega)} \, (\Omega -
  \omega(r,\theta,\Omega))^2 \bigr)^{1/2} \, d t \, .
\label{eq:12.1bk}
\end{eqnarray}    
The equatorial orbit, which is the circular path with the maximum
possible distance from the center of a gravitating body, is obtained
from \eref{eq:12.1bk} by determining the extremum of the functional $J(r)$
associated with \eref{eq:12.1bk}, that is,
\begin{eqnarray}
  J(r) \equiv \int d t \; \bigl( {\rm e}^{2\,\nu(r,\theta,\Omega)} -
  {\rm e}^{2\,\psi(r,\theta,\Omega)} \, (\Omega -
  \omega(r,\theta,\Omega))^2 \bigr)^{1/2} \, .
\label{eq:12.2bk}
\end{eqnarray}
Applying the extremal condition $\delta J(r)=0$ to this functional
leads to 
\begin{eqnarray}
\delta \int d t \; \bigl( {\rm e}^{2\,\nu(r,\theta,\Omega)} - {\rm
  e}^{2\,\psi(r,\theta,\Omega)} \, (\Omega -
\omega(r,\theta,\Omega))^2 \bigr)^{1/2} = 0 \, ,
\label{eq:delta.int}
\end{eqnarray}
from which it follows that \citep{Weber:1999book} 
\begin{equation}
  \int d t \, \delta r \;
  \frac{\nu_{,r} \, {\rm e}^{2\,\nu} - \left(
        \psi_{,r} \, (\Omega - \omega) - \omega_{,r} \right) (\Omega -
        \omega)\,  {\rm e}^{2\,\psi}} {\bigl( {\rm e}^{2\,\nu} -
          {\rm e}^{2\,\psi} (\Omega - \omega)^2 \bigr)^{1/2} } = 0 \, .
\label{eq:12.5bk}
\end{equation}
For the sake of brevity, we suppress all arguments here and in the
following.  The subscripts $_{,r}$ on the metric functions and the
frame dragging frequency in \eref{eq:12.5bk} denote partial
derivatives with respect to the radial coordinate, $r$. Next, we introduce
the orbital velocity $V$ of a co-moving observer at the star's equator
relative to a locally non-rotating observer with zero angular momentum
in the $\phi$-direction. This velocity is given by
\begin{equation}
  V = {\rm e}^{\psi-\nu} \, (\Omega - \omega) \, .
\label{eq:Veq1}
\end{equation}
This relation is suggested by the expression of the time-like
component $u^t$ of the four-velocity of a mass element rotating in the
equatorial plane,
\begin{eqnarray}
u^t = \frac{dt}{d\tau} = {\rm e}^{-\nu} \; \left( 1 - V^2
\right)^{-1/2} \, ,
  \label{eq:ut}
\end{eqnarray}
where $V$ is given by \eref{eq:Veq}. Substituting \eref{eq:Veq1} into
\eref{eq:12.5bk} then leads to
\begin{eqnarray}
  {\psi_{,r}} \, {\rm e}^{2\,\nu} \, V^2 -
  {\omega_{,r}} \, {\rm e}^{\nu+\psi}\, V -
  {\nu_{,r}} \, {\rm e}^{2\,\nu} = 0 \, ,
\label{eq:12.6bk}
\end{eqnarray}
which guarantees that the integrand in \eref{eq:12.5bk} vanishes
identically for arbitrary variations $\delta r$. \Eref{eq:12.6bk}
represents a quadratic equation for the velocity $V$. The solutions
are given by \citep{friedman86:a} \index{Orbital velocity}
\begin{eqnarray}
V = \frac{\omega_{,r}} {2\,\psi_{,r}}\, {\rm e}^{\psi-\nu} \pm \biggl(
\frac{\nu_{,r}} {\psi_{,r}} + \biggl( \frac{\omega_{,r}}
     {2\,\psi_{,r}} \, {\rm e}^{2(\psi - \nu)} \biggr)^2 \biggr)^{1/2} \, .
\label{eq:Veq}
\end{eqnarray}
The general relativistic Kepler frequency, $\Omega_{\rm K}$, is then
obtained from (cf.\ \eref{eq:Veq1}) as
\begin{eqnarray}
\Omega_{\rm K} = {\rm e}^{\nu-\psi} \, V + \omega \, .
\label{eq:OmegaK}
\end{eqnarray}
We note that Eqs.~(\ref{eq:Veq}) and (\ref{eq:OmegaK}) need to be computed
self-consistently together with Einstein's field equations, which
determined the metric functions $\nu$ and $\psi$ and the frame
dragging frequency $\omega$ at an (initially unknown) equatorial
distance. The result of classical Newtonian mechanics for the
Kepler frequency and the velocity of a particle in a circular orbit,
$\Omega_{\rm K} = \sqrt{M/R^3}$ and $V=R \Omega$ respectively, are
recovered from Eqs.~(\ref{eq:Veq}) and (\ref{eq:OmegaK})  by neglecting
the curvature of space-time geometry, the rotational deformation of a
rotating star, and the dragging effect of the local inertial frames.

\begin{figure}[htb]
\centerline{\includegraphics[width=7.5cm]{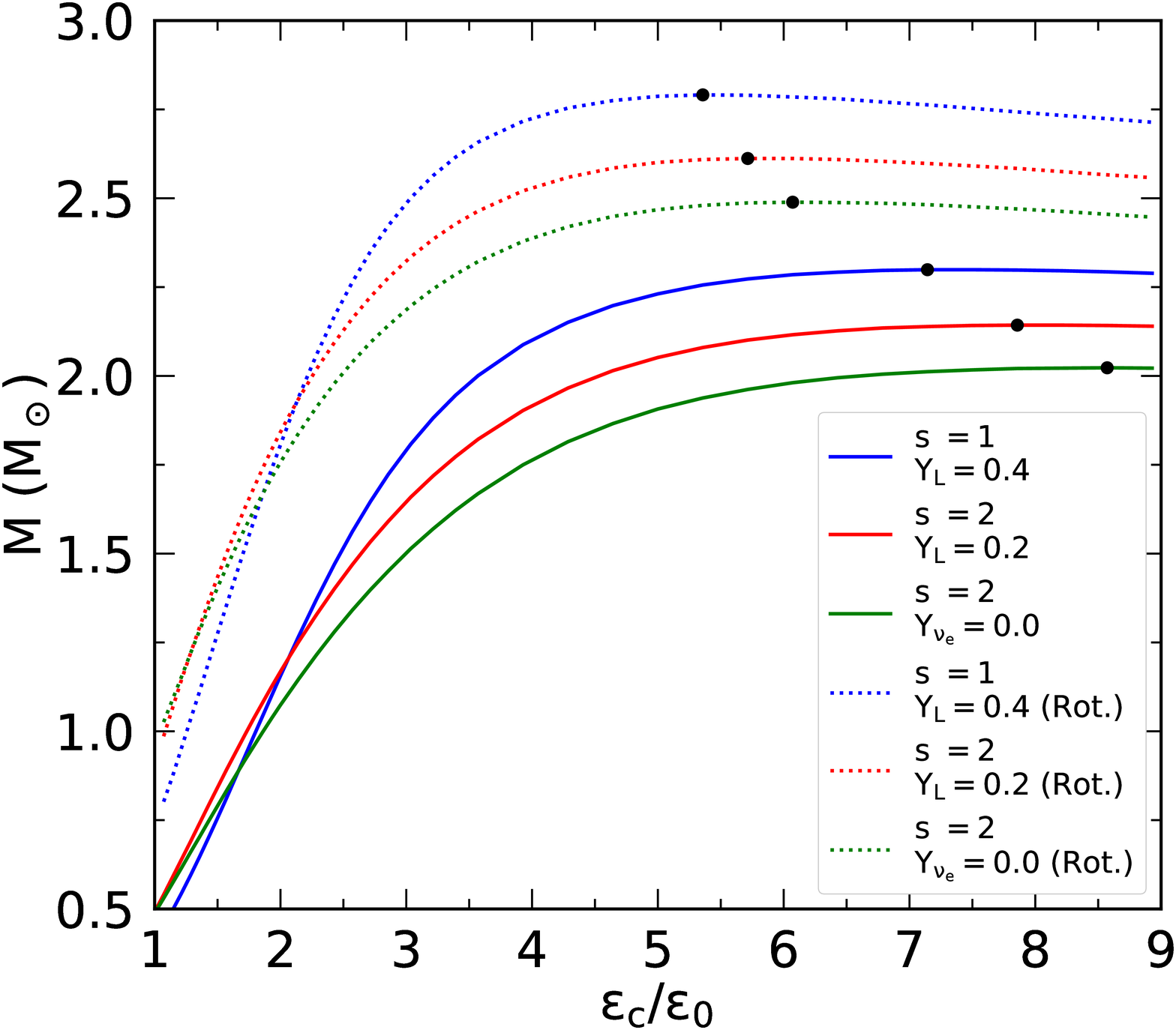}}
\caption{Gravitational mass of PNSs as a function of central energy
  density in units of the energy density of ordinary nuclear matter
  ($\epsilon_0=140$~MeV/fm$^3$) computed for GM1L.}
\label{fig:Mass_Density_GM1L}
\end{figure}

\Fref{fig:Mass_Density_GM1L} shows the impact of rapid rotation on the
gravitational masses of
proto-neutron stars. The solid lines in this figure show the masses of
non-rotating (i.e., TOV) stars. The dashed lines reveal by how much
these masses increase if the stars are rotating at the highest
possible spin rate, which is the Kepler frequency given by
\eqref{eq:OmegaK}. The increase in mass of cold NSs is typically at
the 20\% level, depending on the EOS \citep{friedman86:a}. The same is
the case for the gravitational masses of proto-neutron stars, as can
be inferred from \fref{fig:Mass_Density_GM1L}.  We also note that the
stars' central energy density, $\epsilon_c$, decreases with rotation
speed, because of the additional rotational pressure in the radial
outward direction created by rotation.

As mentioned just above, to find the Kepler frequency $\Omega_{\rm K}$
\index{Kepler frequency} (Kepler period, $P=2\pi/\Omega_{\rm K}$) of a
compact star, Eqs.~(\ref{eq:Veq}) and (\ref{eq:OmegaK}) and are to be
computed self-consistently in combination with the differential
equations for the metric and frame-dragging functions in
\eref{eq:metric.rot}, which follow from Einstein's field equation
\eqref{eq:einstein}. The entire set of coupled equations is to be
evaluated at the equator of the rotating star
\citep{friedman86:a,Weber:1992ApJ}, which is not known at the
beginning of the computational procedure. Here, the results of stars
rotating at the Kepler period are computed in the framework of
Hartle's perturbative rotation formalism \citep{Weber:1992ApJ}.  The
latter constitute a perturbative approach to Einstein's field
equations, that leads to results that are in very good agreement with
those obtained by numerically exact treatments of Einstein's field
equations. This is particularly the case for the mass increase due to
fast rotation and the value of the Kepler frequency.

\begin{figure}[tb]
\centering
{
    \includegraphics[width=7.5cm]{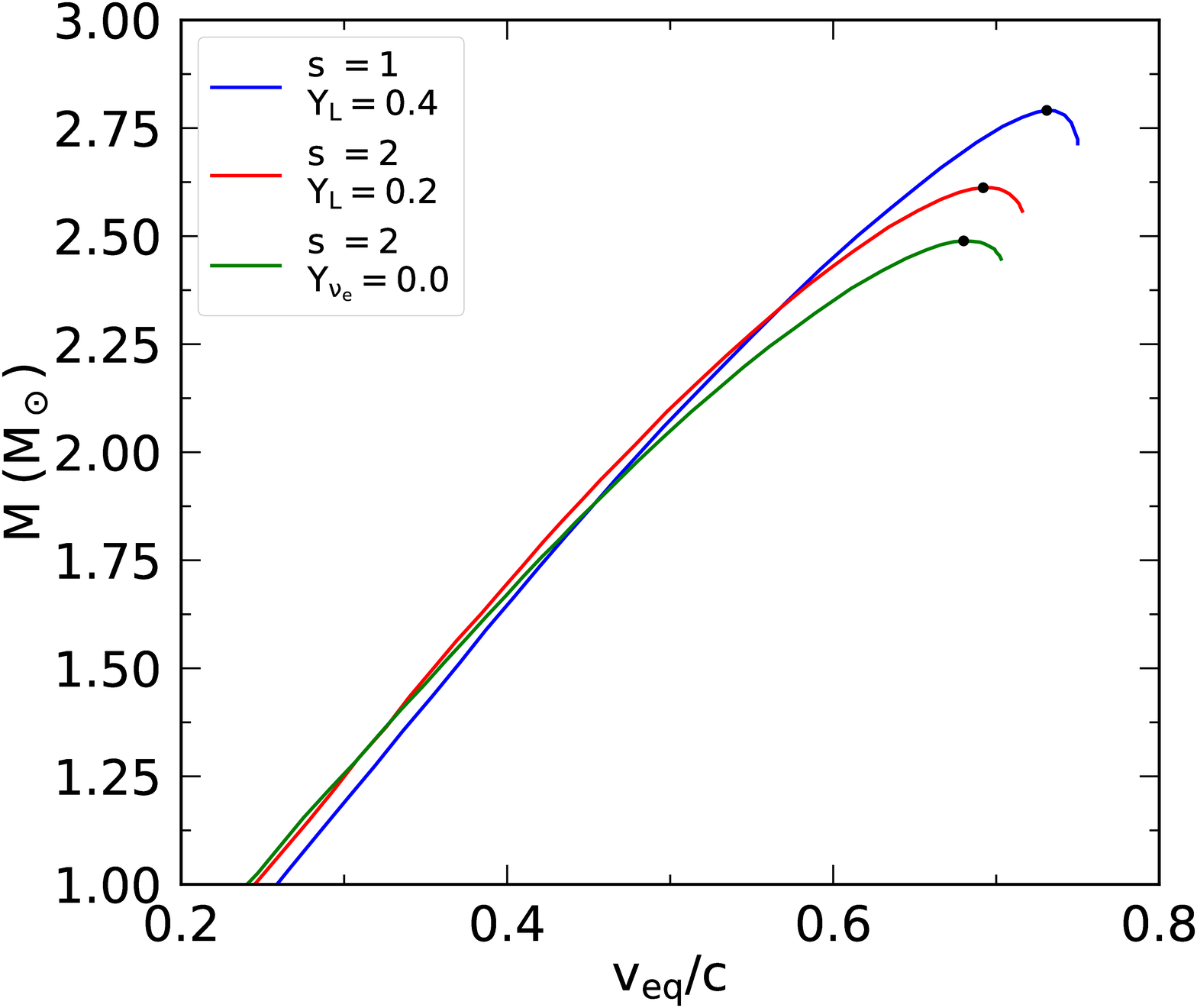}
    \caption{Gravitational mass of rotating PNSs versus equatorial speed for
      GM1L EOS.}
    \label{fig:MvsV_GM1L}
}{
    \includegraphics[width=7.5cm]{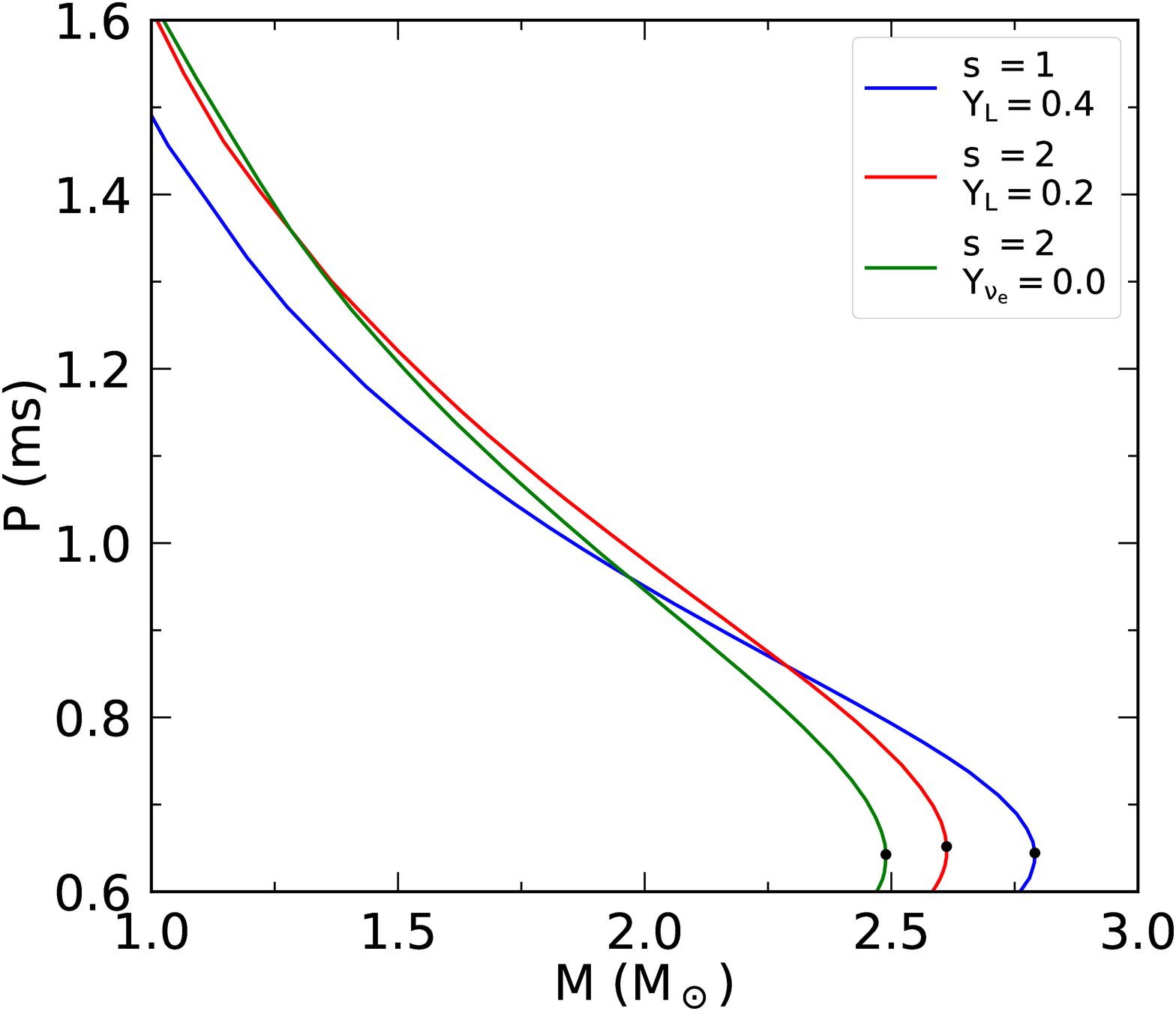}
    \caption{Kepler periods of rotating PNs versus
      gravitational mass for the GM1L EOS.}
    \label{fig:PvsM_GM1L}
}
\end{figure}

\Fref{fig:MvsV_GM1L} shows that the maximum possible rotational
(Kepler) speed at the equator of PNSs is around $0.7\, c$. This value
depends only mildly on the actual stellar composition. One also sees
that the hottest of the three stellar families (green line) terminates
(solid black dot) at an equatorial speed that is smaller than the
speed of the other two sequences. This has its origin in the fact that
$s=2$ and $Y_{\nu_e}=0$ PNSs have the highest temperatures of all three
configurations and thus the biggest radii, so that mass-shedding from
the equator sets in first in these stars.

The Kepler periods of the rotating PNSs of
Figs.~\ref{fig:Mass_Density_GM1L} through \ref{fig:MvsV_GM1L} are
displayed in \fref{fig:PvsM_GM1L}. Based on the results shown in this
figure, we conclude that PNSs posses about the same Kepler periods
as cold neutron stars.

\subsubsection{Gravitational Radiation-Reaction Driven Instabilities}\label{sssec:GRRI}

Besides the absolute upper limit on rapid rotation set by the Kepler
frequency, there are other instabilities that have been shown to set
in at a lower rotational frequency, and which therefore set more
stringent limits on stable rotation
\citep{Lindblom1986,Owen1998,Andersson:1999ApJ,Andersson:2001IJMPD,Lin2021}. They
typically originate from counter-rotating surface vibrational modes,
which at sufficiently high rotational star frequencies are dragged
forward.  In this case, gravitational radiation which inevitably
accompanies the aspherical transport of matter does not damp the
modes, but rather drives them
\citep{Chandrasekhar1970.PRL,Friedman1983,Friedman1983:a}.  Bulk and
shear viscosity play the important role of damping such
gravitational-wave radiation-reaction instabilities \index{GRR driven
  instabilities} at a sufficiently reduced rotational frequency such
that the viscous damping rates and power in gravity waves are
comparable
\citep{Lindblom1977,Andersson:2001IJMPD}. Theoretical
    studies suggest that either the $f$-modes or the $r$-modes
    determine the maximum rotation frequency of neutron stars.

\subsection{The Moment of Inertia}\label{ssec:MOI}

\begin{figure}[htb]
\centering
{
    \includegraphics[width=7.5cm]{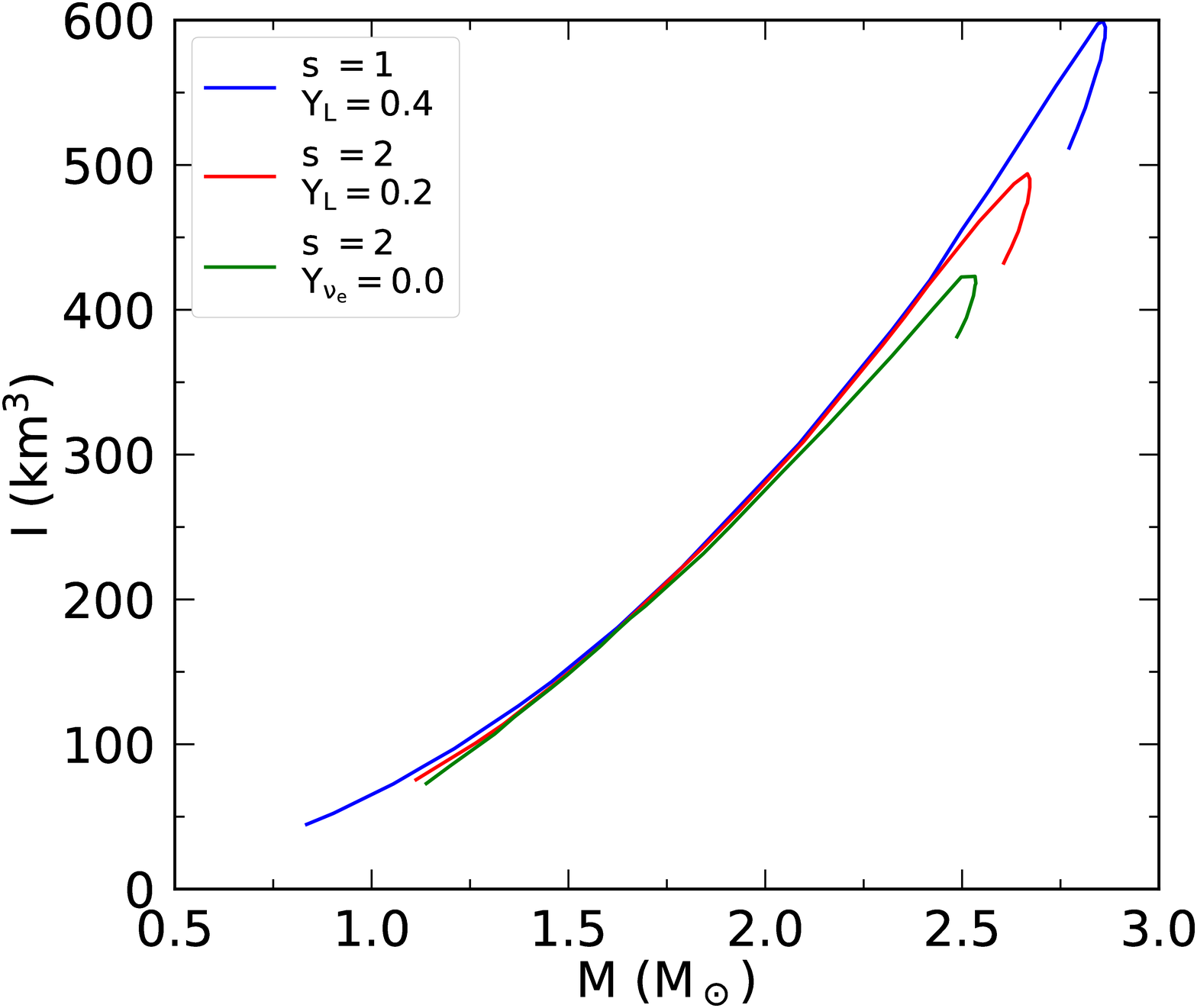}
    \caption{Moment of inertia versus gravitational mass, for the DD2
      parameter set.}
    \label{fig:MOI.DD2}
}{
    \includegraphics[width=7.5cm]{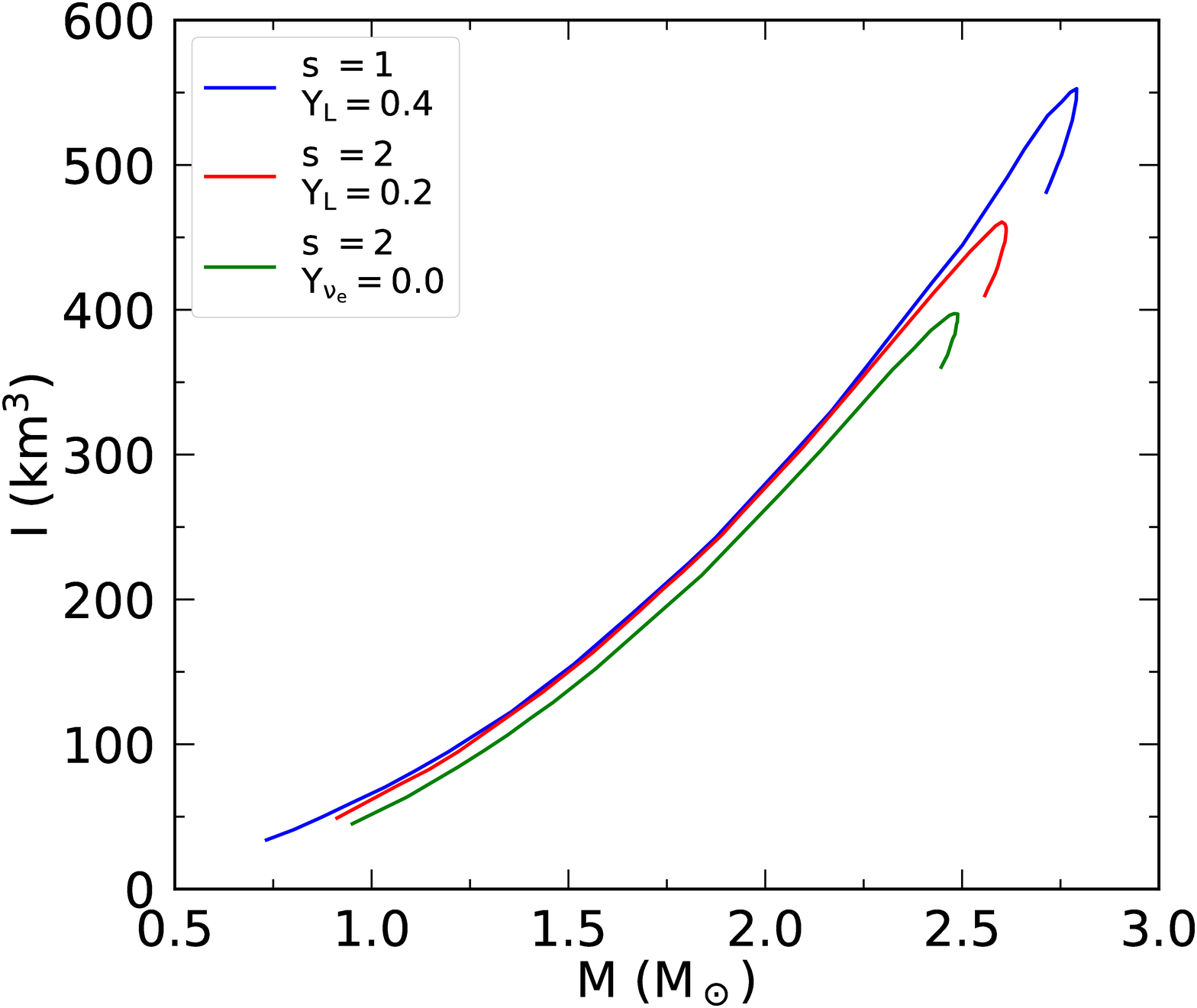}
    \caption{Same as \fref{fig:MOI.DD2}, but for the GM1L parameter
      set.}
    \label{fig:MOI.GM1L}
}
\end{figure}

Another very important stellar quantity, which will be discussed in
this section, is the moment of inertia, \index{Moment of inertia}
$I$. This quantity is given by \cite{Hartle1973:a}
\begin{eqnarray}
  I(\Omega) = \frac{1}{\Omega} \int_A\!\! dr\, d\theta\, d\phi \;
  T_{\,\phi}{^t}(r,\theta,\phi; \Omega)\, \left( - g(r,\theta,\phi;
    \Omega) \right)^{1/2} \, ,
\label{eq:11.64bk}
\end{eqnarray}
where $A$ denotes the region inside of a compact stellar object
rotating at a uniform angular velocity, $\Omega$. The quantity $g$
denotes the determinant of the metric tensor $g_{\mu\nu}$, whose
components can be read off from \eref{eq:metric.rot}. One obtains
\citep{Weber:1999book} 
\begin{equation}
  \sqrt{-g} = {\rm e}^{\lambda + \mu + \nu +\psi} \, .
  \label{eq:minusg}
\end{equation}
  The energy-momentum tensor component $T_{\phi}{}^t$ is given by (see
  \eref{eq:Tmunu.GR})
\begin{equation}
  T_{\phi}{}^t = (\epsilon + P) \; u_\phi \, u^t \, ,
\label{eq:11.TPt}
\end{equation}
with the four-velocities $u_\phi$ and $u^t$ \index{Four-velocity}
given by
\begin{eqnarray}
  u^t &=& \frac{{\rm e}^{-\nu}} {\left( 1 - (\omega - \Omega)^2\,
        {\rm e}^{2\psi -2\nu} \right)^{1/2}}  \, ,
  \label{eq:11.68bk} \\
u_\phi &=& (\Omega - \omega) \; {\rm e}^{2\, \psi} \; u^t \, .
\label{eq:11.69bk}
\end{eqnarray}
Substituting Eqs.~(\ref{eq:11.68bk}) and (\ref{eq:11.69bk}) into
\eref{eq:11.TPt} leads for $T_{\phi}{}^t$ to
\begin{eqnarray}
  T_{\phi}{}^t = \frac{(\epsilon + P)\, (\Omega - \omega)\,
      {\rm e}^{2\,\psi}} { {\rm e}^{2\, \nu} - (\omega - \Omega)^2\,
      {\rm e}^{2\, \psi} } \, .
\label{eq:11.70bk}
\end{eqnarray}
Substituting the expression given in \eref{eq:minusg} and
\eref{eq:11.70bk} into \eref{eq:11.64bk} leads for the moment of
inertia of a rotationally deformed compact stellar object to
\citep{Weber:1999book} \index{Moment of inertia}
\begin{eqnarray}
  I(\Omega) = 2 \pi \! \int_0^\pi \! d\theta \int_0^{R(\theta)} \!\! d r \,
  {\rm e}^{\lambda+\mu+\nu+\psi} \, \frac{\epsilon + P(\epsilon)} {{\rm e}^{2\nu -
  2\psi} - (\omega - \Omega)^2} \, \frac{\Omega - \omega}{\Omega}\, .
\label{eq:11.71bk}
\end{eqnarray}
In Figs.~\ref{fig:MOI.DD2} and \ref{fig:MOI.GM1L} we show the results for the
moment of inertia computed for the DD2 and GM1L nuclear parametrizations.
As can be seen, massive PNSs with $s=1$ and $Y_L=0.4$ posses the
largest moments of inertia, followed by PNS with $s=2$ and $Y_L=0.2$,
and $s=2$ and $Y_{\nu_e}=0.0$. This trend is consistent with the
stiffness of the PNS equations of state shown in
Figs.~\ref{fig:EoS_DD2_combined} through \ref{fig:Pn.GM1L}, the
associated mass-radius relationships shown in
Figs.~\ref{fig:MR.PNS.DD2} and \ref{fig:MR.PNS.GM1L}, and the
mass-central density curves shown
\fref{fig:Mass_Density_GM1L}. Figures~\ref{fig:MOI.DD2} and
\ref{fig:MOI.GM1L} show that $2\, M_\odot$ PNSs, for instance, possess
moments of inertia of around 270~km$^3$. It is interesting to compare
this value with the moment of inertia of a spherical uniform mass
distribution in Newtonian physics, which is given by $I = \frac{2}{5}
M R^2 = \frac{3}{5} (\frac{M}{M_\odot}) (\frac{R}{\rm km})^2 ~{\rm
  km}^3$. This gives values for $I$ for the $2\, M_\odot$ PNSs in
Figs.~\ref{fig:MR.PNS.DD2} and \ref{fig:MR.PNS.GM1L} which are between
175~km$^3$ and 235~km$^3$. These values are on the same order of
magnitude as the general relativistic results, but differ
significantly quantitatively.  Sudden changes $\Delta I$ in the
moments of inertia of (proto-) neutron stars are expected to lead to
sudden changes $\Delta \Omega$ in the spin frequencies of such
objects. Assuming angular momentum conservation during such episodes,
the change in spin frequency may be estimated as
\begin{equation}
  \frac{\Delta \Omega}{\Omega} = - \frac{\Delta I}{I + \Delta I} \, ,
\label{eq:delOmega}
\end{equation}
where $\Omega$ is the frequency of the star before the change in the
moment of inertia. Observed changes in the spin frequencies of
rotating neutron stars (pulsars) range from $10^{-6}$ to $10^{-9}$ and
have been suggested to originate from sudden decreases in the moments
of inertia of such objects
\citep{Lyne1992,Fuentes2017,Manchester2017,Montoli:2022astrophysics}. Similar
arguments have been used by \citet{2015MNRAS.449L..73G} and
\citet{2015MNRAS.453..522M} to explain the anti-glitch
\index{Anti-glitch} observed for the magnetar AXP~1E~2259+586
\cite{2013Natur.497..591A}.  \Eref{eq:delOmega} indicates that already
$|\Delta I| \sim 10^{-6}$~km$^3$ to $10^{-9}$~km$^3$ could suffice to
cause such changes. This is also the reason why the nuclear crust on
hypothetical strange quark stars could explain the observed pulsar
glitches \cite{Glendenning1992}. 

\section{Future Directions of Research}\label{sec:future}

It is often stressed that there has never been a more exciting time in
the overlapping areas of nuclear physics, particle physics,
relativistic astrophysics and astronomy than today. This interest is
stimulated by investments made in international nuclear physics
facilities such as FAIR, FRIB, NICA, CERN, BNL, J-Park and multiple
new instruments for sky surveys that have become operational in recent
years, such as FAST, eROSITA, NICER and the gravitational-wave
detectors LIGO, VIRGO, KAGRA. In particular, the observation of the
first binary neutron star merger, GW170817, using LIGO and VIRGO
\citep{Abbott:2017PRL} have led the scientific community into the new era
named multi-messenger astronomy with gravitational waves.  

Depending on the combined masses of two merging NSs, there are in
principle four possible outcomes to a merger \citep{Chirenti:2019ApJ}:
1) prompt formation of a black hole, 2) formation of a hypermassive NS
(HMNS) \citep{Baumgarte:2000ApJ,Shapiro:2000ApJ}, 3) formation of a
supramassive rotating NS \citep{Falcke:2014AA}, or 4) the formation of a
stable NS.  Numerical relativity simulations have shown that the
threshold masses related to these scenarios depend strongly on the
properties and the EOS of hot and dense NS matter. (For a
comprehensive review of the physics of NS mergers, see
\citet{Baiotti:2017RPP}, and references therein.) The same is true for
the lifetime of HMNSs, which depends strongly on the total mass of the
binary system and, thus, on the nuclear EOS.  The post-merger
emissions are typically characterized by two distinct frequency peaks,
one at high and the other at lower frequencies.  The EOS dependent
high-frequency peak is believed to be associated with the oscillations
of the HMNS produced in a merger, while the low-frequency peak is
understood to be related to the merger process and to the total
compactness (i.e., mass-radius ratio) of the merging objects
\citep{Takami:2015PRD}, which is inexorably linked to the EOS of dense
nuclear matter. 

The EOS of cold and dense nuclear matter is sufficient to describe NS
matter prior to a NS merger.  After contact, however, large shocks
develop which considerably increase the internal energy of the
colliding NSs. Numerical simulations have shown that overall matter in
NS collisions reaches densities that are several times higher than the
nuclear saturation density and temperatures that are roughly as high
as 50~MeV
\citep{Baiotti:2017RPP,Hanauske:2019a.universe,Perego:2019EPJA}. As
shown in this chapter, such extreme conditions of density and pressure
modify the EOS and in particular the baryon-lepton composition of the
matter tremendously.

Very recently, is has been shown that a strong first-order phase
transition in NS mergers may register itself in the gravitational-wave
frequency, $f_{\rm peak}$, and the stellar tidal deformability,
$\Lambda$ \cite{Bauswein:2019.PRL}.  Since both the tidal
deformability during inspiral and the oscillation frequencies of the
post-merger remnant can be determined very reliably
\cite{Faber:2012,Baiotti:2017RPP,Paschalidis:2017,Friedman:2018,Duez:2018},
this finding relates NS merger simulations to the general science
question whether or not phase transitions occur in dense nuclear
matter.  Signatures of possible hadron-quark phase transitions in NS
mergers have also been studied by \citet{Most:2019.PRL}. This study
shows that changes in the pressure of the quark phase can produce a
decisive signature in the post-merger gravitational-wave signal and
spectrum. It was also shown that a hadron-quark phase transition may
lead to a hot and dense quark core which could produce a ring-down
signal different from what is expected for a pure hadronic core.  The
possibility of detecting the hadron-quark phase transition with
gravitational waves has been discussed recently by
\citet{Hanauske:2019b.universe}.

A great deal of experimental, theoretical as well as computational
work will need to be carried out over the coming years to determine a
comprehensive class of state-of-the-art models for the EOS of
ultra-hot and dense nuclear matter for use in binary NS merger
simulations and PNS simulations
\citep{Shen:2011PRC,Rezzolla:2014Hydro,Banik:2014ApJ,
  Hanauske:2019b.universe}.


\noindent
\section*{Acknowledgments}
  This research was supported by the National Science Foundation (USA)
  under Grants No.\ PHY-1714068 and PHY-2012152. MO and IFR-S thank
  CONICET, UNLP, and MinCyT (Argentina) for financial support under
  grants PIP-0714, G157, G007 and PICT 2019-3662. The results of this
  paper contribute to the research projects of the NP3M
  collaboration on the Nuclear Physics of Multi-Messenger Mergers.


\printindex

\end{document}